\documentclass[useAMS,twocolumn,usenatbib]{mn2e}
\pdfoutput=1
\usepackage{times,psfig,epsfig} 
\usepackage{rotating, graphicx}
\usepackage{color}
\usepackage{amssymb, amsmath}
\usepackage[normalem]{ulem}
\usepackage[titletoc]{appendix}

\newcommand{\bdm}{\begin{displaymath}} 
\newcommand{\edm}{\end{displaymath}}
\newcommand{\beq}{\begin{equation}} 
\newcommand{\eeq}{\end{equation}} 
\newcommand{\beqnarr}{\begin{eqnarray}}
\newcommand{\eeqnarr}{\end{eqnarray}}
\newcommand{\bit}{\begin{itemize}} 
\newcommand{\eit}{\end{itemize}} 
\newcommand{\ben}{\begin{enumerate}} 
\newcommand{\een}{\end{enumerate}}
\newcommand{\bfi}{\begin{figure}[htb]} 
\newcommand{\bpfi}{\begin{figure}[p]}
\newcommand{\barr}{\begin{array}}
\newcommand{\earr}{\end{array}}
\newcommand{\bec}{\begin{center}}
\newcommand{\eec}{\end{center}}
\newcommand{\bs}{\begin{sideways}}
\newcommand{\es}{\end{sideways}}

\newcommand{\tmw}{T_{\rm mw}}
\newcommand{\tsl}{T_{\rm sl}}
\newcommand{\mg}{M_{\rm g}}
\newcommand{\yx}{Y_{\rm X}}

\newcommand{\ovisc}{{{\sc nr}}}
\newcommand{\agn}{{{\sc agn}}}
\newcommand{\csf}{{{\sc csf}}}

\newcommand{\lgt}{\textrm{log}_{10}}
 
 \newcommand{\mincir}{\raise
  -2.truept\hbox{\rlap{\hbox{$\sim$}}\raise5.truept \hbox{$<$}\ }}
\newcommand{\magcir}{\raise
  -2.truept\hbox{\rlap{\hbox{$\sim$}}\raise5.truept \hbox{$>$}\ }}
\newcommand{\siml}{\raise
  -2.truept\hbox{\rlap{\hbox{$\sim$}}\raise5.truept \hbox{$<$}\ }}
\newcommand{\simg}{\raise
  -2.truept\hbox{\rlap{\hbox{$\sim$}}\raise5.truept \hbox{$>$}\ }}

\title[Evolution of X-ray scaling relations]{Cosmological hydrodynamical simulations of galaxy clusters: X-ray scaling relations and their evolution}
\author[N. Truong et al]{
N.~Truong$^{1,2}$,
E.~Rasia$^{3,4}$,
P.~Mazzotta$^{1,5}$,
S. Planelles$^{6,7}$,
V.~Biffi$^{4,6}$,
D. Fabjan$^{4,8}$,
  \newauthor 
A.~M.~Beck$^{9}$,
  S.~Borgani$^{4,6,10}$,
K. Dolag$^{9}$,
M. Gaspari$^{*,11}$,
G. L. Granato$^{4}$,
G. Murante$^{4}$,
  \newauthor 
C. Ragone-Figueroa$^{4,12}$,
L. K. Steinborn$^{9}$
  \\~\\
\footnotesize
$^1$ Dipartimento di Fisica, Universit\`a di Roma Tor Vergata, via della Ricerca Scientifica, I-00133, Roma, Italy \\
$^2$ MTA-E\"otv\"os University Lend\"ulet Hot Universe Research Group, P\'azm\'any P\'eter s\'et\'any 1/A, Budapest, 1117, Hungary \\
$^3$  Department of Physics, University of Michigan, 450 Church St., Ann Arbor, MI  48109 \\
$^4$ INAF, Osservatorio Astronomico di Trieste, via Tiepolo 11, I-34131, Trieste, Italy \\
$^5$ Harvard-Smithsonian Center for Astrophysics, 60 Gardner Street, Cambridge, MA 02138, USA \\
$^6$ Dipartimento di Fisica dell' Universit\`a di Trieste, Sezione di Astronomia, via Tiepolo 11, I-34131 Trieste, Italy \\
$^{7}$ Departamento de Astronom\'ia y Astrof\'isica, Universidad de Valencia, c/Dr. Moliner, 50, 46100 Burjassot, Valencia, Spain \\
$^{8}$ Faculty of Mathematics and Physics, University of Ljubljana, Jadranska 19, 1000 Ljubljana, Slovenia \\
$^{9}$ University Observatory Munich, LMU, Scheinerstr. 1, D-81679, Munich, Germany \\
$^{10}$ INFN, Instituto Nazionale di Fisica Nucleare, Trieste, Italy \\
$^{11}$ Department of Astrophysical Sciences, Princeton University, 4 Ivy Ln, Princeton, NJ 08544-1001, USA\\
$^{12}$ Instituto di Astronom\'ia Te\'orica y Experimental (IATE), Consejo Nacional de Investigaciones Cient\'ificas y T\'ecnicas de la Rep\'ublica Argentina (CONICET), \\
\ \ \ \ \ Observatorio Astron\'omico, Universidad Nacional de C\'orboda, Laprida 854, X50000BGR, C\'ordoba, Argentina\\
$^{*}$ Einstein and Spitzer Fellow
}

\begin{document}
\maketitle 

\begin{abstract} 
  We analyse cosmological hydrodynamical simulations of galaxy
  clusters to study the X-ray scaling relations between total masses
  and observable quantities such as X-ray luminosity, gas mass, X-ray
  temperature, and $Y_X$.  Three sets of simulations are performed
   with an improved version of the smoothed-particle-hydrodynamics
  GADGET-3 code. These consider the following: non-radiative gas, star
    formation and stellar feedback, and the addition of
    active-galactic-nucleus (AGN) feedback. 
  We select clusters with $M_{500}>10^{14} M_{\odot} E(z)^{-1}$,
    mimicking the typical selection of Sunyaev--Zeldovich samples. This permits to have a
    mass range large enough to enable robust fitting of the relations even at $z\sim2$.
  The results of the
  analysis show a general agreement with observations.
  The values of the slope of the mass -- gas mass and 
  mass $-$ temperature relations at $z=2$ are 10 per cent lower with respect to $z=0$ due 
  to the applied mass selection, in the former case, and to the effect of early merger in the latter. 
  We investigate the impact of the slope variation on the
  study of the evolution of the normalization.  We conclude that 
    cosmological studies through scaling relations should be limited
    to the redshift range $z=0-1$, where we find that the slope, the
  scatter, and the covariance matrix of the relations are stable.
  The scaling between mass and $Y_X$ is
  confirmed to be the most robust relation, being almost independent
  of the gas physics.  At higher redshifts, the scaling
    relations are sensitive to the inclusion of AGNs which influences
    low-mass systems. The detailed study of these objects will be
    crucial to evaluate the AGN effect on the ICM.
\end{abstract}

\section{Introduction}\label{sec:intro} 

Clusters of galaxies are the largest gravitationally bound
  structures in our Universe and according to the hierarchical process
  of structure formation they are the latest to form.  Due to these
  characteristics, they can provide stringent constraints on the
cosmological parameters (such as the amplitude of the linear
  power spectrum, the amount of matter, and that of dark energy) that
determine the growth rate of structures
\citep[e.g.][]{2011ASL.....4..204B,2015SSRv..188...93P}.

In this respect, one of the most powerful cosmological measurements is the
{\it evolution} of the mass function
\citep{borgani_guzzo,voit2005,2009ApJ...692.1033V,allen.etal.2011}.
However, the measurement of cluster masses via X-ray or
gravitational lensing is complicated and questioned by the presence of
possible biases caused by several factors, such as lack of hydrostatic
equilibrium or triaxiality. For this reason, in order to infer masses
for a {\it large} number of objects, it is preferable to resort to
relations between the total mass and some observable
quantities that are relatively easy to measure, the so-called mass
proxies
\citep{2012MNRAS.424.2086H,takey.etal.2013,2015arXiv151203833G}.  From
an observational point of view, these relations need to be
  calibrated by measuring the total masses, via weak-lensing or X-ray
analyses, for a smaller, but optimal, set of galaxy clusters.

In X-ray studies, the most commonly used mass-proxies are the {\it gas
  mass}, $\mg$, which can be extracted from the surface
brightness profile, the {\it temperature}, $T$, which is solidly
estimated from X-ray spectra with, at least 1000 counts, and
their combination, $Y_X=\mg \times T$ \citep[e.g., ][ for recent and
  detailed studies on X-ray scaling
  relations]{2014MNRAS.437.1171M,mantz.etal.2016}.  The $Y_X$
parameter was first introduced as a mass proxy closely related to the
total thermal content of the intracluster medium (ICM) by
\cite{2006ApJ...650..128K}. In that paper, through the analysis
  of hydrodynamical simulations, the authors proved the advantages of
  this quantity as a mass proxy over gas mass and X-ray
temperature. Specifically, gas mass and temperature react in
  opposite directions to any breaking of self-similarity (see
Section~3.2) caused, for example, by non-gravitational effects.
  In the computation of the product of the two quantities, the
  deviations from the self-similar (SS) behavior compensate each other,
  keeping the $Y_X$ evolution closer to the expected (SS)
one.  In addition, the $M-\yx$ relation is characterized by a
  small scatter. Indeed, the gas mass and temperature respond in
  opposite ways to the effects of, e.g., mergers and
  energy feedback from active galactic nuclei (AGN)
  \citep{2006ApJ...650..128K, 2011MNRAS.416..801F}.  In case of an
encounter, the gas mass immediately increases while the temperature,
at first, decreases due to the presence of the smaller and
thus colder structure \citep{poole.etal.2007,rasia.etal.2011}. The
feedback by AGNs reduces the gas mass by expelling some gas from the
core and, at the same time, it heats the ICM.  These opposite and
  compensating responses of the two quantities make their product,
  $Y_X$, independent of the dynamical state or on the central AGN
activity (see, however, \citealt{lebrun.etal.2014} for a different
conclusion).  For all these reasons, $Y_X$ has
  been widely adopted in cosmological applications of galaxy clusters.

Another X-ray measurement frequently represented in the analysis
  of scaling relations is the {\it X-ray luminosity} because it can
easily be derived from few tens of net photon counts \citep[e.g.][and
  references therein]{giles.etal.2017}. The $L-T$ relation has been
 historically important because from its first determination it
was clear that it provides information on the physics of the cluster
core \citep[e.g.,][]{fabian.etal.1994} and on the phenomena of
feedback by stars or AGNs \citep[e.g.,][]{maxim98,2012MNRAS.421.1583M}.
These connections are also the origin of the large scatter of
this relation as well as of the associated $M-L_X$ relation.
Due to this characteristic, the appeal of $L_X$ as mass-proxy is
limited. However, the inverted relation, i.e., between the luminosity
and the mass (the $L_X-M$ relation), still plays an important role in
establishing the selection function of X-ray surveys
\citep[e.g.][]{nord.etal.2008, 2009ApJ...692.1033V,allen.etal.2011},
since it determines the connection between the survey flux limit and
the minimum mass that can be observed at a given redshift.

 \vspace{0.2cm}

From this discussion, it is clear that in the past 10-15
  yr, scaling relations have been largely studied in observational
  samples.  Up to date, their analysis has rarely been extended beyond
  $z\sim 0.5-0.6$ (\citealt{2011A&A...535A...4R,2012MNRAS.421.1583M,
    2013SSRv..177..247G}).  
  The collection of high-redshift systems will grow thanks to future
  optical missions like {\it
    eROSITA}\footnote{http://www.mpe.mpg.de/eROSITA}\citep{2012arXiv1209.3114M},
  {\it Euclid}\footnote{http://www.euclid-ec.org}
  \citep{laureijs.etal.2011},
  LSST\footnote{http://www.lsst.org}\citep{Ivezic.etal.2008}, and
    to millimetric surveys such as SPT-3G \citep{benson.etal.2014} and CMB-S4. 
   These identify clusters 
  through the small distortions of 
  the cosmic microwave background (CMB) radiation caused by the 
  inverse Compton scattering of the CMB photons that interact with the ICM 
  electrons. The phenomenon, called Sunyaev--Zeldovich (SZ) effect,
  has already enabled the detection of a good number of objects at $z \ge 1$
  \citep{menanteau.etal.2013,bleem.etal.2015}.
  Once the
  clusters will be detected, a possible follow-up will be provided by
  current or future X-ray observatories, first of all, {\it
    Athena}\footnote{http://www.the-athena-x-ray-observatory.eu}.  The
  selection functions characterizing the samples from these future
  surveys are very different one from the others
  \citep{weinberg.etal.2013, ascaso.etal.2017}: the SZ-selected
  samples, such as those of the South Pole Telescope (SPT) or the Atacama Cosmology Telescope (ACT), 
  extend to less massive objects
  at higher redshifts and, thus, are the most suitable for
  high-redshift searches. The limiting mass of the mentioned optical
  surveys, instead, will be almost constant up to $z\sim1$ and then
  will grow at earlier epochs. The efficiency of optical detection is,
  therefore, expected to drop at $z=1-1.2$. Finally, the forecast for
  {\it eROSITA} limits the cluster discovery at $z\sim 0.8-1$ because
  of the dimming of the X-ray emission at large distances.

\vspace{0.2cm}
Over the last decade, the theoretical community has also spent a
  significant effort in the modelling of scaling relations by taking
  advantage of hydrodynamical simulations (see
\citealt{2011ASL.....4..204B} for a review). Special attention
was dedicated to the effects of feedback from stars
\citep{nagai.etal.2007} and AGNs
\citep{short_thomas_2009,puchwein.etal.2008,2010MNRAS.408.2213S,2014MNRAS.438..195P,lebrun.etal.2014,pike.etal.2014,martizzi.etal.2014,hahn.etal.2015,gaspari.etal.2014},
to the evolution of the relations up to $z\sim1$ 
\citep{2012ApJ...758...74B,2011MNRAS.416..801F,planelles.etal.2016,lebrun.etal.2016},
and to developing a theoretical framework to exploit the
simultaneous analysis of multiple signals
\citep{2010ApJ...715.1508S,evrard.etal.2014}.
This paper, based on a set of simulations that includes a new
  model for gas accretion on to supermassive black holes and for the
  ensuing AGN feedback, and an improved implementation of 
  hydrodynamics, extends the analysis of scaling relations out to
$z=2$ in view of what future observational facilities will provide.
 We particularly focus on the selection function typical of SZ
  surveys which is demonstrated to be effective in finding high-$z$ objects
  (see Section 2.1).
 
This work analyses a set of simulations that have been shown to
naturally form cool-core (CC) and non-cool-core (NCC) clusters
\citep{2015ApJ...813L..17R}.  In particular, we have already shown for
these simulations that entropy, gas density, temperature, thermal
pressure, and metallicity profiles of the two populations of clusters
reproduce quite well observational results \citep{2015ApJ...813L..17R,
  planelles.etal.2016,biffi.etal.2017}.  For this reason, we expect
that in our simulated clusters a balance between  the simulated
  level of radiative cooling, which forms stars, and  the
  included amount of AGN feedback, which heats the gas, is reached as
the systems evolve and interact with the cosmological
environment.  However, although the agreement with observations is
  remarkable, there still are several limitations affecting the
  simulations.  For example, the processes linked to the stellar
  population and the BH activity are treated with sub-grid models, and
  some phenomena, such as kinetic feedback by AGNs, magnetic fields, dust production and disruption, or metal
  diffusion, are not implemented into the code yet.  As specified in
  \cite{2015ApJ...813L..17R}, this model should, therefore, be
  intended as {\it effective}.  On
  the same note, as we will discuss in the next section, our sample is
  not a volume-complete sample.  For this reason, the emphasis of our
  discussion is directed on the effect of the physics and on the
  evolutionary trends of the relations rather than on the precise values of the parameters of the best-fitting
  relations.  
 
 \vspace{0.3cm}
 
 The paper is organized as follows: in Section 2, we provide a short
 description of the simulated sample and motivate our sample
   selection.  Section 3 presents the computation of ICM structural
 quantities, the mass-proxy relations, the luminosity-based relations,
 and fitting methods. In Section 4, we examine the validity of our
 simulated data by comparing to observations at low ($z\leq0.25$) and
 intermediate ($z_{\rm median}\approx0.5$) redshifts. Section 5 is
 dedicated to exploring the evolution of scaling relations from $z=0$
 to $z=2$. Finally, a summary of the results and conclusions are given
 in Section 6.

All the quantities for the scaling relations are evaluated at
$R_{500}$ defined as the radius of the sphere whose mean density is
500 times the critical density of the Universe at the considered
redshift. In general, $M_{\Delta}$ is the mass of the sphere of radius
$R_{\Delta}$ and density $\Delta$ times the critical density of the
Universe at the proper redshift. The virial radius is expressed
accordingly to \cite{1998ApJ...495...80B}. For our cosmology,
$\Delta_{\rm vir} \approx 93$ at $z=0$.  Throughout the paper, the
symbol $\lgt$ indicates the {\it decimal} logarithm and the
uncertainty at 1 $\sigma$ on the best-fitting parameters represents the
68.4 per cent confidence maximum-probability interval.

\label{sec:cmrr}

\vspace{0.3cm}

\section{Simulations}
\label{Simulation}
Our analysis is based on three sets of simulations of galaxy clusters
with varying sub-grid physics. These are selected from a parent
DM-only cosmological volume of 1 $h^{-3}$ Gpc$^3$
\citep[][]{2011MNRAS.418.2234B}\footnote{{We define $h\equiv
    H_0/(100\ \rm{km/s/Mpc})=0.72$, where $H_0$ is the Hubble
    constant.}} and re-simulated at higher resolution and with the
inclusion of baryons.  We compute the scaling relations at eight
  different times corresponding to $z=0, 0.25, 0.5, 0.6, 0.8, 1, 1.5,$
  and $z=2$. 

 The re-simulated Lagrangian regions are chosen around the 24 most
  massive clusters with mass
  $M_{FoF}$\footnote{Friends-of-Friends (FoF) refers to
    the algorithm in which a pair of particles are considered to
    belong in the same group or object (i.e. friends) when their
    separation distance is smaller than a given linking length . In
    our simulations, the linking length is equal to 0.16 in unit of
    the mean separation of Dark-Matter particles. }
    $>1\times10^{15}h^{-1}M_\odot$ plus 5 isolated groups with
  $M_{200}=[1-4]\times10^{14}h^{-1}M_\odot$. 
  The Lagrangian regions surrounding each cluster are chosen to be large 
  enough that no contaminating low-resolution DM particle is found out to 
  five virial radii from the centre of each cluster. 
   Their particles, identified at redshift $z=0$, are
traced back to redshift $z\sim 70$,  which is about 50 Myr earlier 
than the starting redshift of the parent DM simulation to ensure the validity of the Zeldovich approximation.
The particle number is
increased to achieve a better spatial and mass resolution,
furthermore, the baryonic component is added. The initial conditions
for the re-simulations are produced by a zoomed-initial technique
(ZIC, \citealt{1997MNRAS.286..865T}). We refer to
\cite{2011MNRAS.418.2234B} for a full description of the re-simulation
technique.  The resimulations are carried out with an improved version
of the GADGET-3 smoothed-particle-hydrodynamics (SPH) code
\citep{springel05} where we included a number of improvements as described in \cite{2015arXiv150207358B}.  In short,
these allow the SPH method to perform better in hydrodynamical
standard tests including weak and strong shocks, gas mixing, and
self-gravitating clouds.

 The cosmological setting is a $\Lambda CDM$ model with cosmological
parameters consistent with the WMAP-7 constraints
(\citealt{komatsu.etal.2011}): $\Omega_m=0.24$ ,
$\Omega_{\Lambda}=0.76$, $n_s=0.96$ for the primordial spectral index,
$\sigma_8=0.8$ for the amplitude of the power spectrum of the density
fluctuations, and $H_0=72$ km s$^{-1}$ Mpc$^{-1}$ for the Hubble
parameter.  When comparing our models to observational data --
presented in Section 4 -- we rescale the latter to the simulated
cosmology.

The Plummer-equivalent gravitational softening of the DM particles is
set to 3.75 $h^{-1}\ \rm{kpc}$ in physical units up to $z=2$ and in
comoving units at higher redshifts.  The gravitational softening of
the gas ($3.75\ h^{-1}\ \rm{kpc}$), stars ($2\ h^{-1}\ \rm{kpc}$), and
black hole particles ($2\ h^{-1}\ \rm{kpc}$) are fixed in comoving
coordinates at all redshifts. The minimum value of the smoothing lengths is
limited to 0.1 per cent of the gravitational softening. For the
computation of SPH quantities related to the gas, we employ the
Wendland $C^4$ interpolating kernel with $200$ neighbours (see
\citealt{2015arXiv150207358B} for more details).  The mass of the DM
particle is $8.47 \times 10^{8} \ h^{-1} M_{\odot}$ and the initial
mass of the gas particle is $1.53 \times 10^{8} \ h^{-1} M_{\odot}$.

We analyse three sets of simulations. These have the same initial
condition for the 29 regions, but they differ in the astrophysical processes included.
Comparing the results obtained from
  the three sets allows us to qualitatively assess their
  origin. Different results among the three samples imply that the
  scaling relations are affected by the astrophysical phenomena
  diversely implemented in the three sets.  Viceversa, if the results
  are consistent, then, the behaviour of the scaling relations is
  determined by gravity, which drives
  the interactions with the environment and 
  the large-scale structures, and by the hydrodynamical forces that, hence, take place.
  
  In the following, we describe the
three sets, tagged as \ovisc, \csf, and \agn, from the simplest
  to the most complex:
 
{(i) \ \ovisc\ (Non-radiative)}. These simulations are carried
out with the same code used by \cite{2015arXiv150207358B}. The main
variations with respect to the {\it standard} GADGET code include the following: the
choice of a higher order Wendland $C^4$ kernel function; a
time-dependent artificial viscosity scheme; a thermal diffusion term
(or artificial conduction) that improves the treatment of contact
discontinuities and promotes fluid mixing.  The performance of this
code with respect to other particle- or grid-based codes is described
in \cite{sembolini.etal.2016b} and \cite{sembolini.etal.2016a}.  These
works show how the thermodynamical properties of the ICM for the
clusters simulated with the improved GADGET version (G3-XArt in those
papers) are quite similar to those produced by grid codes, AREPO, and
the most modern SPH schemes (\citealt{sembolini.etal.2016a}).

{(ii) \ \csf\ (Cooling, star formation and stellar
  feedback)}. The radiative runs consider metal-dependent radiative
cooling rates accordingly to \cite{2009MNRAS.399..574W}, where 15
different elements (H, He, C, Ca, O, N, Ne, Mg, S, Si, Fe, Na, Al, Ar,
Ni) are followed; the effect of a uniform UV/X-ray background
radiation \citep{2001cghr.confE..64H}; the feedback by supernovae
(SN), as originally prescribed by \cite{2003MNRAS.339..289S} with a
mass loading parameter equal to 2; the chemical model by
\cite{2007MNRAS.382.1050T} to account for the metal enrichment from SN
II, SN Ia, and asymptotic giant branch (AGB) stars (for further
details see \citealt{2014MNRAS.438..195P,biffi.etal.2017}).  Kinetic
feedback from the outflows driven by supernova is included. The wind
velocity is set equal to 350 km s$^{-1}$.

{(iii) \ {\agn}.}  This set of simulations is the same as the
\csf\ one, but with the addition of AGN feedback. This feedback
channel is modelled following \cite{steinborn.etal.2015}, who improved
the original model by \cite{springel.dimatteo.hernquist.2005}. In the
new model, we consider both radiative and mechanical feedback
generated from gas accretion on to black hole, both being released into
the surrounding gas as thermal energy.  The radiative and mechanical
efficiencies depend on the (Eddington-limited) accretion rate and the
black hole mass, allowing for a smooth transition between radio- and
quasar-modes. The coupling efficiency between the energy radiated from
the black hole and the gas is expressed though the factor
$\epsilon_f=0.05$. In addition, the model separately treats the
accretion of cold gas and hot gas.  Only for the accretion of the cold
gas, we boost the Bondi rate by a factor of 100, so as to mimic the
effect of the cold accretion mode, as discussed by
\cite{gaspari.etal.2017}. \\ The thermodynamical and chemodynamical
properties, and the dynamical state of the simulated clusters of this
sample are also presented in \cite{2015ApJ...813L..17R},
\cite{villaescusa.etal.2016}, \cite{biffi.etal.2016},
\cite{planelles.etal.2016}, and \cite{biffi.etal.2017}.  The
  \agn\ simulations of the main clusters of the 29 regions are
  presented in \cite{2015ApJ...813L..17R}. The entropy and iron
  profiles of the CC and NCC populations are shown to agree with observational data. The observed anticorrelation
  between the core entropy and core enrichment level is also
  reproduced \citep{biffi.etal.2017}. The different behaviour between
  the two classes is confirmed in the pressure profiles: in the
  central part of the CC systems the pressure is higher by an amount
  that reflects the observational gap measured from SPT and Bolocam
  data \citep{planelles.etal.2016}. The increase in CC central thermal
  pressure is larger than the deepening of the gravitational potential
  arising from a more pronounced adiabatic contraction. As a consequence,
  the bias in the hydrostatic-equilibrium masses measured in the core
  ($R<R_{2500}$) of CC objects is found to disappear or to be
  negligible \citep{biffi.etal.2016}.

\subsection{The Sample}

\begin{figure}
\centering
{\includegraphics[width=0.5\textwidth]{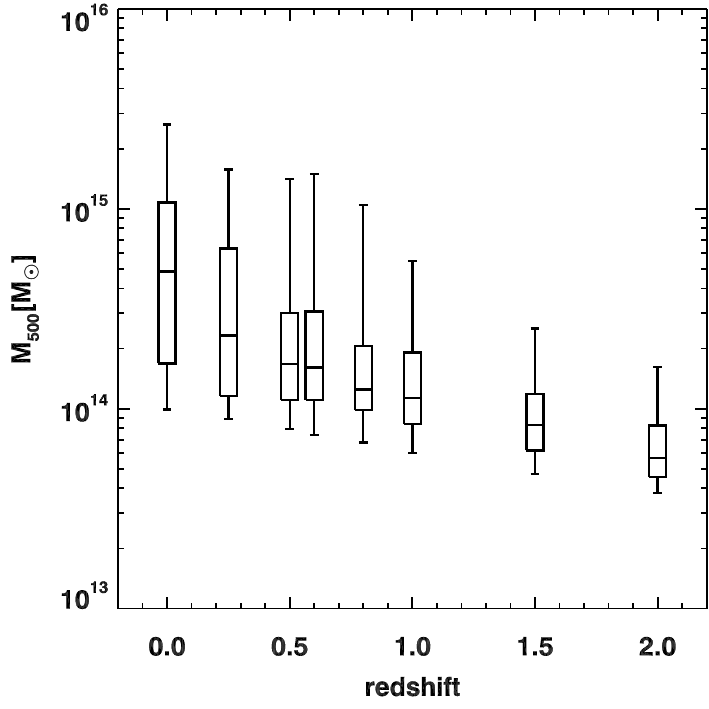}}
\caption{The vertical bars indicate the mass range for the
  \agn\ sample at the various redshifts: z= $0, 0.25, 0.5, 0.6, 0.8,1,
  1.5,$ and 2. For each redshift, the upper limit 
    corresponds to the mass of the most massive cluster, while the lower limit is 
    equal to $7 \times 10^{13}E(z)^{-1} h^{-1} M_{\odot}$.
  The respective number of halos is 58, 75, 93, 88, 91,
  74, 60, and 36. The bottom, top, central lines of the rectangle represent the
  25th, 75th, and 50th percentiles of the mass distribution.}
   \label{fig:fig1}
\end{figure}

The sample includes all the objects in the high-resolution Lagrangian
regions with $M_{500} > 10^{14} E(z)^{-1} M_{\odot}$ where
$E(z)=H(z)/H_0=(\Omega_M\times(1+z)^3+\Omega_{\Lambda})^{1/2}$.  The
number of clusters in the \agn\ run selected at the redshifts of
interest and the corresponding mass range are presented in
Fig.~\ref{fig:fig1}.  In the following we will comment on the
  most important implications related to the sample selection.

  \subsubsection{The limitation on 29 Lagrangian regions}
  The majority of the simulated regions (24 to be precise) are
  centred around massive clusters.  Therefore, we could expect that a
  good fraction of our smallest objects lies in a particularly rich
  environment. This condition could influence some of their
  properties. In particular, these systems might be subject to gas
depletion or over-heating. We check whether any of these two
conditions are present in our sample by comparing the results of the
\ovisc\ set with those obtained by \cite{lebrun.etal.2016}, who
analysed a cosmological box. We selected the NR runs to avoid the
  comparison between samples simulated with different ICM
  prescriptions. We found that their best-fitting relations are
passing through the middle of the distribution of our clusters in both
the ($M-\mg$) and ($M-\tsl$) planes and at both redshifts, $z=0$ and
$z=1.5$. We conclude that our smallest objects are not
  particularly gas poor or unusually hot for their mass, and thus
  are representative of the object population in the lowest
  mass bins.  To further prove this point, we study the behaviour of
  the five groups at the centre of the remaining Lagrangian
  regions. These objects do not have a massive cluster in their
  vicinity and have a small mass, two of them are close to our
  limiting mass. All five groups are located in the middle of the
  distribution of our entire sample in all of the runs. 
  
\subsubsection{The upper mass limit}
Another concern, linked to the restriction of the volume
  analysed, is connected to the representativeness of our most massive
  systems at higher redshifts. 

The {\it upper mass limit} shown in Fig.~1 represents the mass of
  the most massive system found at each redshift. This value might
  also be affected by the restriction of our study to 29 Lagrangian
  regions especially at high redshift. Indeed, in our study,
we automatically include all the progenitors of our $z=0$ sample and
exclude all objects outside the high-resolution resimulated
  region.  It is expected that most of the progenitors of massive
objects are still massive at $z\sim1$ or above. Although, it might be
that a halo considered very massive at $z=2$ stops its growth and
  remains of about the same mass at $z=0$ \citep[see Fig. 7
    in][]{Muldrew.etal.2016}.  This could happen if the accretion was
  as fast and intense as to conglomerate into the object most of the
  surrounding material and to create an under-dense region all
  around. That cluster, which was one of the most massive ones at high
  redshift, will become a relatively small object with respect to the
  entire cluster population at $z=0$. Because of that, it most likely
  could be excluded by our selection based on $z=0$ masses. This
    explains the concern of missing some of the most massive $z=2$
  clusters.
    
To evaluate the extent and consequences of this aspect, we
computed the predicted number of clusters above a certain mass at
  different redshifts by using HMFcalc \citep{murray.etal.2013}. We
consider the functional form of the Halo Mass Function proposed by
\citealt{watson.etal.2013} using the formulation that includes the
redshift evolution. We further set the cosmology according to the
  cosmological parameters of our simulations. We forecast
  that the parent box, with a volume of (1 $h^{-1}$ Gpc)$^3$, at
$z=2$, $z=1.5$, and $z=1$ should have at least four objects with mass
$M_{500}$, respectively, above $9.12 \times10^{13} h^{-1} M_{\odot}$,
$1.66 \times10^{14} h^{-1} M_{\odot}$, and $3.07 \times10^{14} h^{-1}
M_{\odot}$. In our sample, above the same mass limits there are three
clusters at $z=2$ and $z=1$ and two systems at $z=1.5$.
  Considering the Poisson errors, these numbers are consistent with
  the expectation.
 
 Therefore, we conclude that even if we are not considering a
  volume-limited sample at $z>1$, there is a statistically good
  representation of massive objects among the clusters selected within
  the 29 Lagrangian regions. This allows us to study the scaling
  relations over a sufficiently large mass range, which spans from a
  factor of 10 at $z=1$ to $\sim 5$ at $z=2$.  

\subsubsection{The lower mass limit and the variation with redshift}
 The {\it lower mass limit} of our selection reproduces the same
  dependence on $E(z)$ as of the SZ-selected clusters \citep[e.g. see
    fig. 6 from][for the SPT sample]{bleem.etal.2015}.  Our choice is
  aimed at maximizing the statistical size of our simulated sample and
  enlarging the $z>1$ mass range in order to robustly derive the
  scaling relations in single redshift bins.
   
Indeed, applying to our 29 Lagrangian regions the selection functions
typical of X-ray or optical surveys, whose lower mass limits are,
respectively, increasing and nearly constant with redshift, would have
returned a poor statistics, especially at high redshift.  Looking at
specific future surveys of clusters, we recall that the limiting mass of the
selection function of {\it eROSITA} \citep{borm.etal.2014}
rapidly grows with redshift from $10^{14} M_{\odot}$ at $z=0.2$ to $4
\times 10^{14} M_{\odot}$ at $z=0.7$, while that of {\it Euclid}
\citep{sartoris.etal.2016} will be almost constantly equal to $1.1
\times 10^{14}$ from $z=0$ to $z\sim 1$ and will grow afterwards.
Considering the rapid decline of the cosmological mass function, these
missions will not cover as large mass range as future SZ surveys.

Our high-redshift samples, i.e. at $z=1$, $z=1.5$, and $z=2$, include
74, 60, and 36 objects, respectively, numbers comparable to studies on {\it local}
observed scaling relations, and, as previously said, extend in mass by
a factor ranging from almost 5, at $z=2$, to 10, at $z=1$.
  
The smallest system at $z=2$ contains more than $3.5 \times 10^4$
particles providing good estimates of global quantities. We do not
extend the sample to smaller systems, even if we would have obtained
numerically robust global measures, because, again, {\it none} of the
planned missions will reach such small masses.

 As we will further discuss in the next sections, our choice will
  have some impact on the computed evolution of the scaling relations.
  In fact, our selection excludes at $z=0$ the smallest groups of
  galaxies, that are known to cause a break of the power-law fitting
  of the $L_X-T$ and $L_X-M$ relations. We will, therefore, model
  these relations as single power laws, thus avoiding a more
  complicated parametrization (see, however, the detailed analysis
  performed by \citealt{lebrun.etal.2016}).  The same benefit is
  nevertheless not present at high redshifts, when the SZ selection is
  indeed sampling smaller mass systems, which are affected by a
  drastic reduction of the gas fraction \citep{dai.etal.2010}. The
  scaling relations involving the gas mass can still be fitted by a
  single power law -- since less massive objects will be present at
  $z>1$ but the overall sample population will be different (see
  Section 5.1).

\section{Method of Analysis}
\label{Method of Analysis}

\subsection{Computing ICM Quantities}

In the following, we briefly describe how we compute the relevant
quantities from our simulated data sets.

 
\smallskip
\noindent {\it Masses.} The total mass, $M$, is calculated by summing
the contribution of all the species of particles (dark matter, gas,
and stars) within $R_{500}$. For the gas mass, $\mg$, we sum the hot
gas component. In the case of multicomponent
particles\footnote{The gas particles can be multi-phase, carrying
  information on both the hot and cold gas. The cold phase provides a
  reservoir for stellar formation.}, we include all particles
  containing less than 10 per cent of cold gas, and therefore no
star-forming particles. In each region, the mass of the hot gas
contained in all the particles with a cold gas fraction larger than 10
per cent is less than 0.01 per cent of the total hot gas of the
region, and thus is negligible.
 
 
\smallskip
\noindent {\it Temperature.} We consider both the mass-weighted
temperature and the spectroscopic-like temperature
\citep{2004MNRAS.354...10M}.  To compare with observations (Section 4)
we consider the same aperture used in the observational samples
($R/R_{500} < 1$), while to derive our results (Section 5) we exclude
the contribution of the core, defined as the region within 15 per cent
of $R_{500}$ ($0.15 < R/R_{500} < 1$).

The mass-weighted temperature is provided by
     \begin{equation}
      \tmw=\frac{\sum_i m_i T_i}{\sum_i m_i},
      \label{eq:tmw}
     \end{equation}
where $m_i$ and $T_i$ are the hot-gas mass and temperature of the
$i^{th}$ gas particle.  The spectroscopic-like temperature is
introduced to ease the comparison with X-ray observations and the
formula was derived considering the non-flat response of the
instruments on board of {\it Chandra} and {\it XMM--Newton}:
\begin{equation}
\tsl= \frac{\sum_i \rho_i m_i T_i^{0.25}}{\sum_i\rho_i m_i T_i^{-0.75}}, 
\label{eq:tsl}
\end{equation}
where $\rho_i$ is the particle gas density \citep[][see also
  \citealt{vikh2006}]{2004MNRAS.354...10M}. For this computation, we
used only particles emitting in the X-ray band with $T_i> 0.3$ keV.


\smallskip
\noindent {\it The parameter $Y_{X}$.} As previously said, this
parameter is equivalent to the product of the gas mass and the
core-excised temperature within $R_{500}$ and it is a powerful proxy
for the total thermal content of the ICM because it is almost
insensitive to the physical processes included in simulations
\citep{2010ApJ...715.1508S,2011MNRAS.416..801F,2012ApJ...758...74B,sembolini.etal.2014}
and to the dynamical status of the clusters
\citep{poole.etal.2007,rasia.etal.2011,kay.etal.2012}.  $Y_X$ is
derived from X-ray observations, and therefore we adopt the core-excised
spectroscopic-like temperature in its expression:
   \begin{equation}
    Y_{X} = \mg\times \tsl. 
    \label{eq:yx}
   \end{equation}
   
 As specified above, in Section~4 the observational quantities, which we compare to, are available only within
the fixed aperture of $R_{500}$ \citep[][and its erratum]{2013ApJ...767..116M}, 
therefore, {\it exclusively} in that section and in Figures~2 and 3, we compute $Y_X$ without excising the core.

 
\smallskip
\noindent {\it X-ray Luminosity.}  The bolometric luminosity is
computed by summing the contribution of the emissivity, $\epsilon_i$,
of all gas particles within the sphere of radius $R_{500}$:
 \begin{equation}
 L = \sum_{i}\epsilon_i= \sum_i n_{e,i}n_{H,i}\Lambda(T_i,Z_i)\Delta V_i, 
 \label{eq:lx}
 \end{equation}
 where $n_{e,i}$, $n_{H,i}$ are number densities of electrons and
 hydrogen atoms, respectively, $\Delta V_i = m_i/\rho_i$ is the
 particle's volume and $\Lambda$ is the interpolation of the cooling
 function pre-calculated in a fine grid of temperatures and
 metallicities  starting from the values of temperature, $T_i$,
   and global metallicity, $Z_i$, of each gas particle\footnote{We recall that, in our simulations, the ratios of the abundances of the
 elements such as oxygen or silicon over the iron abundance are typically close to the solar value 
 \citep{biffi.etal.2017}. Furthermore, the influence of boosted single element line does not substantially increase the 
 {\it bolometric} luminosity.}.
  The cooling-function tables are created by assuming the APEC model
 \citep{2001ApJ...556L..91S} in XSPEC and by integrating over the
       $[0.01-100]$ keV energy band. 

\subsection{The scaling relations and the self-similar prediction}

The total mass of a cluster can be related to the various ICM
quantities presented in the previous sections through simple power-law
models.  \cite{kaiser86} analytically derived the functional shapes of
the expected scaling relations under the assumption of virial
equilibrium between the kinetic and thermal energy of a galaxy cluster
and its gravitational potential \citep[a recent extension of the
  formalism is provided in][]{ettori2015}.  According to this model,
called SS, the cluster total mass is the only parameter
that defines both thermal and dynamical properties of the ICM (e.g.,
\citealt{2013SSRv..177..247G} and references therein).  Clusters with
different mass are simply the scaled-up or -down version of each
other. The self-similarity stems from the fact that there is no
preferred scale in the problem as gravity and the initial power
spectrum are scale-free (or SS). In this context, the total
mass is the only variable of the problem. For this reason, the scaling
relations are often presented with the mass as independent variable,
especially in theoretical and numerical studies. However, from a
practical and observational perspective, the scaling relations are
adopted to derive the total mass of the clusters once the mass proxy
quantities are measured. We will resort to this presentation for all
relations linking the total mass to a mass-proxy ($\mg$, $T$, and
$Y_X$). On the other hand, we will also consider the $L-T$ and $L-M$
relation with the bolometric luminosity as dependent variable because
both relations are observationally used to derive the flux limit
corresponding to a certain mass or temperature.

In the following, we will describe the mass-proxy relations (1, 2, and
3) and the luminosity-based relations (4 and 5) together with their
expected dependence with redshift parametrized as power of $E(z)$
valid for a $\Lambda CDM$ background cosmology (see e.g.,
\citealt{2011ASL.....4..204B}).  Note that in \cite{kaiser86}, the
redshift evolution was modelled in terms of powers of $(1+z)$ as
  expected for an Einstein-de-Sitter Universe.  In all relations, we
treat the normalizations, $C$, as constants, although in general they
depend on the internal structure of the system.

\smallskip
\noindent (1) {\it The $M-\mg$ relation.} The total mass and the gas mass are linearly related:
  \begin{equation}
   M = C_{\mg} E(z)^{0} \mg, 
   \label{eq:mmg}
  \end{equation}
without any dependence on redshift.


\smallskip
\noindent (2) {\it The $M-T$ relation.} The total mass is related to the temperature according to:
\begin{equation}
 M =  C_{T} E(z)^{-1} T^{3/2}. 
 \label{eq:mt}
\end{equation}


\smallskip
\noindent (3) {\it The $M-\yx$ relation.} From Eqs. \ref{eq:yx}, \ref{eq:mmg}, and \ref{eq:mt}, the self-similar scaling of $\yx$ with cluster mass is
 given by
 \begin{equation}
  M = C_{\yx} E(z)^{-2/5} \yx^{3/5}. 
  \label{eq:myx}
 \end{equation}

 
\smallskip
\noindent (4) and (5) {\it The luminosity relations:} $L-T$ and $L-M$. Assuming thermal bremsstrahlung emission,  the bolometric luminosity can be related to the total mass  by
 \begin{equation}
  L \propto E(z)^2 T^{1/2} f_{gas}^2 M, \label{8}
  \end{equation}
  where $f_{gas}\equiv {\mg}/M\equiv1/C_{M_g}$ is the gas fraction. Given the expression of the $M-\mg$ and $M-T$ relations (equations \ref{eq:mmg} and \ref{eq:mt}), the self-similar form for the $L-T$ relation is expressed as
 \begin{equation}
 L = C_{LT}E(z)T^2, 
 \label{eq:lt}
 \end{equation}
 and for the $L-M$ relation as
 \begin{equation}
 L = C_{LM}E(z)^{7/3}M^{4/3}.
 \label{eq:lm}
 \end{equation}
 
 \subsection{Fitting Method}
 
\begin{table}
\caption{For each scaling relation listed in column 1, we report the
  key parameters of Equation (11): the pivot point, $X_0$ (2nd
  column), the self-similar slope, $\beta$ (3rd column), and the
  self-similar evolution parameter, $\gamma$ (4th column).}
\begin{center}
 \begin{tabular}{|c|c|c|c|}
 \hline
Relation & $X_0$ & $\beta$ & $\gamma$\\
\hline
$M-\mg$ & $1\times10^{13}\ h^{-1} M_\odot $ &1&0\\
\hline
$M-T$ & $3\ \textrm{keV}$                              &3/2&-1\\
\hline
$M-\yx$ & $3\times10^{13}\ h^{-1} M_\odot  \rm{keV}$ &3/5&-2/5\\
\hline
$L-\tsl$ & $3\ \textrm{keV}$                             &2&1\\
\hline
$L-M$ & $1\times10^{14}\ h^{-1} M_\odot $     &4/3& 7/3\\
\hline
  \end{tabular}
 \end{center}
 \label{tabella1}
 \end{table}

To study the evolution of the parameters of the scaling relations, we
fit them through a generic expression:
  \begin{equation}
   \lgt F = \lgt C + \gamma \lgt E(z) + \beta \lgt (X/X_0),
   \label{11}
  \end{equation}
 $F$ can either represent the total mass, $M_{500}$, or the X-ray
  luminosity, $L_X$. In the first case, $X$ is assumed to be one of
  the mass proxies: $\mg$, $\tmw$, $\tsl$, $\yx$. In the second case,
  $X$ stands either for the spectroscopic-like temperature or for the
  total mass.  The value of each pivot point, $X_0$, is listed in
Table~\ref{tabella1}.  These are independent from $z$ to facilitate
the study of the evolution of the normalization. The values of the
pivot points are close to the median values of the variables ($\mg$,
$\tmw$, $\tsl$, $\yx$, and $M$) of the entire \agn\ sample that
includes the objects from all the redshifts.  At first, we fix
$\gamma$ to the SS expectation values (also listed in
Table~\ref{tabella1}) and we derive $C$ and $\beta$ at each {\it
  independent} redshift (Sections~5.1 and 5.2). Later, to study the
evolution of the normalization, we let $\gamma$ free to vary
(Section~5.3).

\vspace{0.5cm}

We used three different algorithms implemented in IDL routines to fit
the data.  Two of them are robust statistical methods commonly used in
recent observational studies \citep[see][for a review]{sereno2016}
while the last method is more appropriate to analyse data from
numerical sets where the two variables are independently derived and
their calculation does not have any associated error. We found that
the best-fitting parameters derived from the three techniques agree
within 1 $\sigma$. Therefore, we will show the exact values obtained
only from the third method with only the exception of the results in
Section~5.3 where we are forced to use the first program (see
below). The IDL routines employed are as follows:

\noindent{(i)} {\texttt{linmix\_err.pro}} adopts a Bayesian
approach described in \cite{2007ApJ...665.1489K} to investigate the
parameter space and to perform the linear regression in logarithmic
space. The routine is applied to the single linear regression with
$C$, $\beta$, and the intrinsic scatter $\sigma$ as free
parameters. This method allows us to treat the intrinsic scatter,
$\sigma$, estimated via the method of Monte Carlo Markov Chains, 
as a free parameter.  {\it Exclusively}, when we treat
$\gamma$ as an additional free parameter (Section~5.3), we use the
\texttt{mlinmix\_err.pro} routine to adopt a fitting function
  that is not a simple power law.

\noindent{(ii)} {\texttt{bces.pro}} is a least-squares bisector
method \citep{1990.ApJ.364.104} that applies a linear regression that
accounts for any possible correlation between the errors associated with
the two variables and the intrinsic scatter in the data
\citep{1996.apj.470.706}.

\noindent{(iii)} {\texttt{robust\_linefit.pro}} (with the bisector
flag switched on) uses a two-variable linear regression and does not
make any distinction between dependent or independent variables. As
the others, it controls the influence of outliers. For all these
characteristics, we consider this as the most suitable approach for
our analysis.  The best-fitting parameters derived with this method
and the previous one do not have any error associated. To estimate
their uncertainty, we apply a bootstrapping method with $10^5$
iterations. The best-fitting parameters and their uncertainties are
the means and standard deviations of the distributions derived from
this technique.

\section{Comparison between simulated and observed scaling relations}

\begin{figure*}
\begin{center}
{\includegraphics[width=0.49\textwidth]{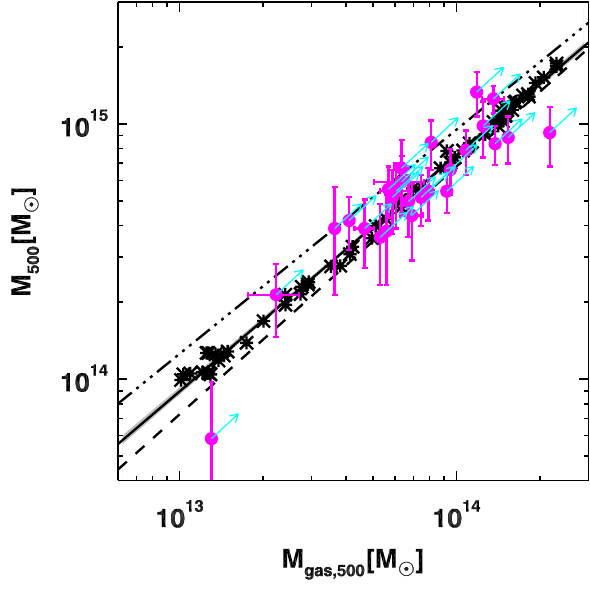}}
{\includegraphics[width=0.49\textwidth]{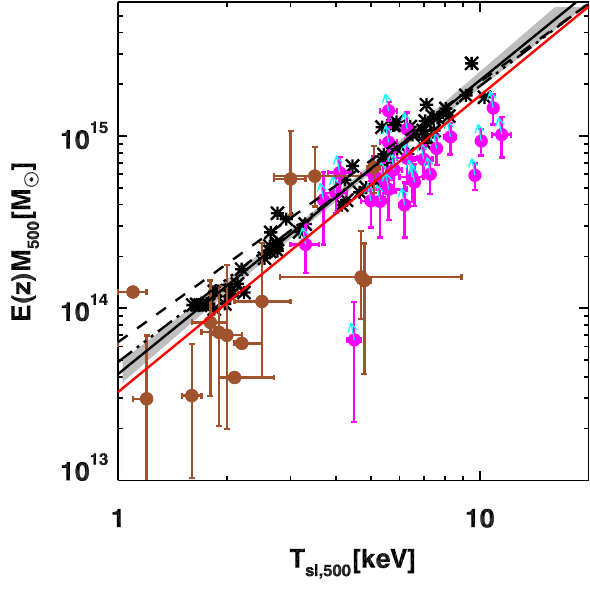}}
{\includegraphics[width=0.49\textwidth]{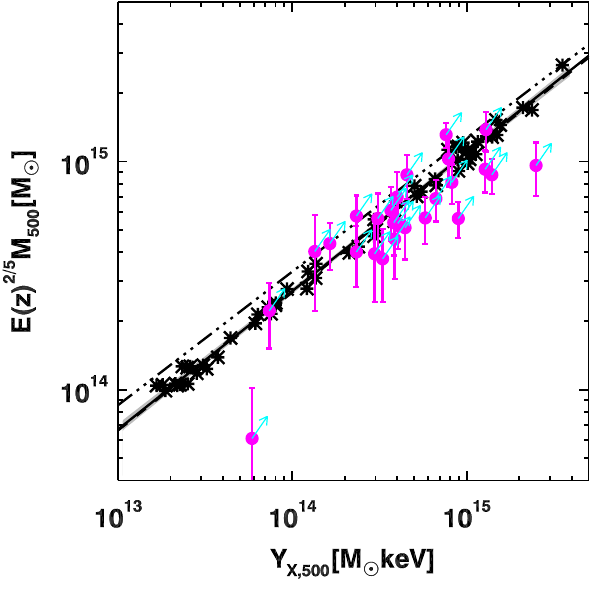}}
{\includegraphics[width=0.49\textwidth]{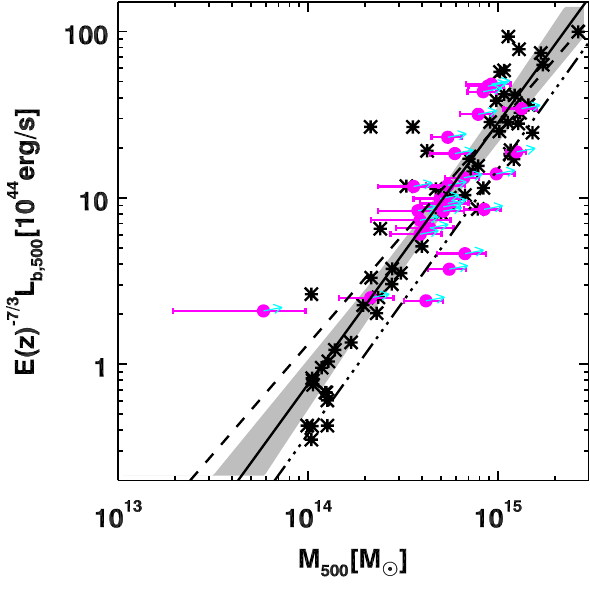}}
\end{center}
\caption{Comparison between scaling relations at $z=0$ and those
  derived from local observations ($z<0.25$). In clock-wise order, we
  plot the $M-\mg$, $M-\tsl$, $L-M$, and $M-\yx$ relations. The
  observational data are taken from
  \protect\citet{2013ApJ...767..116M} and \protect\citet{mah_erratum} (magenta), and
  \protect\citet{2015arXiv151203857L} (brown).  They are shown with 1
  $\sigma$ error bar. The cyan arrows associated with the Mahdavi et al.
  points represent the change of the quantities after correcting (i)
  the total mass by 25 per cent as suggested by Hoekstra et al. (2015)
  and (ii) the other quantities by the amount estimated within our
  \agn\ sample (see the text for details). In each panel, the solid black
  line represents the best-fitting relation of the \agn\ sample shown with
  black asterisks; the grey shaded area is the associated 1 $\sigma$
  scatter around the best fit; the dashed and dashed-dotted lines
  represent the best-fitting relations of the \ovisc\ and \csf\ runs,
  respectively.  The luminosities are bolometric, none of the
  simulated and observed quantities is core-excised, and the observed
  data are rescaled to the cosmology adopted in the simulation.}
 \label{fig:1}
\end{figure*}
\begin{figure*}
\begin{center}
{\includegraphics[width=0.49\textwidth]{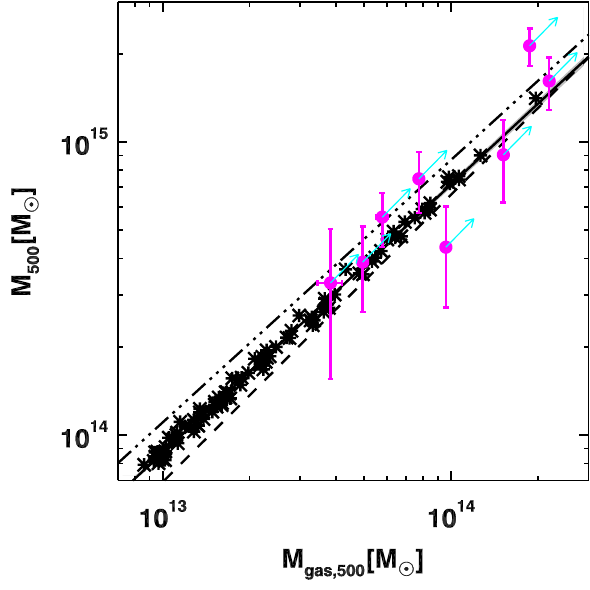}}
{\includegraphics[width=0.49\textwidth]{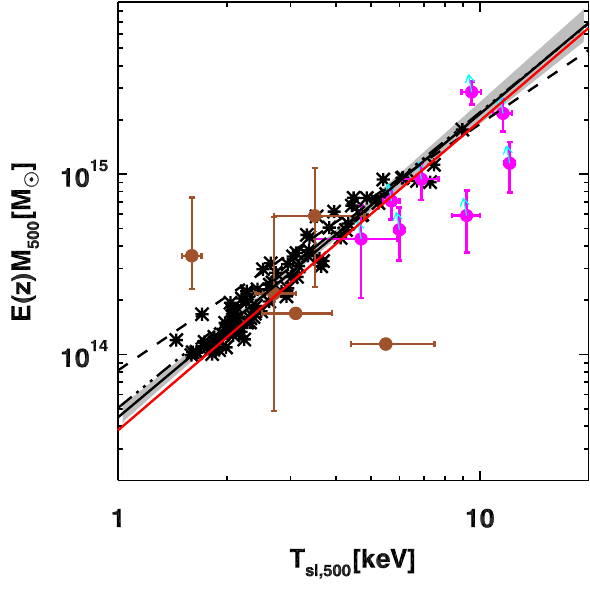}}
{\includegraphics[width=0.49\textwidth]{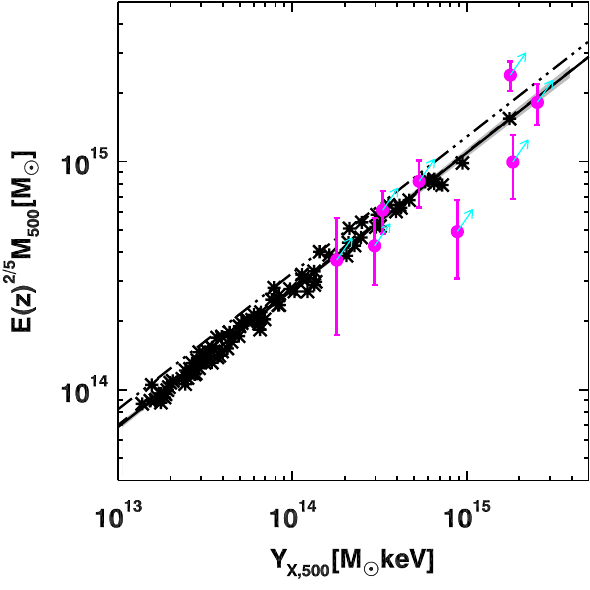}}
{\includegraphics[width=0.49\textwidth]{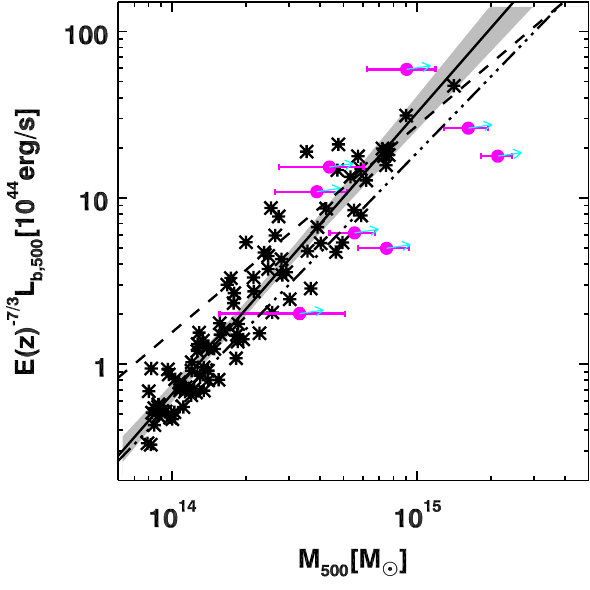}}
\end{center}
\caption{Comparison between simulated scaling relations at $z=0.5$ and
  those derived from intermediate redshifts observations. We selected
  the clusters from $z=0.42$ and $z=0.55$ with a median value equal to
  $0.52$ from \protect\cite{2013ApJ...767..116M} and \protect\citet{mah_erratum} and the objects with
  redshift between 0.43 and 0.52 with a median value equal to 0.45
  from \protect\citealt{2015arXiv151203857L}. In clockwise order, we
  plot the $M-M_{g}$, $M-T_{sl}$, $L-M$, and $M-\yx$
  relations. Symbols and lines are identical to those of Fig.~2.}
  \label{fig:2}
\end{figure*}

\subsection{Observational data sets}
In this section, we qualitatively compare our theoretical predictions for the
scaling relations to some observational results.  For the mass-proxy
relations, due to uncertainties associated with the amplitude of the
X-ray hydrostatic mass bias derived both in observations
\citep{vonderlinden.etal.2014,maughan.etal.2015,khatri.gaspari.2016}
and simulations
\citep{2007ApJ...655...98N,rasia.etal.2012,biffi.etal.2016}, we prefer
to refer to masses estimated via the gravitational lensing
technique. More specifically, we refer to \cite{2015arXiv151203857L}
and \cite{2013ApJ...767..116M}.  Nevertheless, we remark that
potential biases in weak-lensing mass measurements are also present
\citep{meneghetti.etal.2010, becker_kravtsov, bahe.etal.2012}. Indeed,
the weak-lensing masses of the last sample were corrected in a
subsequent study from the same group \citep{hoekstra.etal.2015}.

\cite{2015arXiv151203857L} performed weak-lensing analysis of 38
clusters out of the 100 brightest clusters of the XXL survey
(\citealt{pierre.etal.2016}) to derive their lensing masses using the
Canada--France--Hawaii Telescope Lensing Survey (CFHTLenS,
e.g. \citealt{2012MNRAS.427..146H}; \citealt{2013MNRAS.433.2545E})
shear catalogue. The X-ray temperatures of the 38 clusters are
measured by the $XMM--Newton$ telescope in a central region (within
$300$ kpc). To compare with them, we also calculated the
spectroscopic-like temperature within the same aperture.

The second sample contains 50 galaxy clusters in the redshift range
$0.155<z<0.55$. The optical data are taken from the Canadian Clusters
Comparison Project (CCCP; see \citealt{2013ApJ...767..116M,
  hoekstra.etal.2015}), while the X-ray properties are based on the
combined data from the $Chandra$ Observatory and the $XMM-Newton$
telescope. In the following and in the figures, 
we consider the values taken from the tables provided in the erratum by \cite{mah_erratum}.

\subsection{Comparison to observations at local and intermediate redshifts}

We present the comparison between our \agn-simulated scaling
relations and those obtained from observational samples. We stress that this
comparison can only be qualitative since all samples are differently selected. The
simulations at $z=0$ are associated with observed clusters at $z<0.25$
in Fig.~\ref{fig:1}, while the $z=0.5$ simulated set is matched to a
sample of objects with redshift between 0.42 and 0.55 with median
values equal to 0.5 in Fig. \ref{fig:2}.  Each figure shows four
scaling relations: $M-M_{g}$, $M-\tsl$, $M-\yx$, and $L-M$. In this
Section, all properties are measured within $R_{500}$ without any core
excision in order to be consistent with the observed quantities. For
the comparison, the observational measurements of $M$, $\mg$, $\yx$
and $L$, which depend on $h^{-1}, h^{-5/2}, h^{-5/2}$, and $h^{-2}$, respectively,
have been rescaled according to the value of the Hubble parameter of
the simulation.

\vspace{0.2cm}

In general, we observe a good consistency between simulations and
observations for the three relations $M-\mg$, $M-\yx$, and $L-M$,
especially considering that the \cite{2013ApJ...767..116M}
measurements of the masses should be increased by $\sim25$ per cent
accordingly to the revised work from the CCCP group
\citep{hoekstra.etal.2015}. In the most recent paper, the authors
provide new measurements after accounting for several corrections of key
sources of systematic errors in the cluster mass
estimates. Unfortunately, they did not present the updated values of
the X-ray quantities which might change because of the different
radius related to the new mass profile. Using our \agn\ sample, we
estimate that a variation in mass of 25 per cent corresponds, on
average, to an increase of gas mass, temperature, and bolometric
luminosity of about 25, $-2$, and 5 per cent, respectively. These changes are
represented by the cyan arrows in the figures.

The good agreement in the normalization of the $M-\mg$ relation
assures that the simulated gas fraction is realistic over the mass
range investigated.  However, our \agn--simulated clusters appear to
be colder than those at the same mass obtained from observations. At
fixed mass, the temperatures of \cite{2013ApJ...767..116M} are higher
than our simulated values by about $30$ per cent ($21$ per cent) at
$z=0$ ($z=0.5$). This discrepancy is reduced to 12 per cent once we
considered the {\it corrected} values of total mass
\citep{hoekstra.etal.2015} and the estimated correction for the
temperature.  In the low-mass range, we also observed a temperature
discrepancy of $\sim24\%$ in comparison to \cite{2015arXiv151203857L}
at $z=0$. We recall that the temperature in this work (brown circles
in the figures) is measured within an aperture of $300$ kpc rather
than within $R_{500}$.  The data points should then be compared with
the simulated $M-T_{\rm sl,300kpc}$ relation, shown in the figures
with a red line.  Comparing with this line, we found 1$\sigma$ 
consistency between our best-fitting relation and  that of \cite{2015arXiv151203857L}'s.
In the luminosity--mass plane, our
\agn\ simulated clusters lie in the same region of the observed
data.  We notice, however, that at fixed mass, some of the
  observed clusters are less luminous than all our simulated
  clusters. This leads to an offset of 30 per cent in the
normalization.  A possible explanation could be an overly-peaked
  gas distribution in the simulated sample. However, the gas fraction
  profiles of our simulated clusters are in good agreement with
  observations \citep{simionescu.etal.2017} and we believe that this
  effect, if in place, could play only a minor role. The majority of
this discrepancy, instead, is likely due to the different
sampling choices, related for example to the dynamical state of
the clusters and to a diverse procedural treatment of the
  simulated and observed data. In our analysis, indeed, we do not
mask any sub-clumps present within $R_{500}$.   For example,
  we verify that two simulated objects
  that in Fig.~2 are a factor 6-8 more luminous than the overall (either
  observed or simulated) population are experiencing a major merger
  with a substructure that already crossed $R_{500}$ of the main
  haloes. The luminosity of these two clusters is, thus, boosted
  \citep{torri.etal.2004} by the additional
  contribution of the large merging system, which would be removed in
  any observational analysis.

\vspace{0.2cm}

\begin{figure}
\centering
\includegraphics[width=0.5\textwidth]{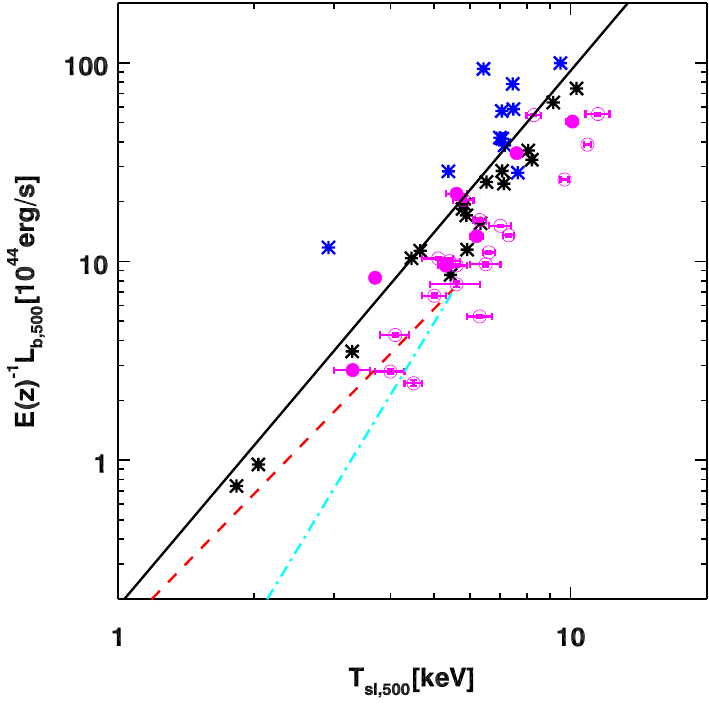}
\caption{The $z=0$ \agn\ simulated luminosity-temperature relation is
  shown in blue and black asterisks, respectively, for cool-core and
  non-cool-core systems \citep{2015ApJ...813L..17R}. For comparison,
  the cool-core and non-cool-core clusters observed by
  \citet{2013ApJ...767..116M} are reported with filled and open
  magenta circles, respectively.  The solid line is the best-fitting
  simulated $L-T$ relation, and the two dashed lines represent the
  results taken from \citet{gaspari.etal.2014} referred to two modes
  of AGN feedback: self regulated (dashed-red) and thermal quasar
  blast (dashed cyan).}
\label{fig:a1}
\end{figure}

To investigate in a deeper manner the influence of the central
regions, we compare the $L-T$ relation measured in the two categories
of cool-core and non-cool-core (Fig.~\ref{fig:a1}) clusters. The two
simulated classes are taken from \cite{2015ApJ...813L..17R} and refer
to the main haloes of the 29 re-simulated regions.  The distinction
between the two classes was established on the basis of the pseudo-entropy level of the objects, which is derived through the following
expression:
\beq
\sigma_K=
\frac{(T_{IN}/T_{OUT})}{(EM_{IN}/EM_{OUT})^{1/3}},
\eeq
where the spectroscopic like temperature, $T$, and the emission
measure, $EM$, are computed in the $IN$ region, $r/R_{180}<0.05$, and
in the $OUT$ region, $0.05<r/R_{180}<0.15$. We apply the cut of
$\sigma_K<0.55$ to define CC clusters. Those with larger values are
classified as NCC clusters.
 
 Overall, the simulated $L-T$ relation is in line with the observed
 scaling, in particular in terms of the {\it slope}. There is an
 offset in the normalization of slightly less than 50 per cent, which
 is mainly produced by the combination of simulated lower temperatures
 and slightly higher luminosities. Nevertheless, it is reassuring
 that, as found in observations, the simulated CCs (blue) tend to have
 larger luminosities than the NCC (black) systems due to a denser core
 that produces a more peaked surface brightness profile.
  To illustrate the effect of the physics of the AGN feedback model, we overplot in
 Fig. 4 the results from very high-resolution ideal hydrodynamical
 simulations which can isolate the impact of two major modes of AGN
 feedback (cf. \citealt{gaspari.etal.2014}), namely tightly
 self-regulated AGN feedback (dashed red line) and a thermal quasar
 blast (dashed--dotted cyan line). The self-regulated AGN feedback
 (typically mediated via chaotic cold accretion on to the super massive
 black hole,  \citealt{gasparo.sadowski.2017}) tends to preserve
 the long-term CC structure, only mildly steepening the cluster $L-T$;
 it has been shown to be crucial to reproduce several properties of
 hot haloes, including gently quenching cooling flows (e.g.,
 \citealt{mcnamara.nulsen.2012} for a review). The quasar blast
 instead promotes a drastic overheating/evacuation (already at the
 poor cluster regime), raising the cooling time above the Hubble time
 and rendering most of the low-temperature system NCCs, which is
 inconsistent with data (e.g., \citealt{sun.etal.2009},
 \citealt{hudson.etal.2010}, \citealt{mcdonald.etal.2013}). In
   this case, the luminosity is rapidly growing with the temperature
   ($L\propto T^{3.8}$). Our sample never experiences such a steep
   relation, not even at $z \geq 1$. The presence of cool cores in
   combination with a reasonable $L-M$ relation implies that a
   regulation of the cooling-heating balance is effective in our
   simulations since the collapse of the systems.  Our effective AGN subgrid model seems to be
 closer to the gentler self-regulated AGN feedback evolution,
 preventing a strong evacuation of the central gas. We remark that
 preserving quasi thermal equilibrium of hot haloes, as observed, is a
 major and difficult constraint to obtain in simulations, which often
 display either overcooling (e.g., \citealt{hahn.etal.2015}) or
 overevacuation (e.g., \citealt{puchwein.etal.2008}).

\section{Evolution of ICM Scaling Relations}

\begin{figure*}
\centering
\includegraphics[width=0.33\textwidth]{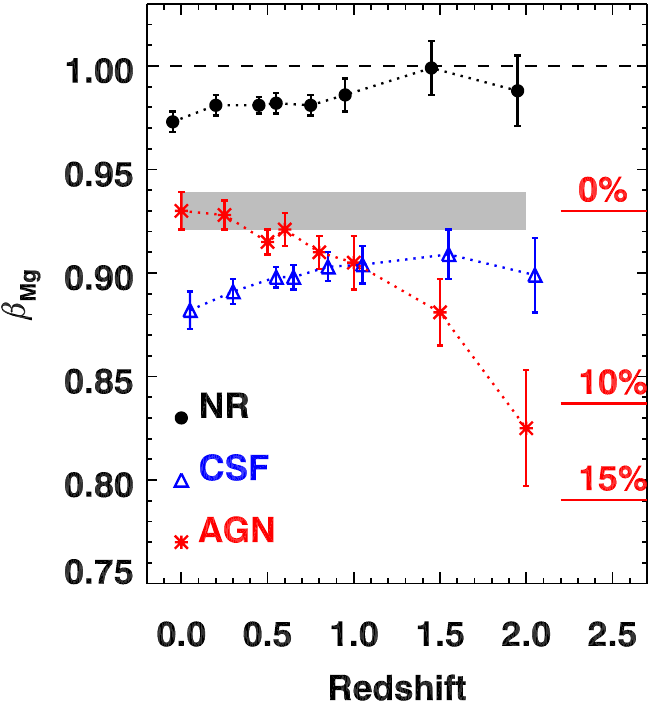}
\includegraphics[width=0.33\textwidth]{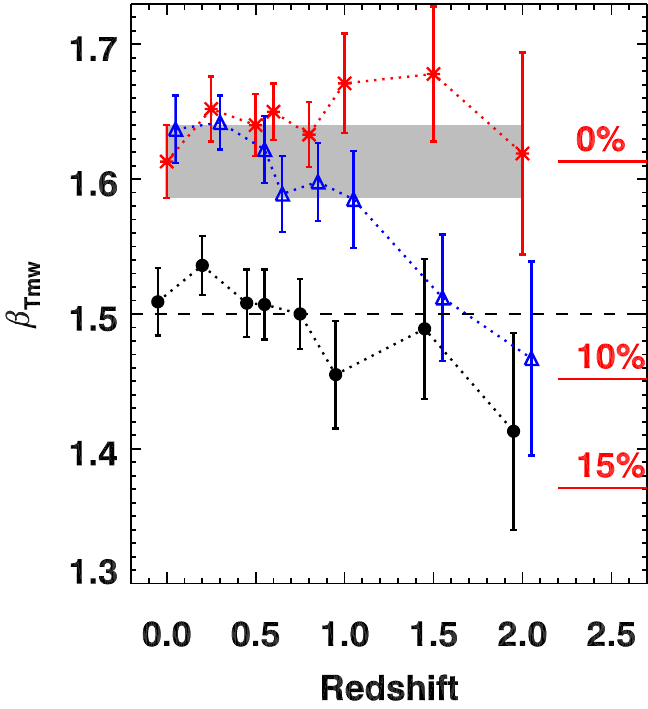}
\includegraphics[width=0.33\textwidth]{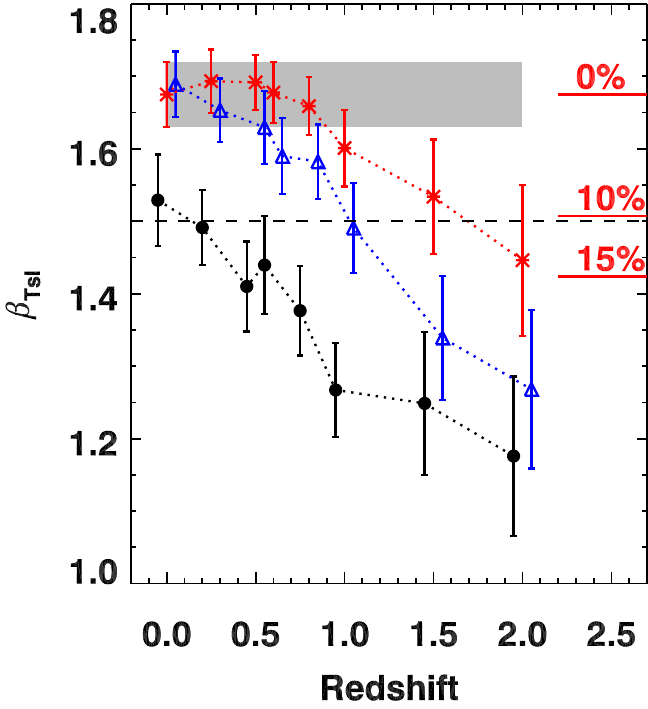}
\includegraphics[width=0.33\textwidth]{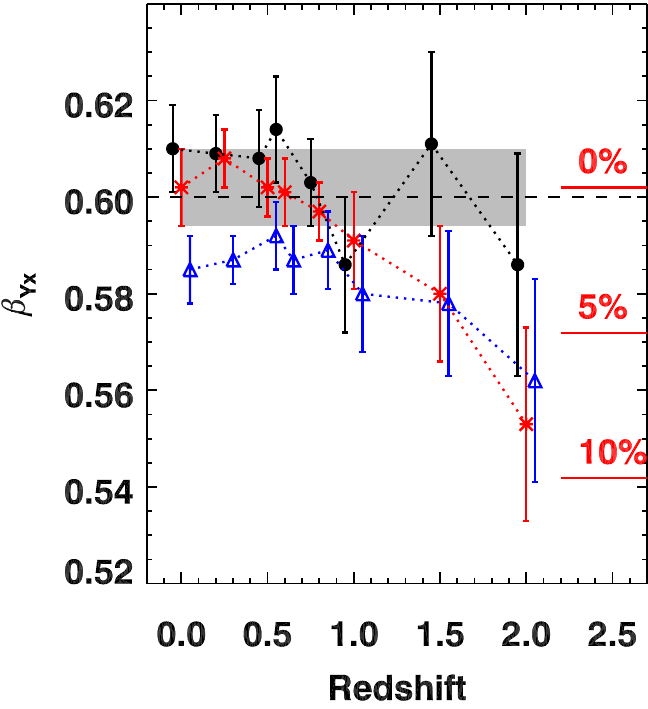}
\includegraphics[width=0.33\textwidth]{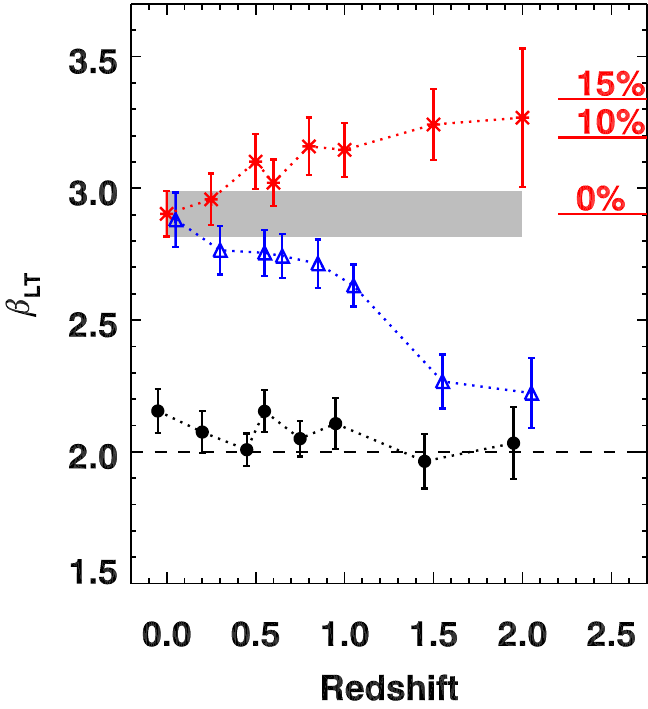}
\includegraphics[width=0.33\textwidth]{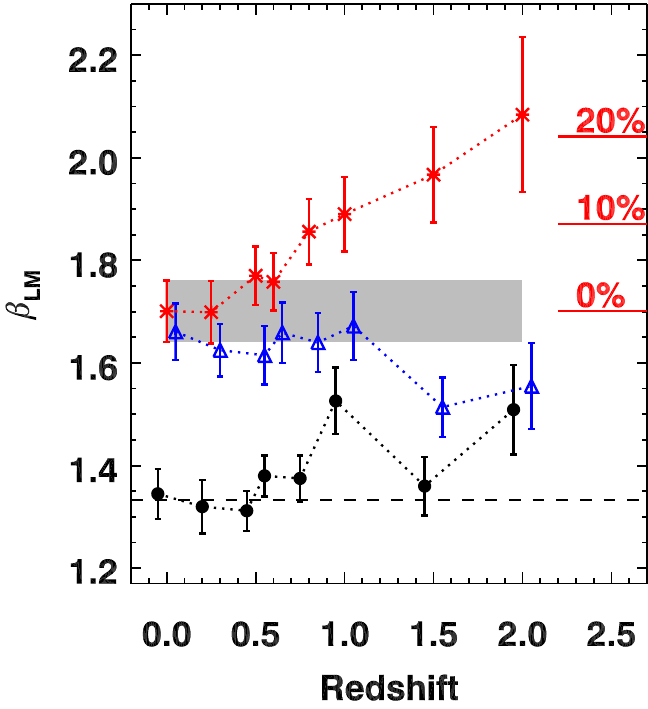}
\caption{We show the slope value for $M-\mg$, $M-\tmw$, $M-\tsl$,
  $M-Y_{X}$, $L-\tsl$, and $L-M$ shown from left to right, top to
  bottom, respectively, as a function of redshift. In each panel, the
  \ovisc, \csf, and \agn\ runs are represented by the black circles,
  open blue triangles, and red asterisks. The dashed black lines
  represent the self-similar evolution of the slopes. The error bars
  represent 1 $\sigma$ of the $\beta$ parameter. The grey shaded area
  shows its value at $z=0$ for the \agn\ sample. In this run, we
  notice that the slopes at redshifts $z<1$ agree within 1$\sigma$
  with the $z=0$ value, while at higher redshifts they can
  significantly deviate. Some variations from the $z=0$ value of the
  \agn\ sample are reported with red horizontal lines on the right
  part of the plots.}
\label{fig:3}
\end{figure*}

In this section, we explore the evolution from $z=0$ up to $z=2$ of
the six scaling relations: $M-\mg$, $M-\tmw$, $M-\tsl$, $M-\yx$,
$L-\tsl$, and $L-M$.  We will discuss in more detail the scaling
relations involving the gas mass and temperature since all the others
are tightly connected to these.

We remind that in this section all the temperatures are obtained
excluding the central region ($<0.15$ $R_{500}$).  We will comment
  on the effect of the exclusion/inclusion of the core but we
  anticipate that excising the core produces a minimal
variation. Indeed, the temperature difference is below 1 per cent in
our \agn\ sample and below 2 per cent in observational samples
\citep{2012MNRAS.421.1583M} once we compare the temperature measured
in the entire sphere within $R_{500}$ or excising the inner core
region. Nevertheless, we decided to follow the standard choice, made
to avoid the influence of the uncertainties related to the status of
the core {(i.e., presence of CCs or NCCs)} on the evolution study.

We apply the fitting procedures described in Section~3.4 and we use
the fitting function expressed by Equation~(\ref{11}), and the pivot
points listed in Table~1. The parameter of the evolution, $\gamma$, is
here fixed to the SS expectations introduced in
Section~3. The results, in terms of best-fitting parameters and
1$\sigma$ uncertainties, are related to the \texttt{robust\_linefit}
method and are reported in Tables 2 and 3 for all relations, ICM
physics treatments, and for different redshifts.

\subsection{The Slope}

The evolution of the slopes of the scaling relations, $\beta$, is
shown in Fig.~\ref{fig:3} for all six scaling relations and ICM
physics. For the following discussion, we remind that the 
expected values of the fitting parameters for the SS predictions are listed in Table~1.

\medskip

\noindent(1) {\it The $M-\mg$ relation.} \\ We observe that the three
runs produce shallower gas slope than the SS prediction
($\beta=1$). The NR value is mildly lower (2-3 per cent) than the SS
value mostly due to the fact that smaller mass systems can lose a
  fraction of their gas as a consequence of violent major encounters. 
This finding is not related to the sample selection applied as we
explained in Section~2.1, indeed, also in the non-radiative
cosmological box of \cite{lebrun.etal.2016} they found that the
$\mg-M$ relation has a slope equal to 1.02, which corresponds to 0.98
once the relation is inverted to be compared with ours.  We confirm
previous results from the literature that compared radiative and
non-radiative runs and find that the slope in the radiative runs is
significantly smaller ($10-15$ per cent) than one
\citep{2010ApJ...715.1508S,battaglia.etal.2013} due to the conversion
of part of the hot gas into stars by the process of radiative cooling
which is more efficient in low massive systems.

 The $M-\mg$ relation is approximately constant over time for the
\ovisc\ run (see Table~2) confirming the expectations: the total mass
and the gas mass grow simultaneously. Indeed, dark matter and gas
increase in mass by the same fraction when the mass growth happens via
slow accretion (due to constant ratio between the densities of the gas
and DM components in the cluster outskirts, e.g.,
\citealt{rasia.etal.2004}) or via major mergers (due to a relatively
constant gas fraction in system of comparable mass, e.g.,
\citealt{2013MNRAS.431.1487P,eckert.etal.2015}, and reference
therein). For this reason, the slope of the \ovisc\ run is very close to 1 
and does not significantly evolve with time.

\begin{figure}
\centering
\includegraphics[width=0.5\textwidth]{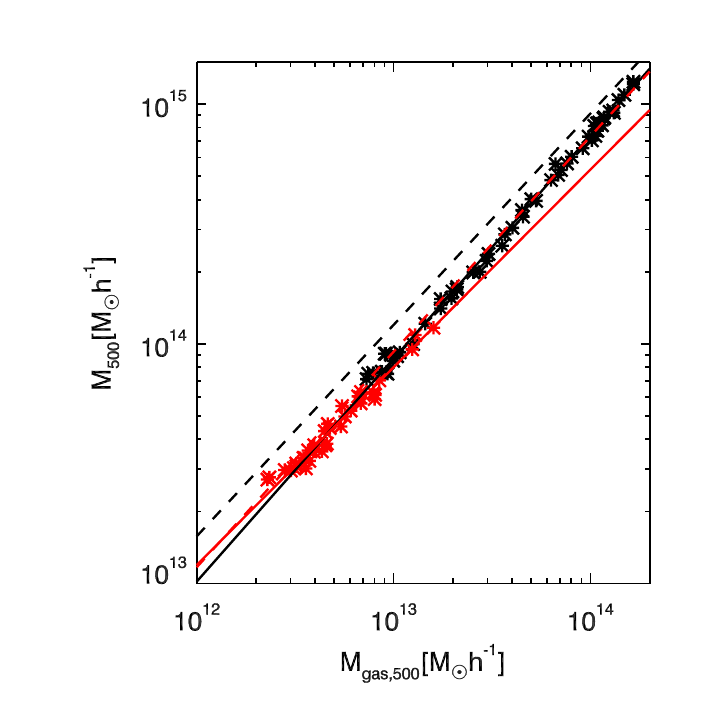}
\caption{$M-\mg$ relation for the AGN clusters at $z=0$ (black
  asterisks) and at $z=2$ (red asterisks). Their best-fitting
  relations are plotted with solid lines together with the best
  fitting relations of the \csf\ sample at $z=0$ (black-dashed line)
  and at $z=2$ (red-dashed line). While for the \csf\ clusters there
  is a net evolution in the normalization, the \agn\ relation is only
  experiencing a change in the slope.}
\label{fig:mmg}
\end{figure}

 As for the radiative simulations, the \csf\ set presents a regular
 shift of the $\mg-M$ relation towards higher normalization as
 redshift decreases (see the shift from the red dashed line for $z=2$
 clusters to the black dashed line for $z=0$ objects in
 Fig.~\ref{fig:mmg}), but there is no drastic change in the slope value (see also Fig.~\ref{fig:3}).
  This is caused by the continuous reduction of
 the gas content for the active stellar production that consumes some
 of the hot gas. On the other hand, when the AGN feedback has been
 effective for some time, it maintains a higher level of hot
 gas. Clearly, the AGN feedback is able to effectively balance the
 radiative cooling and thus to prevent the overcooling and the
 consequent removal of gas from the hot phase, which characterize the
 \csf\ runs. The impact of AGN feedback, however, depends on the cluster mass, and
 thus introduces a modification of the slope as we will further discuss in the following.

\begin{figure*}
\centering
\includegraphics[width=0.95\textwidth]{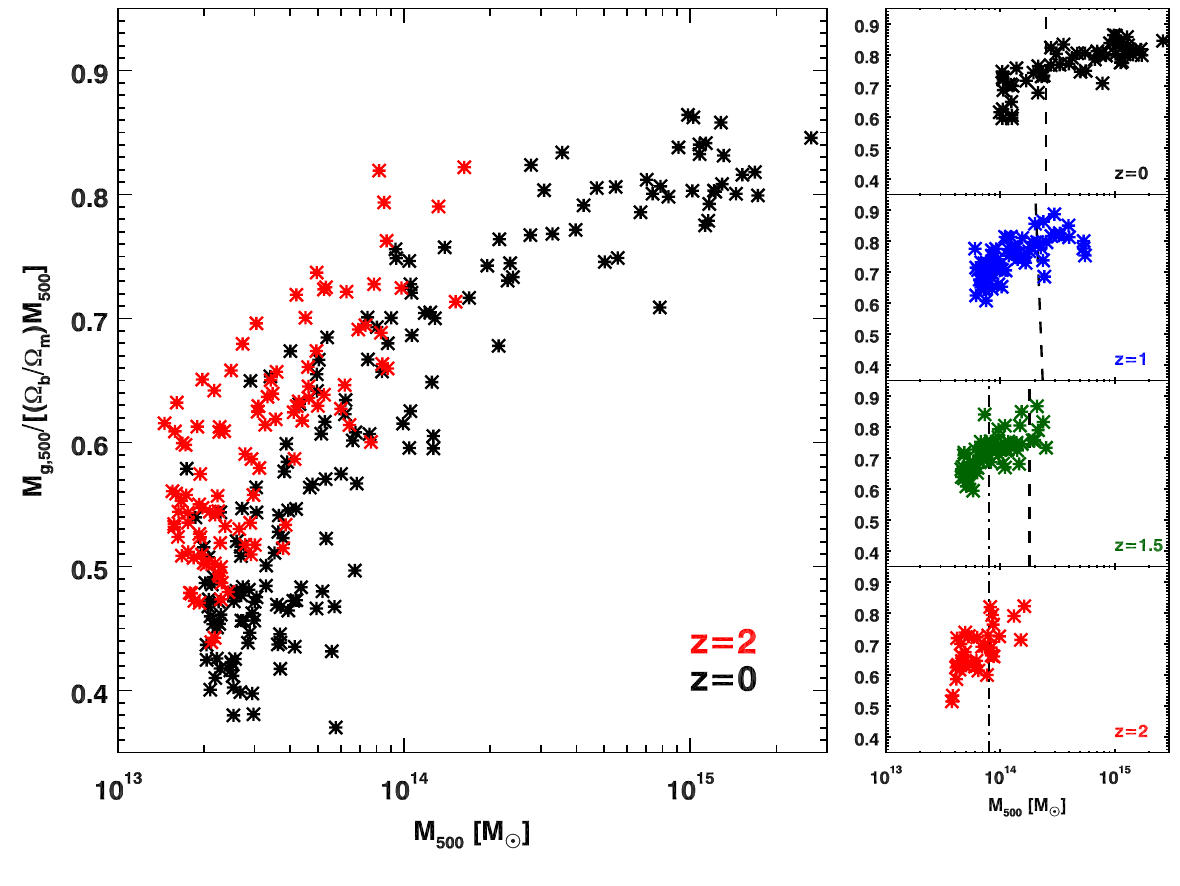}
\caption{In the large panel we show the baryon fraction of the
  \agn\ runs at $z=0$ (black) and $z=2$ (red) normalized for its
  cosmic value. All clusters identified in the 29 Lagrangian regions
  with mass $M_{500}$ above $2 \times 10^{13} M_{\odot}$ are presented
  to display the rapid decrease in gas fraction in low mass
  systems. In the smallest panels at the right, we separately show the
  objects at $z=0,1,1.5$ and $z=2$, but only those above the mass
  limit chosen as threshold ($M_{500}=10^{14} E(z)^{-1} M_{\odot}$).
  We indicate the mass limits of the current SPT selection with the
  dashed-line and that of the future SPT-3G with the dotted-dashed
  line.}
  \label{fig:fg}
\end{figure*}

Looking at Fig.~\ref{fig:3}, we notice that no significant evolution
is measurable in the \agn\ runs from $z=0$ to $z\sim1$ (in agreement
with previous analyses by \citealt{2011MNRAS.416..801F} and
\citealt{battaglia.etal.2013}). At $z=1$, indeed, $\beta$ is just 2.5
per cent lower than at $z=0$.  However, the value of the \agn\ slope,
$\beta_{\mg}$, decreases at $z=1.5$ and even more at $z=2$ to a
maximum difference of $11\%$ with respect to $z=0$.  This discrepancy
is statistically robust and significant at more than 2$\sigma$. To
explain the origin of this behaviour we refer to Fig. \ref{fig:fg}
where we show the trend of the baryon fraction with the total mass at
different redshifts and thus we enhance both the dependence in time
of the $M-\mg$ relation as well as the impact of the SZ cluster
selection which varies with $z$ (see Section 2.1). In the large panel,
we show all the clusters with mass above $2 \times 10^{13} h^{-1}
M_{\odot}$ at $z=0$ in black and $z=2$ in red. In the smallest panels,
we plot the samples considered in this work at four different
redshifts: $z=0,1,1.5,$ and $z=2$. As known from observations, the gas
fraction is almost constant for masses $M_{500}>(2-3) \times 10^{14}
M_{\odot}$ while it decreases with decreasing total mass, below that
limit. This trend seems to be present throughout the cosmic time and
not much variation is detected, except a slightly higher value of the
gas fraction at the highest redshifts at fixed mass.  As previously pointed out, the dependence of the baryon fraction with the cluster
total mass is a particular feature generated by the AGN feedback. In its absence,
 the baryon fraction has a constant value from groups to clusters 
 \citep[e.g.,][]{2010ApJ...715.1508S,2013MNRAS.431.1487P}. For this reason, the SZ-like selection 
does not impact the evolution of the slopes of either the \ovisc\ or the \csf\ runs. On the other hand, the $z=2$ \agn\ sample
is almost entirely on the declining part of the relation, while the
local-universe sample contains a large number of clusters that are
located in the {\it plateau} region.  In other words, the slopes of
the $M-\mg$ relations at $z=0$ and $z=2$ are influenced by the mass
range covered by the sample of clusters. This depends on the SZ-like
selection and it might affect also the current and future SZ
analysis. Indeed, the vertical lines in the smallest panels of
Fig.~\ref{fig:fg} approximately show the mass limits of the selection
function of SPT and SPT-3G.

\medskip

\noindent (2) {\it The $M-\tmw$} and {\it $M-\tsl$ relations.}
\\ Non-radiative runs show a $M-\tmw$ slope consistent at
1$\sigma$ with the SS-predicted value of $\beta=3/2$ for all redshifts
below 1. In the radiative simulations, instead, the efficiency of the
process of radiative cooling, which cools the dense gas to produce
stars, depends on the system mass, being stronger in the low-mass
systems. The removal of this low entropy gas from the hot phase leads
to higher temperatures in groups, and thus to a steeper $M-T$ slope
in the \csf\ and \agn\ runs.

As expected, we do not find a significant difference in the
  values of the slope of the \agn\ $M-T$
  scaling relations obtained by {\it including the core} in the
  computation of the temperatures. The slopes, indeed,
  are consistent within 1$\sigma$ with those reported in Table ~2
   and have a maximum difference of 2.5 per cent 
at $z=0$ and $z=0.25$, otherwise they change by less than
  $\sim$1.5 per cent. Most importantly, they present the same trend
  shown in Fig.~5 and detailed in the following.

The slopes of both temperature relations, $M-\tmw$ and $M-\tsl$, drop
at high redshifts. The values of $\beta_{\tsl}$ at $z=2$ are reduced
with respect to the $z=0$ slope by $16$, $18$, and $16$ per cent in
the \ovisc, \csf, and \agn\ runs, respectively. The decrease in the
$M-\tsl$ slope is present for all the prescriptions of the ICM
physics. This indicates that the origin of the variation is due to
macroscopic events, linked to the global evolution of the clusters.
The same trend is found in the clusters extracted from the
cosmological box (of size $640 h^{-1}$ Mpc) of the Magneticum
simulations\footnote{http://c2papcosmosim.srv.lrz.de/map/find. }
\citep{dolag.etal.2016,ragagnin.etal.2016} at redshifts $z=0$ and $z
\approx 1, 1.5, 2$. As a further confirmation that this finding
  does not depend on the SZ-like selection, we explore the slope
  evolution for the $M-\tsl$ relations derived from all objects that
  at each redshift are above $M_{500}>10^{13} M_{\odot}$.  We find the
  same result as presented in Fig.~5.

\begin{figure}
\centering
\includegraphics[width=0.5\textwidth]{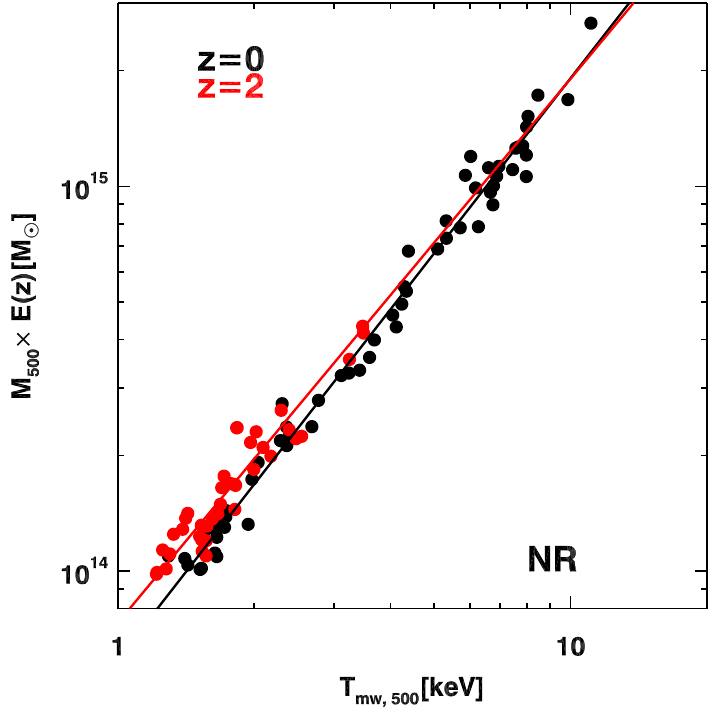}
\caption{Redshift evolution of the $M_{500}-\tmw$ scaling
  relation. The result is shown for the NR simulation at redshifts
  $z=0$ (black) and $z=2$ (red). The solid lines are the best-fitting
  relations corresponding to each redshift.}
\label{fig:4}
\end{figure}

To facilitate the explanation of this result, we plot in
Fig. \ref{fig:4} the $M-\tmw$ relation for the \ovisc\ clusters at
$z=0$ and $z=2$. We chose this comparison, despite its mild $\beta$
variation (less than 10 per cent), to better describe the variation of
the thermal content (linked to $\tmw$ rather than $\tsl$) as
consequence of gravitational interactions more than of radiative
physics. As expected, the kinetic energy of the hot gas is not yet
converted in thermal energy in high-redshift clusters, and therefore
they typically exhibit a lower value of temperature at fixed total
mass, i.e. for $E(z) M < 2 \times 10^{14} h^{-1} M_{\odot}$ the $z=2$
temperatures (red points) are systematically on the left side of the
$z=0$ ones (black points).  The same result was enlightened in fig.~5
of \cite{lebrun.etal.2016}, where they show how the ratio of the
kinetic energy over the thermal energy decreases with time considering
various mass bins of simulated clusters extracted from a cosmological
volume. We confirm the same trend of the energy ratio in our
simulations (not shown).  The phenomenon is present in the entire mass
range but it is less pronounced for the three most massive systems at
$z=2$ that, indeed, lie extremely close to the black solid line
representing the $z=0$ scaling relation. These three objects have recently
experienced a major merger.  As a consequence, their mass has doubled
(notice the mass separation from the rest of the sample), and their
gas has been strongly heated by the induced shocks. To secure that our
result was not influenced by the limited number of objects, we
verified which fraction, among the highest mass systems in the $z \sim
2.3 $ MUSIC-2 sample\footnote{http://music.ft.uam.es}
\citep{sembolini.etal.2013, sembolini.etal.2014}, has recently
experienced a major merger. We found that this condition is verified
for 10 systems out of the 11 objects with mass $E(z) M_{500} > 2.86
\times 10^{14} M_\odot$.  We conclude that if a system about that mass
threshold is already present at $z=2$ it is extremely likely (90\%
probability accordingly to the MUSIC sample) that it just went through
a major merger phase that, generating strong shocks, heats the gas
with a temperature enhancement which is greater than the variation of
the total mass elevated by the power $\beta_{T-M}$ \citep[see
  also][]{rasia.etal.2011}.  To summarize, we expect that while small
clusters are still cold for their potential well, the largest objects
are already located in the $z=0$ scaling relation because of shock
heating due to minor and major mergers. This causes a shallower slope
in the $M-T$ relation. The presence of a significant amount of cold
gas in low-mass objects affects more the spectroscopic-like estimate
of the temperature that, indeed, shows a stronger evolution.

The change in slope is less prominent in the \agn\ runs. In this
circumstance, the gas of the smallest systems at $z=2$ is warmer
because of the recent and intense AGN feedback activity. The
phenomenon brings the smaller objects closer to the $z=0$ relation
reducing the amplitude of the slope evolution in the $M-\tsl$ case,
and even cancelling it for the $M-\tmw$ relation.

\medskip

\noindent (3) {\it The $M-\yx$ relation.} \\ All runs at $z<1$ have a
$M-\yx$ slope close to the predicted SS slope of $5/3$ because of the
opposite deviations of $\beta_{\mg}$ and $\beta_{\tsl}$ from their
SS relations as a response to the changes of the ICM
physics. In particular, the \ovisc\ and \agn\ runs are consistent
within 1$\sigma$ to the SS values while the $\beta_{\yx}$ of the
\csf\ runs is $\leq 3$ per cent below.

The slope of the $M-\yx$ relation is constant until $z=1$ and then
shows a mild decrease (between 5 and 10 per cent) for the radiative
runs consistent with previous results by \cite{sembolini.etal.2014}
and by \cite{pike.etal.2014}.  We notice, also, that the variation of
the \agn\ slope between $z=0$ and $z=1$ is less than 2 per cent and
the two values are consistent at 1$\sigma$.  The origin of this
variation can be understood by decomposing the $\beta_{\yx}$ slope
into the two slopes of the $M-\mg$ and $M-\tsl$
\citep{2014MNRAS.437.1171M}:
\begin{equation}
\beta_{\yx}=\frac{1}{1/\beta_{\mg}+1/\beta_{\tsl}}.
\end{equation}
The complete derivation is presented in the Appendix.  The mildness of
the $\beta_{\yx}$ deviation is generated by the fact that none of the
ICM physics runs present a coincident strong variation in both
$\beta_{\mg}$ and $\beta_{\tsl}$. The \ovisc\ and \csf\ runs have the
largest changes in $\beta_{\tsl}$ but they have a constant
$\beta_{\mg}$, viceversa, the change in $\beta_{\mg}$ for the
\agn\ run is accompanied by the mildest drop of the $\beta_{\tsl}$
value.  From equation~13, it is also clear why the value of the
$M-\yx$ slope is independent from the ICM prescriptions. Indeed, by
comparing the \agn\ and \ovisc\ results of the slopes on the $M-\mg$
and $M-T$ relations, we can notice that the \agn\ runs have lower
$\beta_{\mg}$ but higher $\beta_{\tsl}$, because in the \agn\ runs the
smallest systems, at fixed total mass,  have smaller gas mass but higher
temperature with respect to the other ICM physics (see also discussion
in \citealt{2011MNRAS.416..801F}). This feature makes the $M-\yx$
relation the most suitable for cosmological studies that might include
clusters presenting various astrophysical properties \citep[see
  also][]{biffi.etal.2014}.

\medskip

\noindent (4) and (5) {\it $L-\tsl$} and {\it $L-M$ relations.}  \\ Both
radiative runs exhibit significantly steeper
luminosity-temperature and luminosity-mass slopes compared with the SS
values of $2$ and $4/3$, respectively. The deviation is caused by the
removal of dense gas in small clusters and groups due to efficient
radiative cooling (see above).

 By including the core, we notice overall 
 steeper slopes. However, the absolute
  difference in terms of slopes is less than 1 per cent for all
  redshifts with the exception of $z=0,0.25,1.5$ where it is below 3
  per cent.  All the values obtained by including the core are
  in any case consistent within 1$\sigma$ with those in Table~3.

 In the \csf\ runs, the slopes of the $L-\tsl$ and $L-M$ relations
 respectively decreases by $23$ and 6 per cent at $z=2$ with respect to
 $z=0$, while for the \agn\ sample the same grow by 12.5 and 22.5 per
 cent.  These changes can be again explained by the decomposition of
 the luminosity-based relations' slopes:
\begin{equation}
\beta_{\rm LT}=\beta_{\tsl}\bigg(\frac{2}{\beta_{\mg}}-1\bigg)+\frac{1}{2},
\end{equation}
\begin{equation}
\beta_{\rm LM}=\frac{2}{\beta_{\mg}}+\frac{1}{2\beta_{\tsl}}-1.
\end{equation}
In the first case, $\beta_{\mg}$ and $\beta_{\tsl}$ carry a similar
weight leading to a counterbalance between concurrent changes. In the
second case, $\beta_{\mg}$ is the dominant factor, implying a
variation whenever this occurs in the $M-\mg$ relation. Therefore, the
steepening of the $L-M$ relation at $z>1$ is also due to the gas
depletion within the potential well of small mass systems caused by
the gas expulsion caused by the AGN at even higher redshifts.

Another notable feature in the luminosity relations is the separation
of the slope values among the three versions of baryonic physics. The slope has a
SS behavior for the \ovisc\ simulations while $\beta_{\rm
  LT}$ and $\beta_{\rm LM}$ increase by $\sim 35$ and $\sim$ 25 per
cent, respectively, in the radiative runs. The change is consistent
with the previous argument on the gas fractions: feedback
processes reduce the amount of gas in the simulated smallest systems
and, thus, their total luminosity, similarly to what happens in real
systems. The removal of dense gas in the smallest systems, in the
\csf\ sample is more evident at low redshift when the powerful
radiative cooling cannot be regulated only by stellar
  feedback. On the other hand, in the \agn\ runs, the phenomenon is
more prominent at high redshifts because of the stronger AGN
feedback. The gap between the \agn\ normalization is already well
established at $z=2$ implying that at that epoch the stellar and AGN
feedback in our simulated Universe already had a significant impact in
establishing the energetics of clusters.  Indeed, the AGN need to
eject the gas from the progenitors of groups and clusters at $z>2$ for
getting their baryonic content and ICM properties right at lower
redshift \citep[see e.g.][but see also \citealt{pike.etal.2014}]
{mccarthy.etal.2011, biffi.etal.2017}.

\begin{table*}
 \caption{\label{tb1} Best-fitting normalization, slope, and scatter
   of the $M-\mg$, $M-\tmw$, $M-\tsl$, and $M-\yx$ relations for the
   \ovisc, \csf, and \agn\ runs. The parameters are obtained by
   fitting Equation (\ref{11}) with $X_0$ and $\gamma$ given in
   Table~\ref{tabella1}.  The values of the scatter are derived
     by applying Equation (\ref{eq:sigma}) at fixed mass-proxy and at
     fixed total mass.}
 \begin{center}
 \resizebox{\textwidth}{!}{
 \begin{tabular}{|c|cccc|cccc|cccc|}
 \hline
$M-\mg$ & & NR &  & & & CSF & & & & AGN & & \\
\hline
  z  & $\lgt C$ & $\beta$ & $\sigma_M$ & $\sigma_{Mg}$ & $\lgt C$ & $\beta$ & $\sigma_M$ & $\sigma_{Mg}$ & $\lgt C$ & $\beta$ & $\sigma_M$ & $\sigma_{Mg}$ \\
  \hline
0.0 &  $13.854\pm0.005$ & $0.973\pm0.005$ & $0.018$ & $0.018$ &  $14.079\pm0.007$ & $0.882\pm0.009$ & $0.027$ & $0.031$ &  $13.938\pm0.008$ & $0.930\pm0.009$ & $0.024$ & $0.026$\\
0.25 &$13.841\pm0.003$ & $0.981\pm0.005$ & $0.017$ & $0.017$ &  $14.046\pm0.004$ & $0.891\pm0.006$ & $0.025$ & $0.029$ &  $13.926\pm0.004$ & $0.928\pm0.007$ & $0.022$ & $0.024$\\
0.5 &  $13.838\pm0.002$ & $0.981\pm0.004$ & $0.013$ & $0.013$ &  $14.030\pm0.002$ & $0.898\pm0.005$ & $0.020$ & $0.022$ &  $13.928\pm0.003$ & $0.915\pm0.006$ & $0.019$ & $0.021$\\
0.6 &  $13.834\pm0.002$ & $0.982\pm0.005$ & $0.015$ & $0.015$ &  $14.021\pm0.002$ & $0.898\pm0.006$ & $0.021$ & $0.024$ &  $13.922\pm0.004$ & $0.921\pm0.008$ & $0.023$ & $0.025$\\
0.8 &  $13.825\pm0.002$ & $0.981\pm0.005$ & $0.014$ & $0.014$ &  $14.002\pm0.002$ & $0.903\pm0.007$ & $0.020$ & $0.022$ &  $13.915\pm0.003$ & $0.910\pm0.008$ & $0.022$ & $0.024$\\
1.0 &  $13.823\pm0.002$ & $0.986\pm0.008$ & $0.017$ & $0.017$ &  $13.997\pm0.002$ & $0.904\pm0.009$ & $0.020$ & $0.022$ &  $13.912\pm0.003$ & $0.905\pm0.013$ & $0.025$ & $0.027$\\
1.5 &  $13.816\pm0.002$ & $0.999\pm0.013$ & $0.017$ & $0.017$ &  $13.979\pm0.004$ & $0.909\pm0.012$ & $0.020$ & $0.022$ &  $13.910\pm0.004$ & $0.881\pm0.016$ & $0.025$ & $0.028$\\
2.0 &  $13.817\pm0.004$ & $0.988\pm0.017$ & $0.015$ & $0.016$ &  $13.967\pm0.007$ & $0.899\pm0.018$ & $0.022$ & $0.025$ &  $13.902\pm0.010$ & $0.825\pm0.028$ & $0.032$ & $0.038$\\

 \hline  
 \hline
$M-\tmw$ & & NR &  & & & CSF & & & & AGN & & \\
\hline
z  & $\lgt C$ & $\beta$ & $\sigma_M$ & $\sigma_{\tmw}$ & $\lgt C$ & $\beta$ & $\sigma_M$ & $\sigma_{\tmw}$ & $\lgt C$ & $\beta$ & $\sigma_M$ & $\sigma_{\tmw}$ \\
  \hline
0.0 &  $14.349\pm0.007$ & $1.509\pm0.025$ & $0.051$ & $0.034$ &  $14.240\pm0.007$ & $1.637\pm0.025$ & $0.046$ & $0.028$ &  $14.301\pm0.007$ & $1.613\pm0.027$ & $0.051$ & $0.031$\\
0.25&  $14.352\pm0.006$ & $1.536\pm0.022$ & $0.052$ & $0.034$ &  $14.244\pm0.006$ & $1.642\pm0.020$ & $0.049$ & $0.030$ &  $14.293\pm0.006$ & $1.652\pm0.024$ & $0.052$ & $0.031$\\
0.5 &  $14.379\pm0.006$ & $1.508\pm0.025$ & $0.050$ & $0.033$ &  $14.272\pm0.005$ & $1.622\pm0.025$ & $0.044$ & $0.027$ &  $14.313\pm0.005$ & $1.640\pm0.023$ & $0.042$ & $0.026$\\
0.6 &  $14.388\pm0.006$ & $1.507\pm0.026$ & $0.053$ & $0.035$ &  $14.279\pm0.005$ & $1.589\pm0.028$ & $0.055$ & $0.034$ &  $14.312\pm0.004$ & $1.650\pm0.021$ & $0.040$ & $0.024$\\
0.8 &  $14.376\pm0.005$ & $1.500\pm0.026$ & $0.051$ & $0.034$ &  $14.281\pm0.005$ & $1.598\pm0.029$ & $0.050$ & $0.032$ &  $14.310\pm0.004$ & $1.633\pm0.024$ & $0.040$ & $0.025$\\
1.0 &  $14.386\pm0.008$ & $1.455\pm0.040$ & $0.061$ & $0.042$ &  $14.297\pm0.006$ & $1.585\pm0.036$ & $0.060$ & $0.038$ &  $14.317\pm0.005$ & $1.671\pm0.037$ & $0.048$ & $0.029$\\
1.5 &  $14.417\pm0.014$ & $1.489\pm0.052$ & $0.058$ & $0.039$ &  $14.331\pm0.009$ & $1.512\pm0.047$ & $0.051$ & $0.034$ &  $14.341\pm0.010$ & $1.678\pm0.050$ & $0.046$ & $0.027$\\
2.0 &  $14.398\pm0.021$ & $1.413\pm0.073$ & $0.056$ & $0.039$ &  $14.320\pm0.014$ & $1.467\pm0.072$ & $0.046$ & $0.031$ &  $14.335\pm0.015$ & $1.619\pm0.075$ & $0.041$ & $0.025$\\
 \hline  
$M-\tsl$ & & NR &  & & & CSF & & & & AGN & & \\
\hline
 z  & $\lgt C$ & $\beta$ & $\sigma_M$ & $\sigma_{\tsl}$ & $\lgt C$ & $\beta$ & $\sigma_M$ & $\sigma_{\tsl}$ & $\lgt C$ & $\beta$ & $\sigma_M$ & $\sigma_{\tsl}$ \\
  \hline
0.0 &  $14.416\pm0.016$ & $1.597\pm0.057$ & $0.121$ & $0.076$ &  $14.227\pm0.012$ & $1.731\pm0.048$ & $0.083$ & $0.048$ &  $14.293\pm0.009$ & $1.702\pm0.047$ & $0.081$ & $0.048$\\
0.25 &  $14.426\pm0.012$ & $1.569\pm0.051$ & $0.107$ & $0.068$ &  $14.257\pm0.009$ & $1.700\pm0.035$ & $0.081$ & $0.048$ &  $14.301\pm0.009$ & $1.741\pm0.042$ & $0.078$ & $0.045$\\
0.5 &  $14.476\pm0.015$ & $1.527\pm0.058$ & $0.118$ & $0.077$ &  $14.297\pm0.010$ & $1.701\pm0.048$ & $0.088$ & $0.051$ &  $14.322\pm0.008$ & $1.740\pm0.041$ & $0.065$ & $0.037$\\
0.6 &  $14.498\pm0.017$ & $1.560\pm0.063$ & $0.126$ & $0.081$ &  $14.308\pm0.009$ & $1.656\pm0.049$ & $0.089$ & $0.054$ &  $14.321\pm0.008$ & $1.712\pm0.042$ & $0.068$ & $0.039$\\
0.8 &  $14.457\pm0.014$ & $1.492\pm0.057$ & $0.104$ & $0.070$ &  $14.297\pm0.009$ & $1.654\pm0.055$ & $0.078$ & $0.047$ &  $14.313\pm0.007$ & $1.695\pm0.042$ & $0.060$ & $0.035$\\
1.0 &  $14.447\pm0.019$ & $1.378\pm0.069$ & $0.104$ & $0.075$ &  $14.310\pm0.011$ & $1.573\pm0.066$ & $0.086$ & $0.055$ &  $14.321\pm0.010$ & $1.663\pm0.059$ & $0.068$ & $0.041$\\
1.5 &  $14.475\pm0.026$ & $1.447\pm0.081$ & $0.108$ & $0.075$ &  $14.358\pm0.016$ & $1.494\pm0.069$ & $0.094$ & $0.063$ &  $14.350\pm0.016$ & $1.635\pm0.076$ & $0.072$ & $0.044$\\
2.0 &  $14.421\pm0.027$ & $1.339\pm0.080$ & $0.082$ & $0.062$ &  $14.335\pm0.023$ & $1.425\pm0.089$ & $0.077$ & $0.054$ &  $14.350\pm0.024$ & $1.571\pm0.105$ & $0.064$ & $0.041$\\
 \hline  
 \hline
$M-Y_{X}$ & & NR &  & & & CSF & & & & AGN & & \\
\hline
z  & $\lgt C$ & $\beta$ & $\sigma_M$ & $\sigma_{Y_X}$ & $\lgt C$ & $\beta$ & $\sigma_M$ & $\sigma_{Y_X}$ & $\lgt C$ & $\beta$ & $\sigma_M$ & $\sigma_{Y_X}$ \\
  \hline
0.0 &  $14.064\pm0.009$ & $0.610\pm0.009$ & $0.049$ & $0.080$ &  $14.128\pm0.007$ & $0.585\pm0.007$ & $0.037$ & $0.063$ &  $14.063\pm0.006$ & $0.602\pm0.008$ & $0.037$ & $0.061$\\
0.25 &  $14.064\pm0.006$ & $0.609\pm0.008$ & $0.046$ & $0.076$ &  $14.122\pm0.004$ & $0.587\pm0.005$ & $0.035$ & $0.059$ &  $14.059\pm0.004$ & $0.608\pm0.006$ & $0.033$ & $0.055$\\
0.5 &  $14.087\pm0.005$ & $0.608\pm0.010$ & $0.050$ & $0.082$ &  $14.125\pm0.004$ & $0.592\pm0.007$ & $0.037$ & $0.063$ &  $14.066\pm0.003$ & $0.602\pm0.006$ & $0.028$ & $0.046$\\
0.6 &  $14.091\pm0.006$ & $0.614\pm0.011$ & $0.054$ & $0.088$ &  $14.126\pm0.004$ & $0.587\pm0.007$ & $0.039$ & $0.067$ &  $14.066\pm0.003$ & $0.601\pm0.007$ & $0.030$ & $0.050$\\
0.8 &  $14.076\pm0.005$ & $0.603\pm0.009$ & $0.045$ & $0.074$ &  $14.118\pm0.004$ & $0.589\pm0.008$ & $0.034$ & $0.058$ &  $14.063\pm0.003$ & $0.597\pm0.006$ & $0.025$ & $0.043$\\
1.0 &  $14.082\pm0.006$ & $0.586\pm0.014$ & $0.049$ & $0.084$ &  $14.120\pm0.005$ & $0.582\pm0.012$ & $0.040$ & $0.068$ &  $14.066\pm0.004$ & $0.591\pm0.010$ & $0.032$ & $0.054$\\
1.5 &  $14.091\pm0.009$ & $0.611\pm0.019$ & $0.050$ & $0.082$ &  $14.135\pm0.008$ & $0.578\pm0.015$ & $0.043$ & $0.074$ &  $14.085\pm0.007$ & $0.580\pm0.014$ & $0.034$ & $0.059$\\
2.0 &  $14.072\pm0.014$ & $0.586\pm0.023$ & $0.041$ & $0.071$ &  $14.121\pm0.015$ & $0.562\pm0.021$ & $0.041$ & $0.074$ &  $14.086\pm0.012$ & $0.553\pm0.020$ & $0.032$ & $0.057$\\
 \hline  
  \end{tabular}}
 \end{center}
 \end{table*}
 
  \begin{table*}
 \caption{\label{tb6} Similar to Table \ref{tb1} but for the $L-\tsl$
   and $L-M$ relations. The values of $\sigma_L$ are derived at
     fixed $\tsl$ in the upper panel and a fixed total mass in the
     bottom panel. The other measurement of the scatter is obtained at
     fixed luminosity.}
 \begin{center}
 \resizebox{\textwidth}{!}{
 \begin{tabular}{|c|cccc|cccc|cccc|}
 \hline
$L-\tsl$ & & NR &  & & & CSF & & & & AGN & & \\
\hline
 z  & $\lgt C$ & $\beta$ & $\sigma_L$ & $\sigma_{\tsl}$ & $\lgt C$ & $\beta$ & $\sigma_L$ & $\sigma_{\tsl}$ & $\lgt C$ & $\beta$ & $\sigma_L$ & $\sigma_{\tsl}$ \\
  \hline
0.0 &  $ 0.857\pm0.023$ & $2.155\pm0.083$ & $0.180$ & $0.083$ &  $ 0.154\pm0.033$ & $2.881\pm0.103$ & $0.192$ & $0.067$ &  $ 0.497\pm0.026$ & $2.903\pm0.086$ & $0.189$ & $0.065$\\
0.25& $ 0.897\pm0.018$ & $2.075\pm0.079$ & $0.162$ & $0.078$ &  $ 0.279\pm0.021$ & $2.765\pm0.092$ & $0.180$ & $0.065$ &  $ 0.537\pm0.021$ & $2.958\pm0.098$ & $0.176$ & $0.059$\\
0.5 &  $ 0.988\pm0.015$ & $2.008\pm0.062$ & $0.135$ & $0.067$ &  $ 0.406\pm0.017$ & $2.755\pm0.087$ & $0.164$ & $0.059$ &  $ 0.586\pm0.017$ & $3.101\pm0.104$ & $0.150$ & $0.048$\\
0.6 &  $ 1.032\pm0.020$ & $2.154\pm0.080$ & $0.132$ & $0.061$ &  $ 0.439\pm0.016$ & $2.743\pm0.084$ & $0.155$ & $0.057$ &  $ 0.590\pm0.017$ & $3.021\pm0.088$ & $0.147$ & $0.049$\\
0.8 &  $ 1.023\pm0.017$ & $2.050\pm0.067$ & $0.128$ & $0.062$ &  $ 0.487\pm0.016$ & $2.714\pm0.091$ & $0.155$ & $0.057$ &  $ 0.606\pm0.019$ & $3.159\pm0.109$ & $0.156$ & $0.049$\\
1.0 &  $ 1.031\pm0.027$ & $2.107\pm0.097$ & $0.133$ & $0.063$ &  $ 0.525\pm0.015$ & $2.631\pm0.080$ & $0.132$ & $0.050$ &  $ 0.612\pm0.018$ & $3.146\pm0.0102$ & $0.131$ & $0.042$\\
1.5 &  $ 1.008\pm0.032$ & $1.964\pm0.103$ & $0.127$ & $0.065$ &  $ 0.603\pm0.020$ & $2.267\pm0.102$ & $0.135$ & $0.060$ &  $ 0.660\pm0.023$ & $3.242\pm0.134$ & $0.133$ & $0.041$\\
2.0 &  $ 1.001\pm0.045$ & $2.033\pm0.136$ & $0.096$ & $0.047$ &  $ 0.615\pm0.033$ & $2.223\pm0.132$ & $0.088$ & $0.040$ &  $ 0.655\pm0.054$ & $3.268\pm0.262$ & $0.171$ & $0.052$\\
 \hline  
 \hline
$L-M$ & & NR &  & & & CSF & & & & AGN & & \\
\hline
 z  & $\lgt C$ & $\beta$ & $\sigma_L$ & $\sigma_{M}$ & $\lgt C$ & $\beta$ & $\sigma_L$ & $\sigma_{M}$ & $\lgt C$ & $\beta$ & $\sigma_L$ & $\sigma_{M}$ \\
  \hline
0.0 &  $ 0.294\pm0.031$ & $1.345\pm0.049$ & $0.155$ & $0.116$ &  $-0.229\pm0.034$ & $1.661\pm0.055$ & $0.192$ & $0.116$ &  $-0.006\pm0.034$ & $1.701\pm0.060$ & $0.224$ & $0.131$\\
0.25 &  $ 0.332\pm0.024$ & $1.320\pm0.052$ & $0.160$ & $0.121$ &  $-0.129\pm0.022$ & $1.625\pm0.051$ & $0.179$ & $0.110$ &  $ 0.042\pm0.024$ & $1.699\pm0.061$ & $0.194$ & $0.114$\\
0.5 &  $ 0.359\pm0.016$ & $1.312\pm0.039$ & $0.138$ & $0.105$ &  $-0.048\pm0.017$ & $1.615\pm0.057$ & $0.168$ & $0.104$ &  $ 0.056\pm0.016$ & $1.770\pm0.057$ & $0.170$ & $0.096$\\
0.6 &  $ 0.348\pm0.013$ & $1.380\pm0.040$ & $0.118$ & $0.086$ &  $-0.034\pm0.018$ & $1.659\pm0.059$ & $0.170$ & $0.102$ &  $ 0.073\pm0.018$ & $1.758\pm0.056$ & $0.168$ & $0.096$\\
0.8 &  $ 0.400\pm0.012$ & $1.375\pm0.045$ & $0.117$ & $0.085$ &  $ 0.048\pm0.016$ & $1.640\pm0.058$ & $0.156$ & $0.095$ &  $ 0.108\pm0.016$ & $1.856\pm0.064$ & $0.153$ & $0.082$\\
1.0 &  $ 0.388\pm0.016$ & $1.526\pm0.065$ & $0.139$ & $0.091$ &  $ 0.078\pm0.017$ & $1.672\pm0.066$ & $0.145$ & $0.087$ &  $ 0.121\pm0.018$ & $1.890\pm0.073$ & $0.145$ & $0.077$\\
1.5 &  $ 0.370\pm0.019$ & $1.360\pm0.057$ & $0.111$ & $0.082$ &  $ 0.115\pm0.020$ & $1.514\pm0.058$ & $0.114$ & $0.075$ &  $ 0.170\pm0.030$ & $1.967\pm0.093$ & $0.144$ & $0.073$\\
2.0 &  $ 0.434\pm0.033$ & $1.509\pm0.087$ & $0.093$ & $0.062$ &  $ 0.186\pm0.029$ & $1.555\pm0.084$ & $0.102$ & $0.066$ &  $ 0.252\pm0.061$ & $2.084\pm0.151$ & $0.136$ & $0.065$\\
 \hline  
  \end{tabular}}
 \end{center}
 \end{table*}

 \subsection{The Intrinsic Scatter}

\begin{figure*}
\centering
\includegraphics[width=0.33\textwidth]{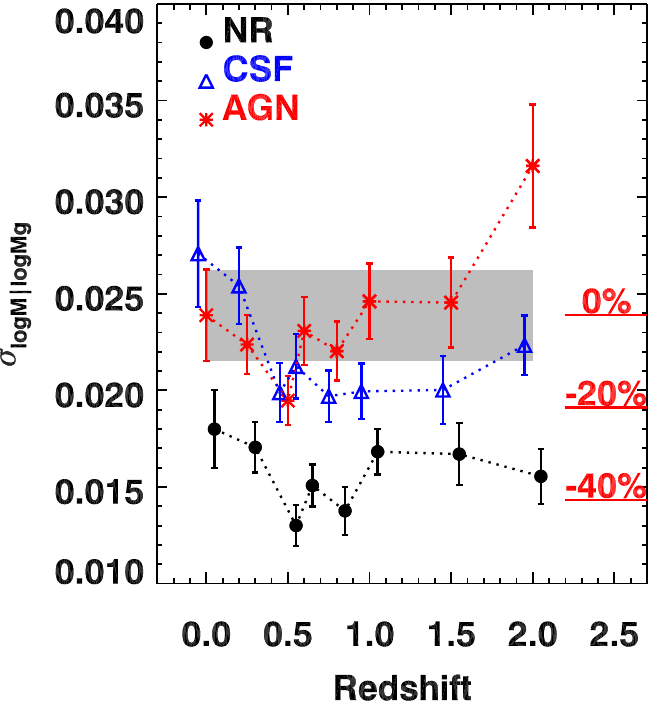}
\includegraphics[width=0.33\textwidth]{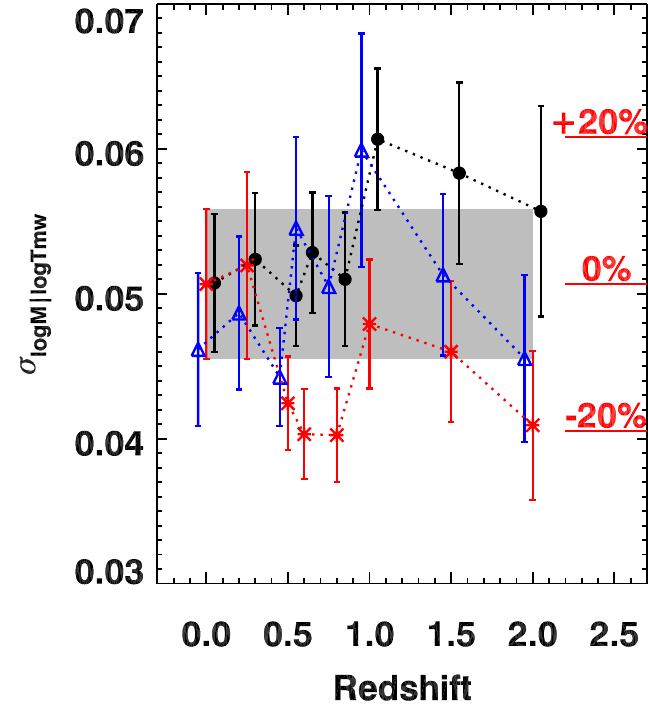}
\includegraphics[width=0.33\textwidth]{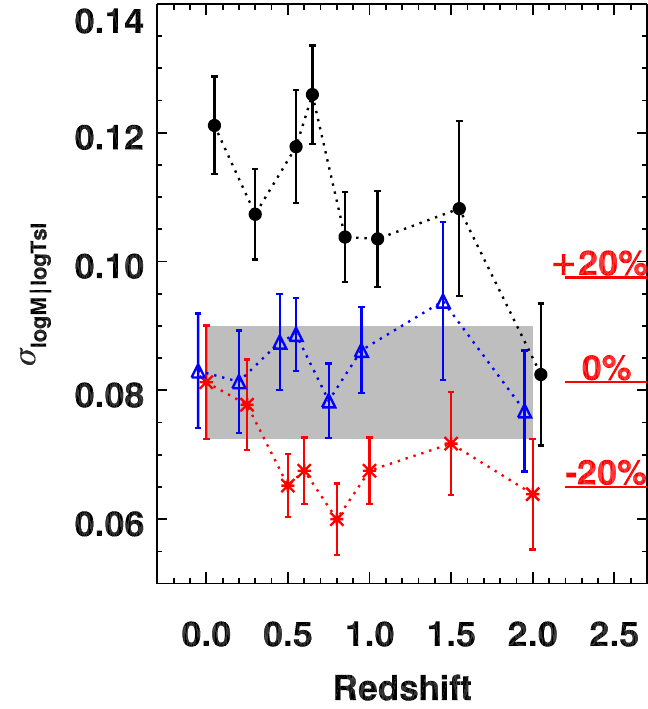}
\includegraphics[width=0.33\textwidth]{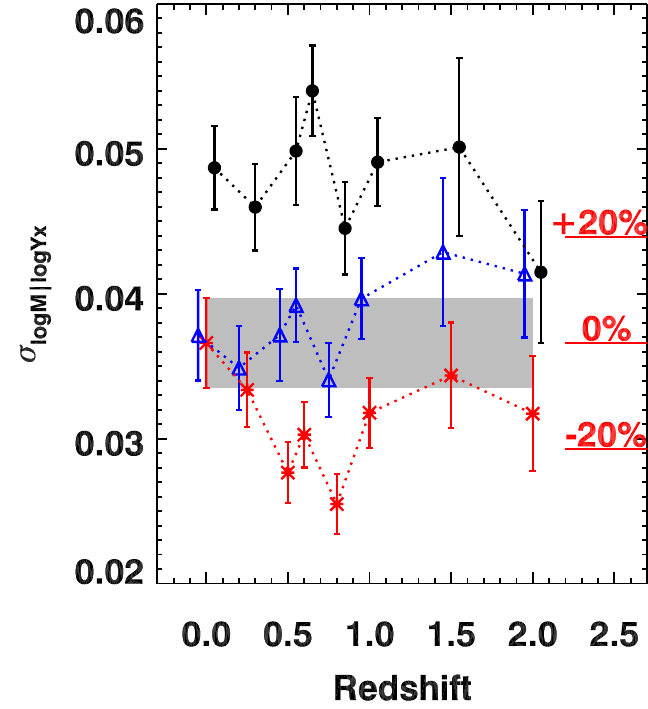}
\includegraphics[width=0.33\textwidth]{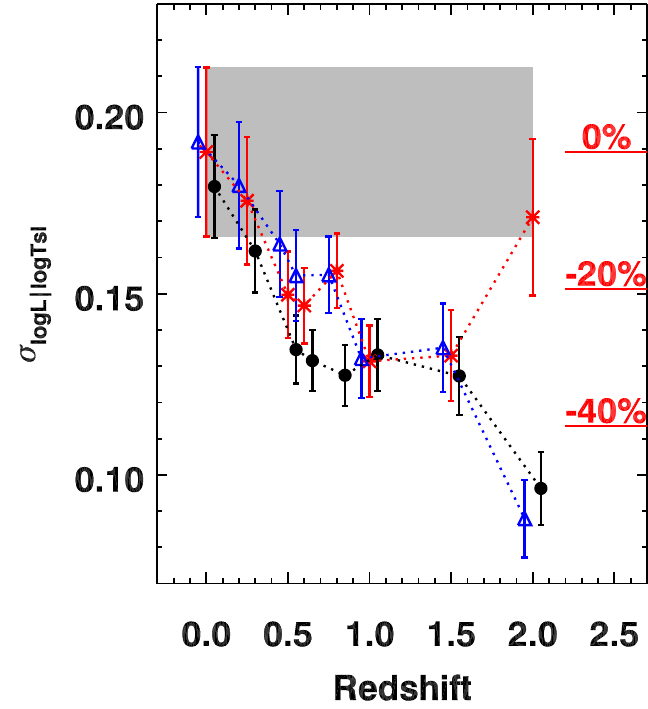}
\includegraphics[width=0.33\textwidth]{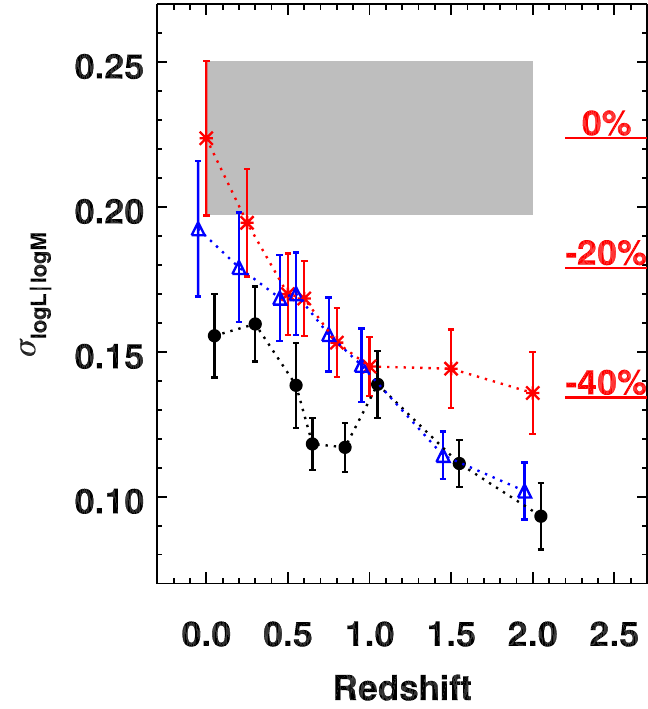}
\caption{The evolution of intrinsic scatter for $M-\mg$, $M-\tmw$,
  $M-\tsl$, $M-Y_{X}$, $L-\tsl$, and $L-M$ is shown (from left to
  right, from top to bottom) as function of redshift and for the three
  simulated runs. The error bars represent the $1\sigma$ error, and
  the grey shaded area stands for the $68\%$ uncertainty of the AGN
  intrinsic scatter at $z=0$.}
\label{fig:5}
\end{figure*}

 We compute the scatter, $\sigma$, defined as the mass variance in the
 decimal logarithm at fixed signal ($\mg, \tmw, \tsl, \yx$) for the
 mass-proxy relations and as the luminosity variance in the decimal
 logarithm at fixed $\tsl$ and $M_{\rm tot}$ for the remaining
 relations.  Using the same syntax of Equation ~11, the scatter can be
 computed as

\begin{equation}
\sigma = \bigg[ \frac{\sum_{i=1}^N [log_{10}(Y_i)-log_{10}(F(X_i))]^2}{N-2}\bigg]^{(1/2)}
\label{eq:sigma}
\end{equation}
where $N$ is the number of clusters in the redshift bin and $Y_i$
represents either the total mass or the bolometric luminosity of each
object.  The resulting scatter is fully consistent with the
  scatter value returned by the routine
  {\texttt{linmix\_err.pro}}.
   The two values are, indeed, only a few per
  cent different between each other, with an absolute maximum
variation below 0.01,
    meaning 
  that this
  quantity is robustly defined for our samples.

In Fig.~\ref{fig:5}, we report its evolution for all our scaling
relations. In addition, we evaluate the scatter measured at fixed mass
or fixed luminosity. Both scatter values can be found in the last two
columns for each physics of Tables 2 and 3.  As often stressed in
  the paper, we remind that our analysis focusses on the {\it
    relative} trend of the scatter rather than its absolute
calibration. Indeed, the latter might depend on the size of the
  sample considered. We do not have hundreds of objects as we would
  expect from selecting all the clusters in the entire cosmological
  box of 1 $h^{-1}$ Gpc, and therefore we are unavoidably limiting
  the presence of outliers. The expectation is that the scatter from
  our sample is biased low. Nevertheless, our goal is to quantify the
  general behaviour of the scatter, investigating its evolution, and
  checking its dependence on the ICM physics.

All mass--proxy scaling relations, with the exception of $M-\tsl$ for
the \ovisc\ runs, present a negligible trend with redshift: the
scatters at $z=1.5$ or $z=2$ are consistent within 1$\sigma$ or 1.5$\sigma$
with the $z=0$ scatter.  The luminosity-based relations, instead,
present a significant variation of the scatter, which decreases with
increasing redshift as can be seen in the figure.

  The $M-\mg$ relation always presents the minimum amount of scatter
  in line with the results from \cite{2010ApJ...715.1508S} and
  \cite{2010MNRAS.401.1670F}. The values are around $2-3$ per cent
  which are 1.5 times smaller than $\sigma_{M|\yx}$ and a factor of
  $2$--$4$ smaller than the scatter of the two temperatures
  ($\sigma_{M|\tmw}=0.05$ and $\sigma_{M|\tsl} =0.08-0.10$,
  respectively).  This is not surprising since the value of
  $\sigma_{M|\yx}$ is consistent with the statistical expectations
  \citep{2010ApJ...715.1508S}: $(\sigma_{M|\yx}/\beta_{\yx})^2 =
  (\sigma_{M|\mg}/\beta_{\mg})^2 + (\sigma_{M|\tsl}/\beta_{\tsl})^2 +
  2 C \sigma_{M|\mg}/\beta_{\mg} \sigma_{M|\tsl}/\beta_{\tsl}$ where
  the correlation factor $C$ between $\mg$ and $\tsl$ is not negative.
  
  The increase of 25-50 per cent with redshift of the \ovisc\ scatter
  of the $M-\tsl$ scaling relation is most likely generated by the
  sequence of minor mergers of smaller and colder substructures. Their
  diffuse gas is efficiently stripped and mixed when the stellar and
  AGN feedback are present, but it is more resilient to be
  incorporated to the main cluster ICM in case of the
  \ovisc\ simulations \citep{dolag.etal.2009}.  The inclusion of
    the core in the \agn\ runs also increases the
    scatter of the $M-\tsl$ by 15-20 per cent, while negligible
    differences (below 3 per cent) are detected for the $M-\tmw$
    relation.

The two luminosity-related scaling relations $L-T$ and $L-M$ have the
highest intrinsic scatter. They reach 0.2 at $z=0$ and decrease to
$0.1$--$0.15$ at $z=2$. The largest variation affects the $L-M$
relation. The increase of the scatter in more recent times indicates
that the luminosity is sensitive to the entire merger history of the
clusters and that the most significant deviations from the global
scaling relation originate from recent ($z<1$) massive mergers.  A
similar trend in the luminosity scatter was recently found by the
Weighing the Giants team \citep{mantz.etal.2016}. The variation of the
$L-M$ scatter with redshift is an important factor that needs to be
considered for cosmological studies. Luckily, the change goes in the
direction of reducing, at higher redshift, the Eddington bias caused
by the different scattering of objects across the threshold of a flux
limited sample
\citep[e.g.,][]{stanek.etal.2006,2012MNRAS.421.1583M}. Finally,
radiative phenomena such as stellar or AGN feedback can also have the
effect of diversifying the systems' luminosity at fixed mass. Indeed,
both the \csf\ and \agn\ runs show a 20--30 per cent higher
$\sigma_{L|M}$ scatter than the \ovisc\ simulations. The scatter
  of the \agn\ runs further increases by 20-40 per cent when the core
  is considered in the computation of the temperature confirming the
  fragility of this scaling relation since its characterization
  depends on many factors as already seen in Section~4.

\begin{figure*}
\centering
\includegraphics[width=0.9\textwidth]{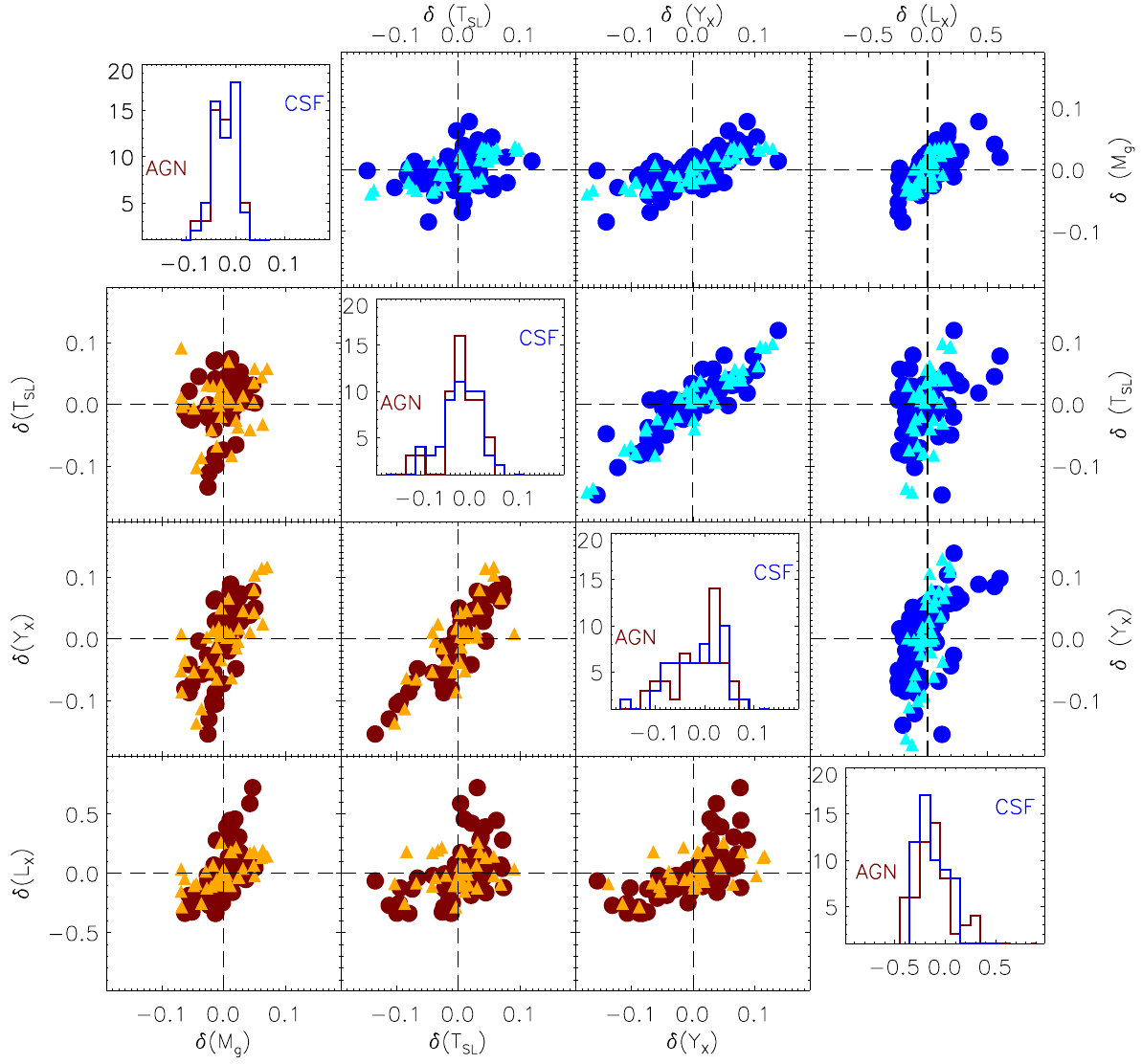}
\caption{ In the diagonal panels, we report the scatter of the
  mass-proxy scaling relations at fixed total mass for the $z=0$
  samples of the \agn\ set (brown line) and \csf\ run (blue line). In
  the panels below and above the diagonal, we represent the covariance
  matrix between the various signals for the \agn\ and \csf\ samples,
  respectively. The circles denote $z=0$ results while the triangles
  refer to the $z=2$ values. The Pearson correlation coefficients for
  $z=0,1,$ and $2$ are listed in Table~4.}
\label{fig:cov}
\end{figure*}

As a second step, we investigated the shape of the deviations of each
signal, $\delta$, from the best-fitting relations {\it at fixed mass}
and their covariance matrix. In Fig.~\ref{fig:cov}, we report the
results for the \csf\ and \agn\ simulations at redshift $z=0$ and
$z=2$, while in Table~\ref{tab:cov} we also list the coefficients
  for the \ovisc\ simulations and for $z=1$. This analysis allows us
to identify the couples of signals with high correlation or
anticorrelation. We quantify this measure via the Spearman's
  rank coefficient, $r$. In Table~4, we highlight the
  signal pairs with $|r| > 0.5$ at a high significance level,
  i.e. when the null hypothesis probability is less than
  $10^{-4}$. These pairs of signals can be jointly used to reduce
the mass scatter with respect to the scatter obtained in the
individual scaling relations
\citep{2010ApJ...715.1508S,ettori.etal.2012,evrard.etal.2014,wu.etal.2015}.
The scatter at fixed mass (shown in the diagonal panels) can be
accurately described by lognormal distributions \citep[see
  also][]{lebrun.etal.2016} at all times.

 Regarding the \agn\ physics, we find that all the couples of X-ray
 quantities present positive correlations at all redshifts with the
 exception of no correlation found between $\delta_{\mg}$ and
 $\delta_{\tsl}$ at all redshifts and between $\delta_{L}$ and
 $\delta_{\tsl}$ at $z=2$. In addition, we notice that the 0.5
   correlation between $\delta_{Y_X}$ and $\delta_L$ at $z=2$ has a
   0.2 per cent probability to be obtained by chance, and therefore
   it is somehow uncertain. The former behaviour is consistent with
 the already-discussed argument about the different time-scales on the
 variation of $\mg$ and $\tsl$ in reaction to mergers and accretion:
 since the two quantities increase at subsequent times we do not
 expect particular correlation.  This is particularly true at $z=2$
 where no correlation between $\delta_{\tsl}$ and $\delta_L$ is also
 expected. Indeed, on the one hand, mergers produce an increase of
 both luminosity and temperature generating a positive correlation
 between the two deviations. On the other hand, strong AGN bursts,
 common phenomena at these redshifts, are likely to cause an increase
 of temperature along with a temporary {\it decrease} of luminosity, due
 to the gas lost as ejected material. This produces a negative
 correlation.  As a confirmation, when the \agn\ are not present the
 correlation at $z=2$ becomes positive and quite strong,
   i.e. $\sim$0.6 in the \ovisc\ run and $\sim$0.7 in the \csf\ case.
 The correlation values that we find between $\delta_L$ and
 $\delta_{\tsl}$ ($r=0.52$ and $r=0.54$) at $z\leq1$ are in good
 agreement with the value $r=0.56\pm0.10$ found by
 \cite{mantz.etal.2016} in a sample containing both relaxed and
 unrelaxed clusters. In this observational paper, the authors claim a
 clean separation between the two classes of systems: CC regular
 systems tend to have higher $\delta_L$ and $\delta_{T}$ than the
 highly disturbed NCC objects in their sample. In our simulations, we
 find that the six CC systems that are X-ray regular have, indeed,
 the largest positive deviation of both quantities.

The $\yx$ parameter shows a strong positive correlation with
  Spearman's coefficient $r>0.5$ with all the parameters
considered. Clusters deviate from the $M-\yx$ mostly due to the
changes in their temperature \citep{rasia.etal.2011} and this is
confirmed by the fact that the highest correlation value is that
between $\yx$ and $\tsl$. Indeed, $r$ is always greater than
  $\sim 0.85$ with the exception of $z=2$ where it still shows a strong
  correlation with $r=0.67$. This result is confirmed even
  when changing the ICM physics.

We confirm that a combination of the luminosity and temperature
($M\propto L\times T$) for local clusters as proposed in
\cite{ettori2013} is a good approach to reduce the scatter with
respect to the scatters of the individual relations $M-L$ and
$M-T$.  In general, the luminosity is well correlated with the
  other quantities with only few exceptions. A lower level of
  correlation is found between the luminosity and the temperature in
  the \ovisc\ runs and no correlation is detected between the
  luminosity and the gas mass at higher redshifts. This is most likely
  related to merging events that significantly boost the luminosity in
  non-radiative simulations.

 The \csf\ and \agn\ runs present similar values of the correlation
 coefficients beside the mentioned difference at high redshift for the
 pair $\delta(\tsl)$ and $\delta(\mg)$ and the associated couple of
 signals $\delta(\tsl)$ and $\delta(L_X)$.

\begin{figure*}
\centering
\includegraphics[width=0.49\textwidth]{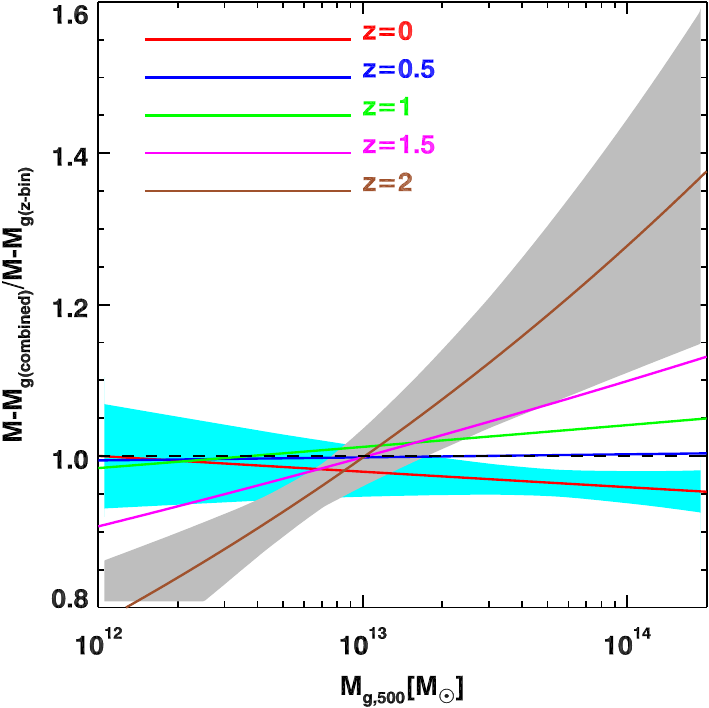}
\includegraphics[width=0.49\textwidth]{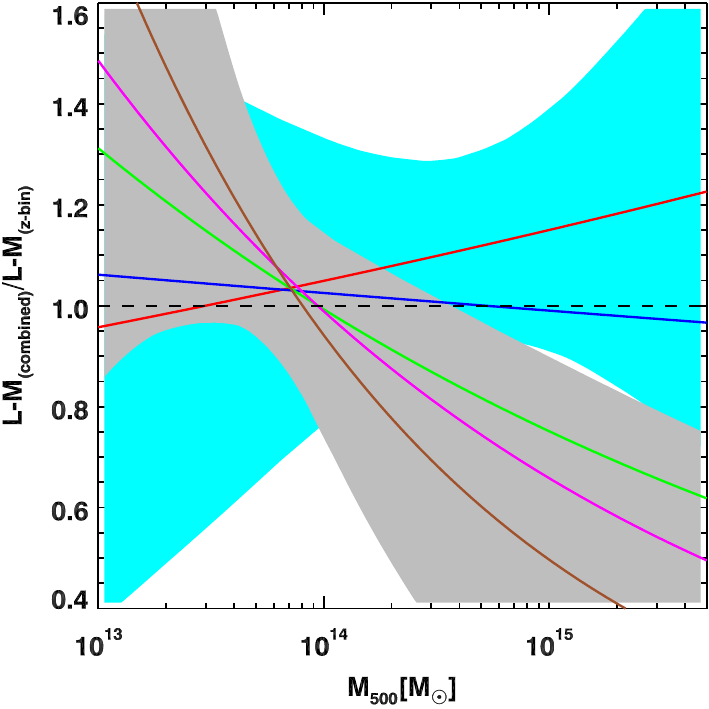}
\caption{ We report, for the \agn\ sample, the ratio for the
    $M-M_{g}$ (left) and $L-M$ (right) between the scaling relations
    of the {\it combined} sample derived via Eq.~11 (best-fitting
    parameters in Table~5) and the scaling relations previously found
    in single {\it redshift bins} (best-fitting parameters in Tables~2
    and ~3).  The shaded area shows the 68.3\% uncertainties for two
    cases: $z=0$ (cyan) and $z=2$ (grey). The red, {\bf blue}, green,
    magenta, and brown lines refer to the specific comparison at the
    $z$ bins equal to $z=0,0.5,1,1.5,$ and $z=2$, respectively.}
\label{fig:7}
\end{figure*}
 
 \begin{table}
   \caption{\label{tab:cov} Spearman's rank correlation coefficients,
     $r$, between two sets of deviations from the best-fitting scaling
     relation at fixed mass and, in parenthesis, the associated
     null-hypothesis values. In bold we indicate the values above
       0.5 with a sufficiently low (below $10^{-4}$) null-hypothesis
       value.}
 \begin{center}
 \begin{tabular}{c|l|l|l}
 {\centering \ovisc} & {\centering $z=0$} & {\centering $z=1$} & {\centering $z=2$}\\
\hline
$\delta(\tsl);\delta(\mg)$ &   0.20  (1 $10^{-1}$) &{\bf 0.50} (3 $10^{-6}$)   & {\bf 0.63} (2 $10^{-5}$)\\
$\delta(\yx);\delta(\mg)$ &   0.40  (2 $10^{-3}$) & {\bf 0.61} (1 $10^{-9}$)  & {\bf 0.75} (3 $10^{-8}$)\\
$\delta(L_X);\delta(\mg)$&  {\bf 0.52}  (2 $10^{-5}$)& 0.42 (8 $10^{-5}$)  &  0.31 (5 $10^{-2}$)\\
$\delta(\yx);\delta(\tsl)$   &  {\bf 0.96}  (3 $10^{-35}$)& {\bf 0.98} (0)           &  {\bf 0.96} (2 $10^{-22}$)\\
$\delta(\tsl);\delta(L_X)$ &  0.34  (8 $10^{-3}$)  & {\bf 0.59} (6 $10^{-9}$)  &  {\bf 0.59} (7 $10^{-5}$)\\
$\delta(\yx);\delta(L_X)$ &  0.42  (8 $10^{-4}$)  & {\bf 0.61} (1 $10^{-9}$)  &  {\bf 0.57} (1 $10^{-4}$)\\
\hline
\hline
{\centering \csf} & {\centering $z=0$} & {\centering $z=1$} & {\centering $z=2$}\\
 \hline
$\delta(\tsl);\delta(\mg)$ &   0.33  (9 $10^{-3}$)  &{\bf 0.53} {(3 $10^{-7}$)}    & {\bf 0.76}  {(2 $10^{-8}$)}\\
$\delta(\yx);\delta(\mg)$ &   {\bf 0.72}  (4 $10^{-11}$)& {\bf 0.75} (4 $10^{-16}$)  & {\bf 0.88} (8 $10^{-14}$)\\
$\delta(L_X);\delta(\mg)$ & {\bf 0.73}  (4 $10^{-11}$)& {\bf 0.67} (3 $10^{-12}$)  &  {\bf 0.76} {(2 $10^{-8}$)}\\
$\delta(\yx);\delta(\tsl)$ &    {\bf 0.87}  (2 $10^{-19}$)& {\bf 0.94} (7 $10^{-41}$)  &  {\bf 0.96} (1 $10^{-21}$)\\
$\delta(\tsl);\delta(L_X)$ &  0.40  (2 $10^{-3}$)  & {\bf 0.59} {(5 $10^{-9}$)}    &  {\bf 0.69} {(1 $10^{-6}$)}\\
$\delta(\yx);\delta(L_X)$ &  {\bf 0.65}  (2 $10^{-8}$)  & {\bf 0.66} (8 $10^{-12}$)  &  {\bf 0.72} {(3 $10^{-7}$)}\\
\hline
\hline
{\centering \agn} & {\centering $z=0$} & {\centering $z=1$} & {\centering $z=2$}\\
\hline
$\delta(\tsl);\delta(\mg)$ &   0.40  {(2 $10^{-3}$)} &0.24 {(4 $10^{-2}$)}   & 0.09 (6 $10^{-1}$)\\
$\delta(\yx);\delta(\mg)$ &   {\bf 0.71}  (5 $10^{-10}$) & {\bf 0.68} (2 $10^{-11}$)  & {\bf 0.72} (6 $10^{-7}$)\\
$\delta(L_X);\delta(\mg)$& {\bf 0.64}  {(5 $10^{-8}$)}  & {\bf 0.62} {(3 $10^{-9}$)}  &   {\bf 0.64} (3 $10^{-5}$)\\
$\delta(\yx);\delta(\tsl)$   &  {\bf 0.90}  (1 $10^{-21}$) & {\bf 0.84} (3 $10^{-21}$)  &  {\bf 0.67} (7 $10^{-6}$)\\
$\delta(\tsl);\delta(L_X)$ &  {\bf 0.54}  {(1 $10^{-5}$)}  & {\bf 0.52} {(2 $10^{-6}$)}  &  0.16 (3 $10^{-1}$)\\
$\delta(\yx);\delta(L_X)$ &  {\bf 0.66}  {(2 $10^{-8}$)}  & {\bf 0.72} (6 $10^{-13}$)  &  0.50 (2 $10^{-3}$)\\
 \end{tabular}
 \end{center}
 \end{table}

 \begin{table*}
   \caption{\label{tb8}Best-fitting parameters from the Bayesian fit of Equation~(11) to the \agn\ data in the redshift range $[0-2]$ with $\gamma$ free to vary.} \begin{center}
 \begin{tabular}{c|rrrr}
 \hline
 & $\lgt C$ &  $\beta$ & $\gamma$  &  $\sigma$\\
 \hline   
 $M-\mg$ &  $13.932\pm0.002$   &  $ 0.916\pm 0.002$ &  $-0.047\pm 0.011$&   $ 0.025\pm 0.001$ \\
 $M-\tmw$  & $14.0112\pm0.003$   &  $ 1.623\pm 0.003$ &  $-0.902\pm 0.017$ & $ 0.045\pm 0.001$\\
  $M-\tsl$& $14.004\pm0.005$ & $ 1.661\pm 0.005$ & $-0.847\pm 0.027$ & $ 0.069\pm 0.002$\\
 $M-Y_{X}$  & $14.058\pm0.002$  & $ 0.597\pm 0.002$ & $-0.314\pm 0.013$ & $ 0.032\pm 0.001$\\
 $L-\tsl$ & $ 0.044\pm0.012$  & $ 2.877\pm 0.012$& $ 1.161\pm 0.063$& $ 0.162\pm 0.005$ \\
 $L-M$ & $ 0.058\pm0.015$  & $ 1.673\pm 0.015$ & $ 2.520\pm 0.077$  & $ 0.176\pm 0.005$  \\
 \hline
 \end{tabular}
 \end{center}
 \end{table*}

\subsection{The Normalization}

The evolution of the normalization is mostly expressed as a power
${\gamma}$ of the factor $E(z)$ \citep[see review on scaling-relation
  evolution by][and references therein]{2013SSRv..177..247G}. In
  other words, the evolution of the normalization is represented as a
  simple upward or downward shift. In this section, we want to
enlighten how this procedure is inaccurate whenever the slope of
  the considered scaling relation varies with time
\citep{branchesi.etal.2007}.

Observationally, we still do not have any indication of evolution of
$\beta$ due to the paucity of observed $z\sim 1$ clusters. The
situation will improve thanks to the collection of high-$z$ data from
SPT-3G \citep{benson.etal.2014} or {\it eROSITA}
\citep{2013SSRv..177..247G}. To provide a forecast on the
  precision that these surveys could reach, there are already ongoing
  studies that extend local samples by including higher redshift
  objects. In addition, Bayesian techniques have been
  developed (see LIRA by \citealt{sereno_ettori2015} and
  \citealt{sereno2016} and also \citealt{andreon.etal.2012}). These
  include a proper treatment for the sample selection
  \citep[e.g.][]{sereno_ettori2015} and the associated selection
  biases such as the Malmquist and Eddington biases \citep[see][for a
    discussion]{allen.etal.2011}. However, the analysis of the data
  cannot yet be performed in redshift bins, but it, typically,
follows two main approaches. Either the slope is computed in a {\it
  local} sample and then fixed to the entire set
(e.g. \citealt{2014MNRAS.444.2723C}) or the fit on the
scaling relations is simultaneously applied to all objects
(e.g. \citealt{2012MNRAS.424.2086H}; \citealt{2015arXiv151203833G}).

In other words, either method assumes a constant slope. To warn about
their application, we focus on the \agn\ $M-\mg$ and $L-M$ relations
because they exhibit the largest slope variation, and therefore they
are the best suited to put in evidence a possible misinterpretation
of the data.  We follow the second observational approach that is the
most used and we build a sample that includes the \agn\-simulated
clusters at all redshifts, from $z=0$ to $z=2$. We fit the
  combined sample of about 600 objects adopting the equation (\ref{11})
where the evolution factor $\gamma$, the slope $\beta$, and the
normalization $C$ are all free to vary, while $E(z)$ is
  represented by a vector whose length is equal to the number of
  clusters and each element is computed at the redshift of
  the corresponding object.  The best-fitting parameters and their 1$\sigma$
errors are reported in Table~\ref{tb8}.

In Fig.\ref{fig:7}, we show the differences between this
  best-fitting relation (of the combined sample) and the relations
  that we previously presented in Tables~2 and 3 and that were
  evaluated in single redshift bins. The brown line is for the
  comparison with the $z=2$ bin and it highlights how much the
  high-$z$ clusters are misrepresented by enforcing a single slope
  across all redshifts.  The most severe differences are present for
the $L-M$ relation where an offset affects not only the highest
redshift bins but also the local measures. The $M-\mg$, whose $\beta$
value is mostly constant up to redshift 1, presents significant
changes only at $z=1.5$.
\vspace{0.2cm}

For the cases where the slope does not substantially vary with time
(in terms of sigma or absolute value) such as $M-T$, $M-\yx$, and
$L-T$, the evolution in their normalization is a well-defined
measurement and the usage of a unique fitting function is well
justified{\footnote All temperatures are measured after excluding the core region. 
 With the inclusion of the cores, the difference in the
  normalization, $C$, of the mass-temperature relations is below 2 per cent for both relations ($\tmw$
  and $\tsl$) with the exception of $-8$ per cent and $+7$ per cent
  variation on the normalization of the $M-\tsl$ relation at $z=1.5$
  and $z=2$, respectively. 
  The differences in the normalization of the $L-\tsl$ relation  reach 7 and 10 per
  cent at $z=0$ and $1.5$ respectively; otherwise they are smaller
  than 5 per cent. In all circumstances, they are consistent within 1$\sigma$
 with the values  of Table~3.}. 
  In particular, we find that the predicted slope and
evolution of both relations are close to the SS expectations (see Table~1)
with the slight tendency of less negative evolution than SS in the
$M-T$ relation ($\gamma = -0.9$ versus $\gamma_{SS}=-1$) and more
positive evolution than SS in the $L-T$ relation ($\gamma=1.2$ versus
$\gamma_{SS}=1$).

The results for the $L-T$ relation from our \agn\ sample, $E(z)^{1.28
  \pm 0.12}$ and $\beta=2.88$, are in line with a positive evolution
$E(z)^{1.64\pm0.77}$ associated with the slope $\beta =3.08 \pm 0.15$
reported by \cite{2015arXiv151203833G}.  This result is, however, in
contrast with the measurements of the bolometric luminosity, $L_{\rm
  bol}\sim5$--$10 \times 10^{44}$ erg s$^{-1}$ of ISCS J1438+3414 at
$z=1.4$ and JKCS041 at $z=2.2$ by \cite{andreon.etal.2011}, of XDCP
J0044.0-2033 at $z=1.579$ by \cite{tozzi.etal.2015}, and of IDCS
J1426.5+3508 at $z=1.75$ by \cite{brodwin.etal.2016}. If local
relations and SS evolution are assumed, all these objects
appeared underluminous for their measured temperature
($4.9^{+3.4}_{-1.6}$ keV, $7.3^{+6.7}_{-2.6}$ keV, $6.7^{+1.3}_{0.9}$
keV, and $7.6^{+8.7}_{-1.9}$ keV, respectively) or viceversa hotter
for their luminosity.  It is difficult to make a significant
comparison due to the low number of objects, the large error bars, and
the impossibility to evaluate any selection effect. It is,
nevertheless, interesting to notice that all these cases are {\it
  exceptionally} massive and almost twice as hot as our most massive
clusters at $z\ge 1.5$.

\section{Comparison with the literature}

 In this section, we compare our results on the evolution of scaling
 relations to some previous numerical works after noting that a
 straightforward comparison is often difficult for the differences in
 terms of cosmological model, implementation of the ICM physics, and
 sample selection.

A number of studies have employed simulations to investigate the
redshift trend of scaling relations, such as
\cite{2010MNRAS.408.2213S}, \cite{2010ApJ...715.1508S},
\cite{2011MNRAS.416..801F}, \cite{2012ApJ...758...74B},
\cite{pike.etal.2014}, \cite{sembolini.etal.2014},
\cite{lebrun.etal.2016}, and \cite{barnes.etal.2016}.

Our results are consistent with \cite{2011MNRAS.416..801F} who found
no significant redshift trend in the slopes of $M-M_{g}$, $M-T_{mw}$,
and $M-Y_{X}$ up to $z=1$. The samples analysed in that paper were
very similar to ours even if we adopted here a more sophisticated
implementation of the \agn\ modelling and of the hydrodynamic code.

\cite{planelles.etal.2016} showed that in our sample there is little
difference between $Y_X$ and $Y_{SZ}$. Therefore, the results
presented in this paper are also in line with
\cite{2012ApJ...758...74B}, \cite{pike.etal.2014}, and
\cite{sembolini.etal.2014}, who found that the $Y_{SZ}-M$ slope remains
almost constant up to $z=1$.

As remarked in \cite{gaspari.etal.2014}, AGN feedback is a purely
inside-out process affecting primarily the small radii; for massive
systems, the binding energy is so large that only the inner core is
affected ($< 0.1 R_{500}$), thus leaving the integrated properties
within $R_{500}$ essentially unaltered. In the cluster regime, it is
thus not surprising that there is no substantial evolution in the
global properties. Comparing with \cite{gaspari.etal.2014} models, our
dominant AGN feedback appears to be gentle, rather than strong and
impulsive as a quasar blast, further limiting major temporal
fluctuations.

Further comparing with \cite{pike.etal.2014}, we acknowledge an
opposite trend with respect to the slope evolution of the $M-\tsl$
relation: while it is decreasing in our case they find a steeper slope
at higher redshifts for all physics that they explored including the
non-radiative case. This implies that their small mass objects are
hotter than ours (at fixed mass); in the paper, the authors suggest 
 that their results might be affected by their sample that
 was composed by only 30 objects randomly selected to equally 
populate five logarithm mass bins in the range
$10^{14}< (M_{200}/h^{-1} M_{\odot})<10^{15}$.

With respect to \cite{lebrun.etal.2016}, we already detailed some
comparisons regarding the \ovisc\ case and stressed that we agree with
their results. The only exception is the evolution that they show for
the $M-\tsl$ relation. In their cosmological box, they also obtain
colder groups at $z=1.5$ with respect to $z=0$ at fixed total mass but
they do detect only a small change in the $M-T$ relation slope. This
could be due to the position of their most massive systems at $z=2$
that apparently are also colder than their local counterpart. The
difference between ours and their results could be ascribed to the
hydrodynamical code or even to the way of measuring the
spectroscopic-like temperature or a different selection (see
  below).

It is more interesting, here, to stress similarities and differences
between the \agn\ models especially on the $M-\mg$ relation since all
the other relations can be related to this and the already discussed
$M-T$ relation.  The $M-\mg$ best-fitting scaling relation proposed by
\cite{lebrun.etal.2016} passes through the points of our \agn\ set;
however, their conclusions are {\it apparently} opposite than ours
since they claim an opposite trend for the slope (note that their
figure is similar to ours but they consider the $\mg-M$ relation
instead of the $M-\mg$ one). This seemingly paradox is due to a
different choice of sample selection. In that paper, the authors
selected all objects with mass above a certain threshold,
$M_{500}>10^{13} M_{\odot}$, which is significantly smaller
than our lower limit at any redshift.  By including the smallest
groups at $z=0$, the authors are sampling the stable population of
clusters together with the population of small groups that exhibit a
quickly decline in gas fraction with total mass (see
  Fig.~\ref{fig:fg}). At $z=0$, their overall $M-\mg$ slope is,
therefore, shallower than ours, as the smallest-mass systems
  weight significantly more in number and dominate in the fitting
  process. Indeed, both their AGN models have $\beta_{\mg} \sim
0.65-0.7$ while our value of $\beta-{\mg}$ is $0.93$.  At $z>1$ the
majority of their objects are characterized by a milder $\mg-M$
relation because the least massive objects retain more gas than the
$z=0$ smallest groups and the most massive objects are not yet
  formed at those times.  On the other hand, in our \agn\ sample we
always consider the most massive objects at each redshift, this
implies that the mass range selected changes with time. Our $z=2$
sample is mostly on the shallower part of the $M-\mg$ relation while
the opposite is true for the $z=0$ sample.

Concerning the evolution of the normalization, we are consistent with
\cite{lebrun.etal.2016} on the positive evolution of the $M-T_{mw}$
relation, which is caused by the incomplete thermalization of high-$z$
clusters. Our result on the positive evolution of the $L-T_{sl}$ also
agrees with the work from \cite{barnes.etal.2016}.  Finally, our study
confirms that there is no significant trend in the intrinsic scatter
of the non-luminosity scaling relations as found in most studies in
literature (but see \citealt{lebrun.etal.2016}). For the $L-T$,
\cite{barnes.etal.2016} also found that the scatter in the luminosity
decreases towards high redshifts. The similar trend for the $L-M$, in
agreement with our result, is also found in their study, yet less
significant.

 \section{Conclusions} 

In this paper, we address the evolution of scaling relations in
simulated galaxy clusters. Our study is based on an 
extended set of cosmological hydrodynamic simulations of clusters.
The simulations
  are carried out with the GADGET-3 code with an upgraded implementation of SPH and 
  including both stellar and AGN feedback. We selected all clusters
  identified within 29 Lagrangian regions above a mass limit
  whose redshift dependence mimics that of the SPT selection 
  function \citep{bleem.etal.2015}.
   Our sample of simulated clusters, by construction, is
  neither volume- nor mass-limited. For this reason, we stress that our
 results are intended to be read as a discussion of general trends
  rather than providing as a precise list of fitting parameters for scaling relations and corresponding scatter.  We
first examine the reliability of our newly performed \agn\ runs by
comparing our predictions to those derived from local ($z<0.25$) and
intermediate redshift ($0.42 \leq z \leq 0.6$)
observations. Subsequently, we investigate the evolution of six
scaling relations: $M-\mg$, $M-\tmw$, $M-\tsl$, $M-Y_X$, $L-\tsl$, and
$L-M$ by comparing the \agn\ model with two other parallel runs
performed with non-radiative physics or with the inclusion of
radiative cooling, star formation, and stellar feedback, but not AGN
feedback. We characterize how the features of the scaling relations
(namely the slope, intrinsic scatter, and normalization) and the
covariance matrix between couples of signals change as a function of
redshift.  We summarize in the following our main results.
\begin{itemize}
\item The scaling relations from simulations at low and intermediate
  redshifts reproduce reasonably well the observed $M-\mg$, $M-\yx$,
  and $L-M$ relations as well as the observed diversity between CC and NCC
  clusters in the $L-T$ relation.  However, our \agn\ model produces
  lower temperatures than observed resulting in a
  normalization shift of $\sim10$ per cent for the $M-T$ relation in
  comparison to observations. A shift of 30 per cent is also present
  in the luminosity mostly due to the sample selection. 

\item From $z=0$ and $z=1$, we do not detect any appreciable change of
  the slopes of the relations with the exclusion of a 4 per cent of
  the $\beta_{\yx}$ for the \ovisc\ model and an $\sim$ $10-15$ per cent
  change of $\beta_{\tsl}$ for the \ovisc\ and \csf\ runs,
  $\beta_{LT}$ for the \csf\ and \agn\ simulations and $\beta_{LM}$
  for the \agn\ set).

\item At higher redshifts, however, all the relations exhibit some
  degree of evolution with the only exception being the luminosity
  relations and the $M-\mg$ relation of the \ovisc\ models which
  remains unchanged.  In the \agn\ runs, the gas slope, $\beta_{\mg}$,
  at $z=2$ is reduced by $\sim10$ per cent with respect to the
  present-time value. This is caused by the effect of intense high-$z$
  AGN activity that has more impact on the lowest mass systems,
    more numerous in the high-redshift bins for the applied SZ-like
    selection. At $z=2$, a shallower slope is found also for the
  $M-T$ relation that declines by $\sim15$ per cent with respect to
  $z=0$.  The evolution of the $M-\tsl$ is not due to the
    selection, but depends on the fact that the smallest groups have a
    systematically lower temperature due to their incomplete
    thermalization at $z=2$.  The decrease of $Y_X$ ($\sim 8$ per
  cent) and increase of $L-M$ ($\sim20$ per cent) slopes can be
  explained by analytically decomposing their slopes in the two
  contributors: $\beta_{\mg}$ and $\beta_{\tsl}$.  Radiative
    processes reduce the hot gas content by removing the coldest and
    densest gas to produce stars. The impact is stronger for
    group-sized objects. This implies that the $M-\mg$ and $M-T$
    relations deviate in different direction from the self-similarity
    scalings. When combined in the $Y_X$ parameter, the
    SS behaviour is recovered
    \citep{2006ApJ...650..128K}.
\item We do not find any significant redshift trend of the scatter in
  any of the mass-proxy relations. Consistent with previous
  theoretical studies, the $M-\mg$ relation has the smallest
  scatter. Instead, the scatter of the two luminosity relations,
  $L-\tsl$ and $L-M$, is the largest, over the redshift range $[0-2]$.
  The $L-M$ scatter increases with the decrease of redshift
  enlightening the significant impact on the X-ray luminosity by
  recent major mergers \citep{torri.etal.2004}. When a merger occurs,
  the luminosity registers a permanent increase
  \citep{rowley.etal.2004}. The scatter of all relations can be
  well described by a lognormal distribution whose widths are mostly
  constant over the redshift up to $z\sim1.5$.
\item  The inclusion of the core, defined as the inner sphere
  within 0.15 $R_{500}$, does not significantly influence
  the best-fitting values of the slopes and normalizations which are
  consistent within 1$\sigma$ to the respective values obtained in
  the core-excised sample. The scatter, however, grows by $15-20$ per
  cent in the $M-\tsl$ relation and by $20-40$ per cent in the $L-\tsl$
  relation.
\item In the \agn\ run, no correlation is evident between the pair of
  deviation in $\mg$ and $\tsl$ at fixed mass and at all redshift. No
  correlation is registered also between $L$ and $\tsl$ at $z=2$. In
  all other cases, positive correlations are found with Pearson
  coefficients always greater than 0.4.
\item Regarding the study of the evolution of the normalization, we
  stress that in situations where the slopes vary with redshift, such
  as the $M-\mg$ and $L-M$ in the \agn\ runs, the evolution on the
  normalization cannot be uniquely established since the trend
    will depend on the pivotal point used to measure it. Owing to the 
    redshift dependence of slope and normalization of scaling relations, we warn
    against the fitting with a single relation data for objects distributed over 
    a wide redshift range. On the other
  hand, we find that the $M-\tmw$ and $L-\tsl$ relations, whose slopes
  have a milder or no evolution, exhibit a negative and positive
  evolution of the normalization (respectively, $\gamma = -0.9$ and $\gamma=1.2$) for the redshift range $[0-2]$, with
  values for the evolution parameters in line with recent
  observational studies and close to the SS
    predictions.

\item Overall, we confirm that the $M-\yx$ relation evaluated from
  $z=0$ to $z=1$ is the best suited for cosmological studies for the
  combination of its properties: in that redshift range, the slope
  does not vary, the evolution of the normalization can be robustly
  determined, the scatter is small and constant over time and most
  importantly the relation is solid, the closest to a
    SS behaviour, independent from the source of feedback
  either from star, SN, or AGNs \citep[but see][for a different
    result]{lebrun.etal.2016}.
\end{itemize}

On the basis of our analysis of the intrinsic variations of simulated
clusters, we also conclude that pushing the study of the scaling
relations to higher redshifts does not seem to be an advantage because
the intense AGN activity, peaking at $z \sim 2$, could have a
significant impact and produce deviations from the SS
behaviour at redshifts $z>1$. This cosmic epoch is still almost an
unexplored territory where the predictions of higher resolution
simulations can help in designing observational strategies for future
missions.  From an observational perspective, the scaling relations
are expected to be calibrated by measuring the mass from weak-lensing
analyses \citep{marrone.etal.2012,hoekstra.etal.2015,
  sereno_ettori2015,mantz.etal.2016}.  However, even if this procedure
is not expected to introduce a significant averaged mass bias over a
large sample of objects \citep{meneghetti.etal.2010,
  becker_kravtsov,rasia.etal.2012} it will likely enlarge the scatter
of the relations \citep{sereno_ettori.2015} since the lensing mass of
single clusters can be both underestimated and overestimated by a
considerable amount. Finally, attention will have to be devoted to
clusters close to the flux limit threshold
\citep{nord.etal.2008}. Certainly, more efforts will need to be
dedicated to reducing the uncertainties on the mass calibration down
to the few per cent level.

\section*{ACKNOWLEDGEMENTS} 
We are greatly indebted to Volker Springel for giving us access to the
developer version of the GADGET3 code; to the entire MUSIC team for
sending their simulated data, and to Dolag's group who built and
updates the Magneticum Pathfinder Webportal. We would like to thank
Gustavo Yepes, Stefano Ettori, and Mauro Sereno for providing useful
discussions and comments.
We acknowledge support from the following: 
the Hungarian Academy of Sciences for the grant Lend\"ulet LP2016-11 (NT);
PIIF-GA- 2013-627474 (ER); NSF AST-1210973 (ER); 
ASI Grant 2016-24-H.0 (PM); 
the {\it Spanish Ministerio de Econom{\'i}a y  Competitividad} (MINECO) grants AYA2013-48226-C3-2-P and AYA2016-77237-C3-3-P (S.P.);
the Generalitat Valenciana, grant GVACOMP2015-227 (SP); 
the {\it Slovenian Research Agency}, research core funding No. P1-0188 (DF); 
the DFG Cluster of Excellence ``Origin and Structure of the Universe'' (AMB);
the DFG Research Unit 1254 `` Magnetisation of interstellar and intergalactic medium'' (AMB);
CONICET, FonCyT and SeCyT-UNC, Argentina (CRF); 
NASA through Einstein PostDoctoral Fellowship Award Number PF5-160137 and Chandra GO7-18121X issued by the Chandra X-ray Observatory Center, which is operated by the SAO for and on behalf of NASA under contract NAS8-03060 (MG).
This work is also supported by the PRIN-MIUR 201278X4FL ``Evolution of cosmic baryons'' funded by the Italian Ministry of Research, by the PRIN-INAF 2012 grant ``The Universe in a Box: Multi-scale Simulations of Cosmic Structures", by the INDARK INFN grant, by the ``Consorzio per la Fisica di Trieste''.
Simulations were performed using Flux HCP Cluster at the University of Michigan and the Galileo Machine at 
CINECA  with CPU time assigned through ISCRA proposals and an agreement with 
the University of Trieste. 
Data, post-processing, and storage have been done on the CINECA facility PICO, granted to us thanks to our expression of interest. 
\bibliographystyle{mnbst}
\bibliography{paper} 

\begin{thebibliography}{129}
\providecommand{\natexlab}[1]{#1}

\bibitem[{{Akritas} and {Bershady}(1996)}]{1996.apj.470.706}
{Akritas} M.G., {Bershady} M.A., 1996, \apj, 470, 706

\bibitem[{{Allen} et~al.(2011){Allen}, {Evrard}, and {Mantz}}]{allen.etal.2011}
{Allen} S.W., {Evrard} A.E., {Mantz} A.B., 2011, \araa, 49, 409

\bibitem[{{Andreon} and {Hurn}(2013)}]{andreon.etal.2012}
{Andreon} S., {Hurn} M., 2013, Statistical Analysis and Data Mining: The ASA
  Data Science Journal, Vol.~9, Issue 1, p.~15-33, 6, 15

\bibitem[{{Andreon} et~al.(2011){Andreon}, {Trinchieri}, and
  {Pizzolato}}]{andreon.etal.2011}
{Andreon} S., {Trinchieri} G., {Pizzolato} F., 2011, \mnras, 412, 2391

\bibitem[{{Ascaso} et~al.(2017){Ascaso}, {Mei}, {Bartlett}, and
  {Ben{\'{\i}}tez}}]{ascaso.etal.2017}
{Ascaso} B., {Mei} S., {Bartlett} J.G., {Ben{\'{\i}}tez} N., 2017, \mnras, 464,
  2270

\bibitem[{{Bah{\'e}} et~al.(2012){Bah{\'e}}, {McCarthy}, and
  {King}}]{bahe.etal.2012}
{Bah{\'e}} Y.M., {McCarthy} I.G., {King} L.J., 2012, \mnras, 421, 1073

\bibitem[{{Barnes} et~al.(2017){Barnes}, {Kay}, {Henson}, {McCarthy}, {Schaye},
  and {Jenkins}}]{barnes.etal.2016}
{Barnes} D.J., {Kay} S.T., {Henson} M.A., {McCarthy} I.G., {Schaye} J.,
  {Jenkins} A., 2017, \mnras, 465, 213

\bibitem[{{Battaglia} et~al.(2012){Battaglia}, {Bond}, {Pfrommer}, and
  {Sievers}}]{2012ApJ...758...74B}
{Battaglia} N., {Bond} J.R., {Pfrommer} C., {Sievers} J.L., 2012, \apj, 758, 74

\bibitem[{{Battaglia} et~al.(2013){Battaglia}, {Bond}, {Pfrommer}, and
  {Sievers}}]{battaglia.etal.2013}
{Battaglia} N., {Bond} J.R., {Pfrommer} C., {Sievers} J.L., 2013, \apj, 777,
  123

\bibitem[{{Beck} et~al.(2016)}]{2015arXiv150207358B}
{Beck} A.M., et~al., 2016, \mnras, 455, 2110

\bibitem[{{Becker} and {Kravtsov}(2011)}]{becker_kravtsov}
{Becker} M.R., {Kravtsov} A.V., 2011, \apj, 740, 25

\bibitem[{{Benson} et~al.(2014)}]{benson.etal.2014}
{Benson} B.A., et~al., 2014, in \emph{Millimeter, Submillimeter, and
  Far-Infrared Detectors and Instrumentation for Astronomy VII}, vol. 9153 of
  \emph{\procspie}, p. 91531P

\bibitem[{{Biffi} et~al.(2014){Biffi}, {Sembolini}, {De Petris}, {Valdarnini},
  {Yepes}, and {Gottl{\"o}ber}}]{biffi.etal.2014}
{Biffi} V., {Sembolini} F., {De Petris} M., {Valdarnini} R., {Yepes} G.,
  {Gottl{\"o}ber} S., 2014, \mnras, 439, 588

\bibitem[{{Biffi} et~al.(2016)}]{biffi.etal.2016}
{Biffi} V., et~al., 2016, \apj, 827, 112

\bibitem[{{Biffi} et~al.(2017)}]{biffi.etal.2017}
{Biffi} V., et~al., 2017, \mnras, 468, 531

\bibitem[{{Bleem} et~al.(2015)}]{bleem.etal.2015}
{Bleem} L.E., et~al., 2015, \apjs, 216, 27

\bibitem[{{Bonafede} et~al.(2011){Bonafede}, {Dolag}, {Stasyszyn}, {Murante},
  and {Borgani}}]{2011MNRAS.418.2234B}
{Bonafede} A., {Dolag} K., {Stasyszyn} F., {Murante} G., {Borgani} S., 2011,
  \mnras, 418, 2234

\bibitem[{{Borgani} and {Guzzo}(2001)}]{borgani_guzzo}
{Borgani} S., {Guzzo} L., 2001, \nat, 409, 39

\bibitem[{{Borgani} and {Kravtsov}(2011)}]{2011ASL.....4..204B}
{Borgani} S., {Kravtsov} A., 2011, Advanced Science Letters, 4, 204

\bibitem[{{Borm} et~al.(2014){Borm}, {Reiprich}, {Mohammed}, and
  {Lovisari}}]{borm.etal.2014}
{Borm} K., {Reiprich} T.H., {Mohammed} I., {Lovisari} L., 2014, \aap, 567, A65

\bibitem[{{Branchesi} et~al.(2007){Branchesi}, {Gioia}, {Fanti}, and
  {Fanti}}]{branchesi.etal.2007}
{Branchesi} M., {Gioia} I.M., {Fanti} C., {Fanti} R., 2007, \aap, 472, 739

\bibitem[{{Brodwin} et~al.(2016){Brodwin}, {McDonald}, {Gonzalez}, {Stanford},
  {Eisenhardt}, {Stern}, and {Zeimann}}]{brodwin.etal.2016}
{Brodwin} M., {McDonald} M., {Gonzalez} A.H., {Stanford} S.A., {Eisenhardt}
  P.R., {Stern} D., {Zeimann} G.R., 2016, \apj, 817, 122

\bibitem[{{Bryan} and {Norman}(1998)}]{1998ApJ...495...80B}
{Bryan} G.L., {Norman} M.L., 1998, \apj, 495, 80

\bibitem[{{Clerc} et~al.(2014)}]{2014MNRAS.444.2723C}
{Clerc} N., et~al., 2014, \mnras, 444, 2723

\bibitem[{{Dai} et~al.(2010){Dai}, {Bregman}, {Kochanek}, and
  {Rasia}}]{dai.etal.2010}
{Dai} X., {Bregman} J.N., {Kochanek} C.S., {Rasia} E., 2010, \apj, 719, 119

\bibitem[{{Dolag} et~al.(2009){Dolag}, {Borgani}, {Murante}, and
  {Springel}}]{dolag.etal.2009}
{Dolag} K., {Borgani} S., {Murante} G., {Springel} V., 2009, \mnras, 399, 497

\bibitem[{{Dolag} et~al.(2016){Dolag}, {Komatsu}, and
  {Sunyaev}}]{dolag.etal.2016}
{Dolag} K., {Komatsu} E., {Sunyaev} R., 2016, \mnras, 463, 1797

\bibitem[{{Eckert} et~al.(2016)}]{eckert.etal.2015}
{Eckert} D., et~al., 2016, \aap, 592, A12

\bibitem[{{Erben} et~al.(2013)}]{2013MNRAS.433.2545E}
{Erben} T., et~al., 2013, \mnras, 433, 2545

\bibitem[{{Ettori}(2013)}]{ettori2013}
{Ettori} S., 2013, \mnras, 435, 1265

\bibitem[{{Ettori}(2015)}]{ettori2015}
{Ettori} S., 2015, \mnras, 446, 2629

\bibitem[{{Ettori} et~al.(2012){Ettori}, {Rasia}, {Fabjan}, {Borgani}, and
  {Dolag}}]{ettori.etal.2012}
{Ettori} S., {Rasia} E., {Fabjan} D., {Borgani} S., {Dolag} K., 2012, \mnras,
  420, 2058

\bibitem[{{Evrard} et~al.(2014){Evrard}, {Arnault}, {Huterer}, and
  {Farahi}}]{evrard.etal.2014}
{Evrard} A.E., {Arnault} P., {Huterer} D., {Farahi} A., 2014, \mnras, 441, 3562

\bibitem[{{Fabian} et~al.(1994){Fabian}, {Crawford}, {Edge}, and
  {Mushotzky}}]{fabian.etal.1994}
{Fabian} A.C., {Crawford} C.S., {Edge} A.C., {Mushotzky} R.F., 1994, \mnras,
  267, 779

\bibitem[{{Fabjan} et~al.(2010){Fabjan}, {Borgani}, {Tornatore}, {Saro},
  {Murante}, and {Dolag}}]{2010MNRAS.401.1670F}
{Fabjan} D., {Borgani} S., {Tornatore} L., {Saro} A., {Murante} G., {Dolag} K.,
  2010, \mnras, 401, 1670

\bibitem[{{Fabjan} et~al.(2011){Fabjan}, {Borgani}, {Rasia}, {Bonafede},
  {Dolag}, {Murante}, and {Tornatore}}]{2011MNRAS.416..801F}
{Fabjan} D., {Borgani} S., {Rasia} E., {Bonafede} A., {Dolag} K., {Murante} G.,
  {Tornatore} L., 2011, \mnras, 416, 801

\bibitem[{{Gaspari} and {S{\c a}dowski}(2017)}]{gasparo.sadowski.2017}
{Gaspari} M., {S{\c a}dowski} A., 2017, \apj, 837, 149

\bibitem[{{Gaspari} et~al.(2014){Gaspari}, {Brighenti}, {Temi}, and
  {Ettori}}]{gaspari.etal.2014}
{Gaspari} M., {Brighenti} F., {Temi} P., {Ettori} S., 2014, \apjl, 783, L10

\bibitem[{{Gaspari} et~al.(2017){Gaspari}, {Temi}, and
  {Brighenti}}]{gaspari.etal.2017}
{Gaspari} M., {Temi} P., {Brighenti} F., 2017, \mnras, 466, 677

\bibitem[{{Giles} et~al.(2016)}]{2015arXiv151203833G}
{Giles} P.A., et~al., 2016, \aap, 592, A3

\bibitem[{{Giles} et~al.(2017)}]{giles.etal.2017}
{Giles} P.A., et~al., 2017, \mnras, 465, 858

\bibitem[{{Giodini} et~al.(2013){Giodini}, {Lovisari}, {Pointecouteau},
  {Ettori}, {Reiprich}, and {Hoekstra}}]{2013SSRv..177..247G}
{Giodini} S., {Lovisari} L., {Pointecouteau} E., {Ettori} S., {Reiprich} T.H.,
  {Hoekstra} H., 2013, \ssr, 177, 247

\bibitem[{{Haardt} and {Madau}(2001)}]{2001cghr.confE..64H}
{Haardt} F., {Madau} P., 2001, in D.M. {Neumann}, J.T.V. {Tran}, eds.,
  \emph{Clusters of Galaxies and the High Redshift Universe Observed in
  X-rays}, p.~64

\bibitem[{{Hahn} et~al.(2017){Hahn}, {Martizzi}, {Wu}, {Evrard}, {Teyssier},
  and {Wechsler}}]{hahn.etal.2015}
{Hahn} O., {Martizzi} D., {Wu} H.Y., {Evrard} A.E., {Teyssier} R., {Wechsler}
  R.H., 2017, \mnras, 470, 166

\bibitem[{{Heymans} et~al.(2012)}]{2012MNRAS.427..146H}
{Heymans} C., et~al., 2012, \mnras, 427, 146

\bibitem[{{Hilton} et~al.(2012)}]{2012MNRAS.424.2086H}
{Hilton} M., et~al., 2012, \mnras, 424, 2086

\bibitem[{{Hoekstra} et~al.(2015){Hoekstra}, {Herbonnet}, {Muzzin}, {Babul},
  {Mahdavi}, {Viola}, and {Cacciato}}]{hoekstra.etal.2015}
{Hoekstra} H., {Herbonnet} R., {Muzzin} A., {Babul} A., {Mahdavi} A., {Viola}
  M., {Cacciato} M., 2015, \mnras, 449, 685

\bibitem[{{Hudson} et~al.(2010){Hudson}, {Mittal}, {Reiprich}, {Nulsen},
  {Andernach}, and {Sarazin}}]{hudson.etal.2010}
{Hudson} D.S., {Mittal} R., {Reiprich} T.H., {Nulsen} P.E.J., {Andernach} H.,
  {Sarazin} C.L., 2010, \aap, 513, A37

\bibitem[{{Isobe} et~al.(1990){Isobe}, {Feigelson}, {Akritas}, and
  {Babu}}]{1990.ApJ.364.104}
{Isobe} T., {Feigelson} E.D., {Akritas} M.G., {Babu} G.J., 1990, \apj, 364, 104

\bibitem[{{Ivezic} et~al.(2008){Ivezic}, {Tyson}, and et~al. {for the LSST
  Collaboration}}]{Ivezic.etal.2008}
{Ivezic} Z., {Tyson} J.A., et~al. {for the LSST Collaboration}, 2008, preprint,
  (arXiv:0805.2366)

\bibitem[{{Kaiser}(1986)}]{kaiser86}
{Kaiser} N., 1986, \mnras, 222, 323

\bibitem[{{Kay} et~al.(2012){Kay}, {Peel}, {Short}, {Thomas}, {Young},
  {Battye}, {Liddle}, and {Pearce}}]{kay.etal.2012}
{Kay} S.T., {Peel} M.W., {Short} C.J., {Thomas} P.A., {Young} O.E., {Battye}
  R.A., {Liddle} A.R., {Pearce} F.R., 2012, \mnras, 422, 1999

\bibitem[{{Kelly}(2007)}]{2007ApJ...665.1489K}
{Kelly} B.C., 2007, \apj, 665, 1489

\bibitem[{{Khatri} and {Gaspari}(2016)}]{khatri.gaspari.2016}
{Khatri} R., {Gaspari} M., 2016, \mnras, 463, 655

\bibitem[{{Komatsu} et~al.(2011)}]{komatsu.etal.2011}
{Komatsu} E., et~al., 2011, \apjs, 192, 18

\bibitem[{{Kravtsov} et~al.(2006){Kravtsov}, {Vikhlinin}, and
  {Nagai}}]{2006ApJ...650..128K}
{Kravtsov} A.V., {Vikhlinin} A., {Nagai} D., 2006, \apj, 650, 128

\bibitem[{{Laureijs} et~al.(2011)}]{laureijs.etal.2011}
{Laureijs} R., et~al., 2011, preprint, (arXiv:1110.3193)

\bibitem[{{Le Brun} et~al.(2014){Le Brun}, {McCarthy}, {Schaye}, and
  {Ponman}}]{lebrun.etal.2014}
{Le Brun} A.M.C., {McCarthy} I.G., {Schaye} J., {Ponman} T.J., 2014, \mnras,
  441, 1270

\bibitem[{{Le Brun} et~al.(2016){Le Brun}, {McCarthy}, {Schaye}, and
  {Ponman}}]{lebrun.etal.2016}
{Le Brun} A.M.C., {McCarthy} I.G., {Schaye} J., {Ponman} T.J., 2016, \mnras

\bibitem[{{Lieu} et~al.(2016)}]{2015arXiv151203857L}
{Lieu} M., et~al., 2016, \aap, 592, A4

\bibitem[{{Mahdavi} et~al.(2013){Mahdavi}, {Hoekstra}, {Babul}, {Bildfell},
  {Jeltema}, and {Henry}}]{2013ApJ...767..116M}
{Mahdavi} A., {Hoekstra} H., {Babul} A., {Bildfell} C., {Jeltema} T., {Henry}
  J.P., 2013, \apj, 767, 116

\bibitem[{{Mahdavi} et~al.(2014){Mahdavi}, {Hoekstra}, {Babul}, {Bildfell},
  {Jeltema}, and {Henry}}]{mah_erratum}
{Mahdavi} A., {Hoekstra} H., {Babul} A., {Bildfell} C., {Jeltema} T., {Henry}
  J.P., 2014, \apj, 794, 175

\bibitem[{{Mantz} et~al.(2016)}]{mantz.etal.2016}
{Mantz} A.B., et~al., 2016, \mnras, 463, 3582

\bibitem[{{Markevitch}(1998)}]{maxim98}
{Markevitch} M., 1998, \apj, 504, 27

\bibitem[{{Marrone} et~al.(2012)}]{marrone.etal.2012}
{Marrone} D.P., et~al., 2012, \apj, 754, 119

\bibitem[{{Martizzi} et~al.(2014){Martizzi}, {Mohammed}, {Teyssier}, and
  {Moore}}]{martizzi.etal.2014}
{Martizzi} D., {Mohammed} I., {Teyssier} R., {Moore} B., 2014, \mnras, 440,
  2290

\bibitem[{{Maughan}(2014)}]{2014MNRAS.437.1171M}
{Maughan} B.J., 2014, \mnras, 437, 1171

\bibitem[{{Maughan} et~al.(2012){Maughan}, {Giles}, {Randall}, {Jones}, and
  {Forman}}]{2012MNRAS.421.1583M}
{Maughan} B.J., {Giles} P.A., {Randall} S.W., {Jones} C., {Forman} W.R., 2012,
  \mnras, 421, 1583

\bibitem[{{Maughan} et~al.(2016){Maughan}, {Giles}, {Rines}, {Diaferio},
  {Geller}, {Van Der Pyl}, and {Bonamente}}]{maughan.etal.2015}
{Maughan} B.J., {Giles} P.A., {Rines} K.J., {Diaferio} A., {Geller} M.J., {Van
  Der Pyl} N., {Bonamente} M., 2016, \mnras, 461, 4182

\bibitem[{{Mazzotta} et~al.(2004){Mazzotta}, {Rasia}, {Moscardini}, and
  {Tormen}}]{2004MNRAS.354...10M}
{Mazzotta} P., {Rasia} E., {Moscardini} L., {Tormen} G., 2004, \mnras, 354, 10

\bibitem[{{McCarthy} et~al.(2011){McCarthy}, {Schaye}, {Bower}, {Ponman},
  {Booth}, {Dalla Vecchia}, and {Springel}}]{mccarthy.etal.2011}
{McCarthy} I.G., {Schaye} J., {Bower} R.G., {Ponman} T.J., {Booth} C.M., {Dalla
  Vecchia} C., {Springel} V., 2011, \mnras, 412, 1965

\bibitem[{{McDonald} et~al.(2013)}]{mcdonald.etal.2013}
{McDonald} M., et~al., 2013, \apj, 774, 23

\bibitem[{{McNamara} and {Nulsen}(2012)}]{mcnamara.nulsen.2012}
{McNamara} B.R., {Nulsen} P.E.J., 2012, New Journal of Physics, 14, 055023

\bibitem[{{Menanteau} et~al.(2013)}]{menanteau.etal.2013}
{Menanteau} F., et~al., 2013, \apj, 765, 67

\bibitem[{{Meneghetti} et~al.(2010){Meneghetti}, {Rasia}, {Merten},
  {Bellagamba}, {Ettori}, {Mazzotta}, {Dolag}, and
  {Marri}}]{meneghetti.etal.2010}
{Meneghetti} M., {Rasia} E., {Merten} J., {Bellagamba} F., {Ettori} S.,
  {Mazzotta} P., {Dolag} K., {Marri} S., 2010, \aap, 514, A93

\bibitem[{{Merloni} et~al.(2012)}]{2012arXiv1209.3114M}
{Merloni} A., et~al., 2012, preprint, (arXiv:1209.3114)

\bibitem[{{Muldrew} et~al.(2015){Muldrew}, {Hatch}, and
  {Cooke}}]{Muldrew.etal.2016}
{Muldrew} S.I., {Hatch} N.A., {Cooke} E.A., 2015, \mnras, 452, 2528

\bibitem[{{Murray} et~al.(2013){Murray}, {Power}, and
  {Robotham}}]{murray.etal.2013}
{Murray} S.G., {Power} C., {Robotham} A.S.G., 2013, Astronomy and Computing, 3,
  23

\bibitem[{{Nagai} et~al.(2007{\natexlab{a}}){Nagai}, {Kravtsov}, and
  {Vikhlinin}}]{nagai.etal.2007}
{Nagai} D., {Kravtsov} A.V., {Vikhlinin} A., 2007{\natexlab{a}}, \apj, 668, 1

\bibitem[{{Nagai} et~al.(2007{\natexlab{b}}){Nagai}, {Vikhlinin}, and
  {Kravtsov}}]{2007ApJ...655...98N}
{Nagai} D., {Vikhlinin} A., {Kravtsov} A.V., 2007{\natexlab{b}}, \apj, 655, 98

\bibitem[{{Nord} et~al.(2008){Nord}, {Stanek}, {Rasia}, and
  {Evrard}}]{nord.etal.2008}
{Nord} B., {Stanek} R., {Rasia} E., {Evrard} A.E., 2008, \mnras, 383, L10

\bibitem[{{Pierre} et~al.(2016)}]{pierre.etal.2016}
{Pierre} M., et~al., 2016, \aap, 592, A1

\bibitem[{{Pike} et~al.(2014){Pike}, {Kay}, {Newton}, {Thomas}, and
  {Jenkins}}]{pike.etal.2014}
{Pike} S.R., {Kay} S.T., {Newton} R.D.A., {Thomas} P.A., {Jenkins} A., 2014,
  \mnras, 445, 1774

\bibitem[{{Planelles} et~al.(2013){Planelles}, {Borgani}, {Dolag}, {Ettori},
  {Fabjan}, {Murante}, and {Tornatore}}]{2013MNRAS.431.1487P}
{Planelles} S., {Borgani} S., {Dolag} K., {Ettori} S., {Fabjan} D., {Murante}
  G., {Tornatore} L., 2013, \mnras, 431, 1487

\bibitem[{{Planelles} et~al.(2014){Planelles}, {Borgani}, {Fabjan}, {Killedar},
  {Murante}, {Granato}, {Ragone-Figueroa}, and {Dolag}}]{2014MNRAS.438..195P}
{Planelles} S., {Borgani} S., {Fabjan} D., {Killedar} M., {Murante} G.,
  {Granato} G.L., {Ragone-Figueroa} C., {Dolag} K., 2014, \mnras, 438, 195

\bibitem[{{Planelles} et~al.(2015){Planelles}, {Schleicher}, and
  {Bykov}}]{2015SSRv..188...93P}
{Planelles} S., {Schleicher} D.R.G., {Bykov} A.M., 2015, \ssr, 188, 93

\bibitem[{{Planelles} et~al.(2017)}]{planelles.etal.2016}
{Planelles} S., et~al., 2017, \mnras, 467, 3827

\bibitem[{{Poole} et~al.(2007){Poole}, {Babul}, {McCarthy}, {Fardal},
  {Bildfell}, {Quinn}, and {Mahdavi}}]{poole.etal.2007}
{Poole} G.B., {Babul} A., {McCarthy} I.G., {Fardal} M.A., {Bildfell} C.J.,
  {Quinn} T., {Mahdavi} A., 2007, \mnras, 380, 437

\bibitem[{{Puchwein} et~al.(2008){Puchwein}, {Sijacki}, and
  {Springel}}]{puchwein.etal.2008}
{Puchwein} E., {Sijacki} D., {Springel} V., 2008, \apjl, 687, L53

\bibitem[{{Ragagnin} et~al.(2017){Ragagnin}, {Dolag}, {Biffi}, {Cadolle Bel},
  {Hammer}, {Krukau}, {Petkova}, and {Steinborn}}]{ragagnin.etal.2016}
{Ragagnin} A., {Dolag} K., {Biffi} V., {Cadolle Bel} M., {Hammer} N.J.,
  {Krukau} A., {Petkova} M., {Steinborn} D., 2017, Astronomy and Computing, 20,
  52

\bibitem[{{Rasia} et~al.(2004){Rasia}, {Tormen}, and
  {Moscardini}}]{rasia.etal.2004}
{Rasia} E., {Tormen} G., {Moscardini} L., 2004, \mnras, 351, 237

\bibitem[{{Rasia} et~al.(2011){Rasia}, {Mazzotta}, {Evrard}, {Markevitch},
  {Dolag}, and {Meneghetti}}]{rasia.etal.2011}
{Rasia} E., {Mazzotta} P., {Evrard} A., {Markevitch} M., {Dolag} K.,
  {Meneghetti} M., 2011, \apj, 729, 45

\bibitem[{{Rasia} et~al.(2012)}]{rasia.etal.2012}
{Rasia} E., et~al., 2012, New Journal of Physics, 14, 055018

\bibitem[{{Rasia} et~al.(2015)}]{2015ApJ...813L..17R}
{Rasia} E., et~al., 2015, \apjl, 813, L17

\bibitem[{{Reichert} et~al.(2011){Reichert}, {B{\"o}hringer}, {Fassbender}, and
  {M{\"u}hlegger}}]{2011A&A...535A...4R}
{Reichert} A., {B{\"o}hringer} H., {Fassbender} R., {M{\"u}hlegger} M., 2011,
  \aap, 535, A4

\bibitem[{{Rowley} et~al.(2004){Rowley}, {Thomas}, and
  {Kay}}]{rowley.etal.2004}
{Rowley} D.R., {Thomas} P.A., {Kay} S.T., 2004, \mnras, 352, 508

\bibitem[{{Sartoris} et~al.(2016)}]{sartoris.etal.2016}
{Sartoris} B., et~al., 2016, \mnras, 459, 1764

\bibitem[{{Sembolini} et~al.(2013){Sembolini}, {Yepes}, {De Petris},
  {Gottl{\"o}ber}, {Lamagna}, and {Comis}}]{sembolini.etal.2013}
{Sembolini} F., {Yepes} G., {De Petris} M., {Gottl{\"o}ber} S., {Lamagna} L.,
  {Comis} B., 2013, \mnras, 429, 323

\bibitem[{{Sembolini} et~al.(2014){Sembolini}, {De Petris}, {Yepes}, {Foschi},
  {Lamagna}, and {Gottl{\"o}ber}}]{sembolini.etal.2014}
{Sembolini} F., {De Petris} M., {Yepes} G., {Foschi} E., {Lamagna} L.,
  {Gottl{\"o}ber} S., 2014, \mnras, 440, 3520

\bibitem[{{Sembolini} et~al.(2016{\natexlab{a}})}]{sembolini.etal.2016a}
{Sembolini} F., et~al., 2016{\natexlab{a}}, \mnras, 457, 4063

\bibitem[{{Sembolini} et~al.(2016{\natexlab{b}})}]{sembolini.etal.2016b}
{Sembolini} F., et~al., 2016{\natexlab{b}}, \mnras, 459, 2973

\bibitem[{{Sereno}(2016)}]{sereno2016}
{Sereno} M., 2016, \mnras, 455, 2149

\bibitem[{{Sereno} and {Ettori}(2015{\natexlab{a}})}]{sereno_ettori2015}
{Sereno} M., {Ettori} S., 2015{\natexlab{a}}, \mnras, 450, 3675

\bibitem[{{Sereno} and {Ettori}(2015{\natexlab{b}})}]{sereno_ettori.2015}
{Sereno} M., {Ettori} S., 2015{\natexlab{b}}, \mnras, 450, 3633

\bibitem[{{Short} and {Thomas}(2009)}]{short_thomas_2009}
{Short} C.J., {Thomas} P.A., 2009, \apj, 704, 915

\bibitem[{{Short} et~al.(2010){Short}, {Thomas}, {Young}, {Pearce}, {Jenkins},
  and {Muanwong}}]{2010MNRAS.408.2213S}
{Short} C.J., {Thomas} P.A., {Young} O.E., {Pearce} F.R., {Jenkins} A.,
  {Muanwong} O., 2010, \mnras, 408, 2213

\bibitem[{{Simionescu} et~al.(2017){Simionescu}, {Werner}, {Mantz}, {Allen},
  and {Urban}}]{simionescu.etal.2017}
{Simionescu} A., {Werner} N., {Mantz} A., {Allen} S.W., {Urban} O., 2017,
  \mnras, 469, 1476

\bibitem[{{Smith} et~al.(2001){Smith}, {Brickhouse}, {Liedahl}, and
  {Raymond}}]{2001ApJ...556L..91S}
{Smith} R.K., {Brickhouse} N.S., {Liedahl} D.A., {Raymond} J.C., 2001, \apjl,
  556, L91

\bibitem[{{Springel}(2005)}]{springel05}
{Springel} V., 2005, \mnras, 364, 1105

\bibitem[{{Springel} and {Hernquist}(2003)}]{2003MNRAS.339..289S}
{Springel} V., {Hernquist} L., 2003, \mnras, 339, 289

\bibitem[{{Springel} et~al.(2005){Springel}, {Di Matteo}, and
  {Hernquist}}]{springel.dimatteo.hernquist.2005}
{Springel} V., {Di Matteo} T., {Hernquist} L., 2005, \mnras, 361, 776

\bibitem[{{Stanek} et~al.(2006){Stanek}, {Evrard}, {B{\"o}hringer},
  {Schuecker}, and {Nord}}]{stanek.etal.2006}
{Stanek} R., {Evrard} A.E., {B{\"o}hringer} H., {Schuecker} P., {Nord} B.,
  2006, \apj, 648, 956

\bibitem[{{Stanek} et~al.(2010){Stanek}, {Rasia}, {Evrard}, {Pearce}, and
  {Gazzola}}]{2010ApJ...715.1508S}
{Stanek} R., {Rasia} E., {Evrard} A.E., {Pearce} F., {Gazzola} L., 2010, \apj,
  715, 1508

\bibitem[{{Steinborn} et~al.(2015){Steinborn}, {Dolag}, {Hirschmann}, {Prieto},
  and {Remus}}]{steinborn.etal.2015}
{Steinborn} L.K., {Dolag} K., {Hirschmann} M., {Prieto} M.A., {Remus} R.S.,
  2015, \mnras, 448, 1504

\bibitem[{{Sun} et~al.(2009){Sun}, {Voit}, {Donahue}, {Jones}, {Forman}, and
  {Vikhlinin}}]{sun.etal.2009}
{Sun} M., {Voit} G.M., {Donahue} M., {Jones} C., {Forman} W., {Vikhlinin} A.,
  2009, \apj, 693, 1142

\bibitem[{{Takey} et~al.(2013){Takey}, {Schwope}, and
  {Lamer}}]{takey.etal.2013}
{Takey} A., {Schwope} A., {Lamer} G., 2013, \aap, 558, A75

\bibitem[{{Tormen} et~al.(1997){Tormen}, {Bouchet}, and
  {White}}]{1997MNRAS.286..865T}
{Tormen} G., {Bouchet} F.R., {White} S.D.M., 1997, \mnras, 286, 865

\bibitem[{{Tornatore} et~al.(2007){Tornatore}, {Borgani}, {Dolag}, and
  {Matteucci}}]{2007MNRAS.382.1050T}
{Tornatore} L., {Borgani} S., {Dolag} K., {Matteucci} F., 2007, \mnras, 382,
  1050

\bibitem[{{Torri} et~al.(2004){Torri}, {Meneghetti}, {Bartelmann},
  {Moscardini}, {Rasia}, and {Tormen}}]{torri.etal.2004}
{Torri} E., {Meneghetti} M., {Bartelmann} M., {Moscardini} L., {Rasia} E.,
  {Tormen} G., 2004, \mnras, 349, 476

\bibitem[{{Tozzi} et~al.(2015)}]{tozzi.etal.2015}
{Tozzi} P., et~al., 2015, \apj, 799, 93

\bibitem[{{Vikhlinin}(2006)}]{vikh2006}
{Vikhlinin} A., 2006, \apj, 640, 710

\bibitem[{{Vikhlinin} et~al.(2009)}]{2009ApJ...692.1033V}
{Vikhlinin} A., et~al., 2009, \apj, 692, 1033

\bibitem[{{Villaescusa-Navarro} et~al.(2016)}]{villaescusa.etal.2016}
{Villaescusa-Navarro} F., et~al., 2016, \mnras, 456, 3553

\bibitem[{{Voit}(2005)}]{voit2005}
{Voit} G.M., 2005, Reviews of Modern Physics, 77, 207

\bibitem[{{von der Linden} et~al.(2014)}]{vonderlinden.etal.2014}
{von der Linden} A., et~al., 2014, \mnras, 443, 1973

\bibitem[{{Watson} et~al.(2013){Watson}, {Iliev}, {D'Aloisio}, {Knebe},
  {Shapiro}, and {Yepes}}]{watson.etal.2013}
{Watson} W.A., {Iliev} I.T., {D'Aloisio} A., {Knebe} A., {Shapiro} P.R.,
  {Yepes} G., 2013, \mnras, 433, 1230

\bibitem[{{Weinberg} et~al.(2013){Weinberg}, {Mortonson}, {Eisenstein},
  {Hirata}, {Riess}, and {Rozo}}]{weinberg.etal.2013}
{Weinberg} D.H., {Mortonson} M.J., {Eisenstein} D.J., {Hirata} C., {Riess}
  A.G., {Rozo} E., 2013, \physrep, 530, 87

\bibitem[{{Wiersma} et~al.(2009){Wiersma}, {Schaye}, {Theuns}, {Dalla Vecchia},
  and {Tornatore}}]{2009MNRAS.399..574W}
{Wiersma} R.P.C., {Schaye} J., {Theuns} T., {Dalla Vecchia} C., {Tornatore} L.,
  2009, \mnras, 399, 574

\bibitem[{{Wu} et~al.(2015){Wu}, {Evrard}, {Hahn}, {Martizzi}, {Teyssier}, and
  {Wechsler}}]{wu.etal.2015}
{Wu} H.Y., {Evrard} A.E., {Hahn} O., {Martizzi} D., {Teyssier} R., {Wechsler}
  R.H., 2015, \mnras, 452, 1982

\end{thebibliography}

\appendix
\section{Decomposition of the slopes of  composite signals}

In this section we derive the features of $M-\yx$, $L-\tsl$, and $L-M$ based on the two fundamental relations: $M-{\mg}$ and $M-\tsl$:

\begin{equation}
M \propto E(z)^{\gamma_{\mg}} \times \mg^{\beta_{\mg}}  \to  \mg \propto E(z)^{-\frac{\gamma_{\mg}}{\beta_{\mg}}}\times M^{\frac{1}{\beta_{\mg}}}
\end{equation}
\begin{equation}
M \propto E(z)^{\gamma_{\tsl}}\times {\tsl}^{\beta_{\tsl}}  \to  \tsl \propto E(z)^{-\frac{\gamma_{\tsl}}{\beta_{\tsl}}}\times M^{\frac{1}{\beta_{\tsl}}}.
\end{equation}

\smallskip
\noindent {\it The $M-\yx$ relation.}
By definition the $\yx$ is the product of gas mass and temperature, and thus
\begin{equation}
Y_{X} = \mg \times \tsl.
\end{equation} 
 From equations (A1)-(A3), one deduces
\begin{eqnarray}
Y_{X}\propto M^{\frac{1}{\beta_{\mg}}}\times E(z)^{-\frac{\gamma_{\mg}}{\beta_{\mg}}}\times M^{\frac{1}{\beta_{\tsl}}}\times E(z)^{-\frac{\gamma_{\tsl}}{\beta_{\tsl}}}.
\end{eqnarray}
Writing $M-Y_X$ in general form
\begin{equation}
M = C_{\yx}\times E(z)^{\gamma_{\yx}}\times\bigg(\frac{Y_X}{Y_{X0}}\bigg)^{\beta_{\yx}},
\end{equation}
the slope and power of evolution of the $M-Y$ are given by
\begin{equation}
\beta_{\yx}=\frac{1}{1/\beta_{\mg}+1/\beta_{\tsl}},
\end{equation}
\begin{equation}
\gamma_{\yx} = \frac{\gamma_{\mg}/\beta_{\mg}+\gamma_{\tsl}/\beta_{\tsl}}{1/\beta_{\mg}+1/\beta_{\tsl}}.
\end{equation}

\smallskip
\noindent {\it The $L-\tsl$ relation.} The general form of $L-\tsl$ is given by
\begin{equation}
L = C_{LT}\times E(z)^{\gamma_{LT}}\times \bigg(\frac{\tsl}{T_0}\bigg)^{\beta_{LT}}.
\end{equation}
Recalling that $L$ is related to the total mass and temperature via equation (\ref{8})
\begin{eqnarray}
L &\propto& E(z)^2f_g^2\times M\times T_{sl}^{1/2}\nonumber\\
&\propto& E(z)^2\times\frac{M_g^2}{M}\times T_{sl}^{1/2}\nonumber\\
&\propto& E(z)^2\times E(z)^{-\frac{2\gamma_{M_g}}{\beta_{M_g}}}\times M^{\frac{2}{\beta_{M_g}}-1}\times T_{sl}^{1/2}.
\end{eqnarray}
In the last derivation  we use equation (A1) to rewrite $M_g$ in terms of $M$. In turn, by substituting $M$ in terms of $T_{sl}$  via equation (A2) one obtains
\begin{eqnarray}
L&\propto& E(z)^{2(1-\frac{\gamma_{M_g}}{\beta_{M_g}})}\times E(z)^{\gamma_{T_{sl}}\times(\frac{2}{\beta_{Mg}}-1)}\times T_{sl}^{\beta_{T_{sl}}(\frac{2}{\beta_{M_g}}-1)+\frac{1}{2}}\nonumber\\
&\propto& E(z)^{2(1-\frac{\gamma_{M_g}}{\beta_{M_g}})+\gamma_{T_{sl}}(\frac{2}{\beta_{M_g}}-1)}\times T_{sl}^{\beta_{T_{sl}}(\frac{2}{\beta_{M_g}}-1)+\frac{1}{2}}.
\end{eqnarray}
 By comparing equation (A10) to equation (A8) one deduces
\begin{equation}
\beta_{LT}=\beta_{\tsl}\bigg(\frac{2}{\beta_{\mg}}-1\bigg)+\frac{1}{2},
\end{equation}
\begin{equation}
\gamma_{LT}=2\bigg(1-\frac{\gamma_{\mg}}{\beta_{\mg}}\bigg)+\gamma_{\tsl}\bigg(\frac{2}{\beta_{\mg}}-1\bigg).
\end{equation}

\smallskip
\noindent {\it The $L-M$ relation.} We define the general form of $L-M$ as
\begin{equation}
L = C_{LM}\times E(z)^{\gamma_{LM}}\times \bigg(\frac{M}{M_0}\bigg)^{\beta_{LM}}.
\end{equation}
 In equation (A9), we rewrite $T_{sl}$ in term of $M$ via equation (A2) and obtain
\begin{eqnarray}
L&\propto& E(z)^{2}\times E(z)^{\frac{-2\gamma_{M_g}}{\beta_{M_g}}}\times M^{\frac{2}{\beta_{M_g}}-1}\times E(z)^{-\frac{\gamma_{T_{sl}}}{2\beta_{T_{sl}}}}\times M^{\frac{1}{2\beta_{T_{sl}}}}\nonumber\\
L&\propto& E(z)^{2(1-\frac{\gamma_{M_g}}{\beta_{M_g}})-\frac{\gamma_{T_{sl}}}{2\beta_{T_{sl}}}}\times M^{\frac{2}{\beta_{M_g}}+\frac{1}{2\beta_{T_{sl}}}-1}.
\end{eqnarray}
From the last equation we can derive the slope and evolution of the $LM$ relation given as
\begin{equation}
\beta_{LM}=\frac{2}{\beta_{\mg}}+\frac{1}{2\beta_{\tsl}}-1,
\end{equation}
\begin{equation}
\gamma_{LM}=2\bigg(1-\frac{\gamma_{\mg}}{\beta_{\mg}}\bigg)-\frac{1}{2}\frac{\gamma_{\tsl}}{\beta_{\tsl}}.
\end{equation}

\end{document}